\documentclass[aps,prx,twocolumn,superscriptaddress,amsmath,amssymb,10pt, epsfig,nofootinbib]{revtex4-2}

\usepackage{appendix}
\usepackage{amssymb, amsmath,amsthm}    
\usepackage{graphicx}   
\usepackage{verbatim}   
\usepackage{color}      
\usepackage{xcolor}
\usepackage{subcaption} 
\usepackage{bbm}        
\usepackage{caption}
\usepackage{tikz}
\usetikzlibrary{tikzmark}
\usepackage{pgflibraryarrows}
\usepackage{pgflibrarysnakes}
\usepackage{lipsum}
\usepackage{mathtools}
\usepackage{dirtytalk}
\usepackage{ae} 
\usepackage[T1]{fontenc}
\usepackage[pdfstartview=FitV,linktocpage,colorlinks=true,linkcolor=darkblue,citecolor=darkblue,urlcolor=darkblue]{hyperref}
\usepackage{epsfig}
\usepackage{enumerate}
\usepackage{varioref}
\usepackage{cleveref}
\usepackage{makeidx}
\makeindex
\usepackage[english]{babel}
\usepackage[normalem]{ulem}
\usepackage{physics}
\usepackage{bbold}
\usepackage{float}
\usepackage{footnote}
\usepackage{gensymb}

\usepackage{algorithm}
\usepackage{algpseudocode}
\algblock{Input}{EndInput}
\algnotext{EndInput}
\algblock{Output}{EndOutput}
\algnotext{EndOutput}

\newcommand{\beq}{\begin{equation}}
\newcommand{\eeq}{\end{equation}}
\newcommand{\beqq}{\begin{equation*}}
\newcommand{\eeqq}{\end{equation*}}
\newcommand\bea{\begin{array}}
\newcommand\eea{\end{array}}
\newcommand\beaa{\begin{array*}}
\newcommand\eeaa{\end{array*}}
\newcommand\beal{\begin{align}}
\newcommand\eeal{\end{align}}
\newcommand\beall{\begin{align*}}
\newcommand\eeall{\end{align*}}

\def\o{{\omega}}
\def\a{{\alpha}}
\def\L{\Lambda}

\def\s{\sigma}
\def\th{\theta}

\def\nleap{n_{\mathrm{leap}}}

\newcommand{\eps}{\varepsilon}

\newcommand{\f}{\frac}

\newcommand{\td}{\tilde}

\def\[{\left[}
\def\]{\right]}
\def\({\left(}
\def\){\right)}
\def\<{\langle}
\def\>{\rangle}

\definecolor{darkblue}{cmyk}{0.9,0.9,0,0}
\definecolor{greennote}{RGB}{0,135,41}

\begin{document}
	\title{Non-Hertz-Millis scaling of the antiferromagnetic quantum critical metal via scalable Hybrid Monte Carlo}
	\author{Peter Lunts}
	\email{plunts@umd.edu}
	\affiliation{Joint Quantum Institute and Department of Physics, University of Maryland, College Park, MD 20742, USA}
	\affiliation{Center for Computational Quantum Physics, Flatiron Institute, 162 5th Avenue, New York, NY 10010, USA}
	\author{Michael~S.~Albergo}
	\thanks{These authors contributed equally}
	\affiliation{Center for Cosmology and Particle Physics, New York University, New York, NY 10003, USA}
	\author{Michael Lindsey}
	\thanks{These authors contributed equally}
	\affiliation{Courant Institute of Mathematical Sciences, New York University, New York, New York 10012, USA}

	\date{\today}
	
\begin{abstract}

A key component of the phase diagram of many iron-based superconductors and electron-doped cuprates is believed to be a quantum critical point (QCP), delineating the onset of antiferromagnetic spin-density wave order in a quasi-two-dimensional metal. The universality class of this QCP is believed to play a fundamental role in the description of the proximate non-Fermi liquid and superconducting phases. A minimal model for this transition is the $\mathrm{O}(3)$ spin-fermion model. Despite many efforts, a definitive characterization of its universal properties is still lacking. Here, we numerically study the $\mathrm{O}(3)$ spin-fermion model and extract the scaling exponents and functional form of the static and zero-momentum dynamical spin susceptibility. We do this using a Hybrid Monte Carlo (HMC) algorithm with a novel auto-tuning procedure, which allows us to study unprecedentedly large systems of $80 \times 80$ sites. We find a strong violation of the Hertz-Millis form, contrary to all previous results. Furthermore, the form that we do observe provides good evidence that the universal scaling is actually governed by the analytically tractable fixed point discovered near perfect ``hot-spot'" nesting, even for a larger nesting window. Our predictions can be directly tested with neutron scattering. Additionally, the HMC method we introduce is generic and can be used to study other fermionic models of quantum criticality, where there is a strong need to simulate large systems.

\end{abstract}

\maketitle

\section{Introduction}

Quantum critical phenomena play an important role in condensed matter physics~\cite{sachdev_2011}. Of particular interest are quantum phase transitions in metals, since they are ubiquitous in strongly correlated materials displaying exotic quantum phenomena, most notably high-temperature superconductivity. 
\\ \indent 
These phase transitions are notoriously difficult to study theoretically~\cite{doi:10.1146/annurev-conmatphys-031016-025531}. This is due to the presence of an extensive number of gapless fermionic modes on the Fermi surface and the strong coupling between these modes and the transition order parameter. These difficulties render nearly all analytical perturbative approaches uncontrolled. On the numerical side, these difficulties are manifested in a large amount of entanglement and nearly ubiquitous sign problems, making it hard for controlled numerical techniques such as tensor networks and quantum Monte Carlo (QMC) to make progress. 
\\ \indent 
The most common of such phase transitions in Nature is the onset of antiferromagnetic (AF) spin-density wave (SDW) order in a metal. This transition exists in many material classes of interest, such as electron-doped cuprates~\cite{doi:10.1146/annurev-conmatphys-031119-050558}, iron-based materials~\cite{Green_Fe_review}, and heavy fermion compounds~\cite{doi:10.1146/annurev-conmatphys-062910-140546}, in which it is believed to be a continuous transition. It is often accompanied by a superconducting `dome,' where the maximal $T_c$ occurs near the putative zero-temperature critical point. This makes the (near-critical) SDW fluctuations a strong candidate for the `glue' of Cooper pairs in those materials.
\\ \indent 
Due to its importance, the theory of this phase transition has received a considerable amount of attention in the last three decades \cite{doi:10.1080/0001873021000057123, PhysRevLett.84.5608,PhysRevLett.93.255702,PhysRevB.82.075128,PhysRevB.91.125136,PhysRevB.95.245109,PhysRevX.7.021010, PhysRevB.100.235104}. Very early on it was believed to be well described by Hertz-Millis theory \cite{PhysRevB.14.1165,PhysRevB.48.7183}, although it was soon realized that, due to the dimensionality, the arguments of Hertz-Millis theory are invalid \cite{doi:10.1080/0001873021000057123, PhysRevLett.84.5608,PhysRevLett.93.255702}. Crucially, the large-N expansion was shown to fail \cite{PhysRevB.82.075128}, which left the theory without a controlled approach until the introduction of the fully local $\epsilon$-expansion \cite{PhysRevB.91.125136, PhysRevB.95.245109}. Using these controlled results as guidance, in Ref. \cite{PhysRevX.7.021010} it was shown that there exists a parameter regime where the theory naturally develops a small control parameter, $w$, which is a ratio of velocities, without the need for any dimensional modification. It is then possible to compute observables perturbatively in $w$, giving the only fully controlled analytical calculation of the universal low-energy data for the unmodified problem. However, the parameter regime, or `basin of attraction,' of this solution could not be determined from the arguments of Ref. \cite{PhysRevX.7.021010}. It is therefore not clear whether this solution exists for physically relevant parameter values, or only in a minuscule slice of parameter space. One of the central goals of this work is to answer this question.
\\ \indent 
In the last decade, the SDW transition in metals has also been studied extensively using numerical techniques. The seminal work of Ref. \cite{Berg21122012} introduced a microscopic two-band model of the effective field theory for this transition, which crucially lacks a sign problem. This has led to many studies of this model with Determinantal Quantum Monte Carlo (DQMC) \cite{PhysRevB.95.035124,PhysRevB.95.174520,PhysRevLett.120.247002,PhysRevLett.117.097002,PhysRevResearch.2.023008,Li2016925,PhysRevB.95.214505,doi:10.1146/annurev-conmatphys-031218-013339,doi:10.1146/annurev-conmatphys-033117-054307}, as well as works that have studied other sign-problem-free models of quantum criticality \cite{PhysRevX.6.031028,doi:10.1073/pnas.1620651114,PhysRevX.10.031053,PhysRevX.7.031058,QMC_nFL_Yang,doi:10.1073/pnas.1901751116,PhysRevB.105.L041111}. 
\\ \indent 
A recent such DQMC work \cite{PhysRevResearch.2.023008} focused on the critical scaling of the spin susceptibility, in particular seeking a comparison to the predictions of Ref. \cite{PhysRevX.7.021010} by tuning the UV value of the nesting parameter, $v$ (explained in detail in Sec. \ref{sec:theory_nesting_fixed_point}) close to the value at the fixed point of Ref. \cite{PhysRevX.7.021010}, $v = 0$. However, at criticality, the spin susceptibility was actually observed to have a (nearly-perfect) Hertz-Millis form, contradicting the theoretical finding \cite{doi:10.1080/0001873021000057123, PhysRevLett.84.5608,PhysRevLett.93.255702} that the Hertz-Millis arguments are not valid in two dimensions. Importantly, the maximal system size studied in Ref. \cite{PhysRevResearch.2.023008} was $V = L \times L = 14 \times 14$. With such a small system size, it is often difficult to convincingly extract long-wavelength behavior of a critical system. Therefore, even though Ref. \cite{PhysRevResearch.2.023008} is the current state-of-the-art for this problem, it is very desirable to revisit the problem at much larger $L$. 
\\ \indent 
Although DQMC provides a numerically exact and unbiased way to study the properties of these phase transitions, it is severely hindered in its ability to simulate systems with large spatial volume $V$ by its computational scaling of at least $ \sim \beta V^3$, where $\beta$ is the inverse temperature (in the case of small fermion density, this can be reduced by exploiting the low-rank structure of the fermionic determinant \cite{doi:10.1021/acs.jctc.8b00996, PhysRevLett.123.136402}). The true scaling may in fact be worse due to the need to take smaller steps in high dimensions in order to maintain a nonvanishing Metropolis acceptance probability, as well as the presence of `critical slowing down,' discussed further below. Improving the computational scaling with respect to $V$ is of great interest in the study of quantum criticality. Indeed, in order to extract scaling properties near a quantum critical point (QCP), it is crucial to be `close enough' to the thermodynamic limit $V \rightarrow \infty$. This `close enough' is never possible to determine \emph{a priori}, and, in principle, due to potential semi-stable fixed points, there is never any reason to expect that it has been reached, unless the observed scaling matches a predicted result.
\\ \indent 
In this paper, we use a different QMC method to study the microscopic model of Ref. \cite{Berg21122012}, namely Hybrid Monte Carlo (HMC), which is sometimes referred to as Hamiltonian Monte Carlo. This method is the main numerical tool in the study of lattice quantum chromodynamics (LQCD). Its primary advantage over DQMC is the potential for improved computational scaling with respect to $V$, which we explain in more detail below. HMC has seen a recent revival in condensed matter physics \cite{PhysRevB.36.8632,PhysRevB.38.12023,PhysRevB.97.085144}. Several works have used it to study the half-filled Hubbard model on various lattices (square, honeycomb, hexagonal) \cite{PhysRevB.97.085144,PhysRevB.98.235129,PhysRevB.100.075141,KRIEG201915,PhysRevB.102.245105,PhysRevB.104.155142,ULYBYSHEV2019118}, electron-phonon models \cite{PhysRevB.97.085144,arxiv.2203.01291}, as well as extended and long-range Hubbard models of graphene \cite{PhysRevLett.102.026802,PhysRevB.78.165423,PhysRevLett.111.056801,arxiv.1204.5424,PhysRevB.89.195429}. Nearly all of these results point to extremely favorable scaling with $V$. However, to the best of our knowledge, HMC has not yet been applied to a model of quantum criticality in the presence of a Fermi surface. 
\\ \indent 
Using our large-scale simulations, we find that the critical theory does in fact strongly deviate from the Hertz-Millis prediction. By tuning the nesting parameter, $v$, closer to the fixed-point value of $v = 0$, we observe a systematic reduction of the dynamical critical exponent $z$ below the value of $z=2$ predicted by Hertz-Millis. The prediction of Ref. \cite{PhysRevX.7.021010} is that $z \rightarrow 1^+$ as $v \rightarrow 0$. Additionally, we find that the momentum dependence of the critical spin susceptibility is $\mathrm{O}(2)$-symmetric at intermediate momenta, as predicted by Hertz-Millis, but at lower momenta the symmetry gets reduced to $\mathrm{C}_4$, as predicted in Ref. \cite{PhysRevX.7.021010}. These two findings provide strong numerical evidence that the criticial point theory is governed by the fixed point of Ref. \cite{PhysRevX.7.021010}, even at values of $v$ that are appreciable. This is summarized in Table \ref{tab:main results}.
\begin{table*}[ht]
\centering
    \begin{tabular}{ |c|c|c|c| } 
     \hline
      & $z(\theta)$ & symmetry & functional form \\ 
     \hline
     This work & 
     \begin{tabular}{@{}c@{}}
     $\in (1.665(29),1.953(35))$
     \\ for $\theta \in (0.5 \degree, 8 \degree)$
     \end{tabular}
     & $ \mathrm{C}_4 $ &  
     \begin{tabular}{@{}c@{}} $ \chi^{-1}(\omega) \sim \abs{\omega}^{\Delta} $ 
     \\ 
     $\chi^{-1}(q_x) \sim \abs{q_x}^{z \Delta}$
     \\
     $\chi^{-1}(\textbf q) \sim \abs{q_x}^{z \, \Delta} + \abs{q_y}^{z \, \Delta} $
     \end{tabular}
     \begin{tabular}{@{}c@{}}
     $z\Delta \in (1.466(14),1.521(15))$ 
     \\ 
     $\Delta \in (0.880(12),0.779(12))$
     \\
     for $\theta \in (0.5 \degree, 8 \degree)$
     \end{tabular}
     \\ 
     \hline
     Ref. \cite{PhysRevX.7.021010} & $z \rightarrow 1^+$ as $\theta  \rightarrow 0$ & $\mathrm{C}_4$ &
     \begin{tabular}{@{}c@{}}
     $ \chi^{-1}(\omega) \sim \abs{\omega}^{\Delta}$
     \\
     $\chi^{-1}(q_x) \sim \abs{q_x}^{z \Delta}$
     \\
     $\chi^{-1}(\textbf q) \sim $ not determined
     \end{tabular}
     \begin{tabular}{@{}c@{}}
     $z\Delta, \Delta \rightarrow 1^+$ as $\theta \rightarrow 0$  
     \end{tabular}
     \\ 
     \hline
     Hertz-Millis & $z = 2 \quad \forall \theta$ & $\mathrm{O}(2)$ & 
     \begin{tabular}{@{}c@{}}
     $ \chi^{-1}(\omega) \sim \abs{\omega} $
     \\
     $\chi^{-1}(q_x) \sim q_x^{2}$
     \\
     $\chi^{-1}(\textbf q) \sim q_x^2 + q_y^2 $
     \end{tabular}
     \\
     \hline
    \end{tabular}
    \caption{\textbf{Summary of results.} The main results of this paper, which pertain to the critical spin susceptibility, $\chi(\omega, \textbf q)$. We list its dynamical critical exponent $z$, spatial symmetry, and functional form, as compared to that predicted in Ref. \cite{PhysRevX.7.021010} and by Hertz-Millis theory, all for the varying nesting angle $\theta$ we study (here we use $\theta$ instead of the nesting parameter $v = \tan(\theta)$ that we use in the rest of the paper). The forms quoted for Ref. \cite{PhysRevX.7.021010} are neglecting the logarithmic flow of $v$.}
    \label{tab:main results}
\end{table*}
Our predictions for the dynamical spin susceptibility can be directly tested with neutron scattering.
\\ \indent
In our largest computations we reach system sizes $N_{\tau} \times L \times L$  given by $N_{\tau} = 200$,  $L = 80$ and $N_{\tau} = 800$, $L = 20$, where $N_{\tau}$ is the number of slices in the imaginary time direction. Going to such large scales turns out to be necessary for extracting accurate critical scaling behaviors at the critical point.
\\ \indent
Now we give an overview of the computational scaling properties of our approach that allow us to achieve these results. Within the HMC algorithm, the number of `integration steps' per effective sample can enjoy scaling as low as $O(d^{1/4})$, where $d = O(\beta V)$ is the dimension of our discretized field to be sampled \cite{NealHMC}. This scaling may be worse in the presence of `critical slowing down' for critical models, yielding a number of integration steps per effective sample of $O(\beta^{1/4 + z_1} V^{1/4 + z_2})$, where potentially $z_1, z_2 > 0$. Even in the presence of criticality, $z_1$ and $z_2$ may be reduced via the choice of the `metric' within HMC, as discussed below. For the critical model, our algorithm in fact achieves $z_2 \approx 0$ and $z_1 \lesssim 0.5$. Each of the integration steps within HMC requires the solution of linear system of size $O(\beta V)$, which constitutes the bottleneck for the algorithm.
Our approach for solving the linear system in fact scales linearly with respect to $V$ for fixed $\beta$ and permits fast GPU implementation. In conjunction with the scaling $z_2 \approx 0$ observed above, this performance yields overall wall clock scaling with exponent approximately $5/4$ with respect to $V$. See Section \ref{sec:performance} for further details.
\\ \indent 
In our implementation we in fact develop several augmentations of the basic HMC algorithm. First, we introduce an auto-tuning procedure, which tunes our hyperparameters in an initial warmup phase.
This procedure is common practice in statistics and industry applications of HMC \cite{10.2307/24308995, osti_1430202} but to the best of our knowledge has not yet been fully applied in condensed matter physics. Moreover, relative to such works, our translation-invariant physical setting allows us to tune the aforementioned HMC metric with operations that scale linearly in $\beta V$, up to log factors.
\\ \indent 
In the Results Section,
we introduce the low-energy effective theory of the SDW transition, both in the continuum and on the lattice, we review the theoretical results near perfect nesting of Ref. \cite{PhysRevX.7.021010} and derive its predictions for the observables that are computed numerically in this work, and we show our main results. In the Methods Section we give a brief review of HMC, followed by a detailed presentation of our implementation and its numerical performance.

\section{Results}
\label{sec:results}

\subsection{Continuum effective action}
\label{sec:continuum S_eff}

\begin{figure}[!ht]
    \centering
    \includegraphics[width=0.35\textwidth]{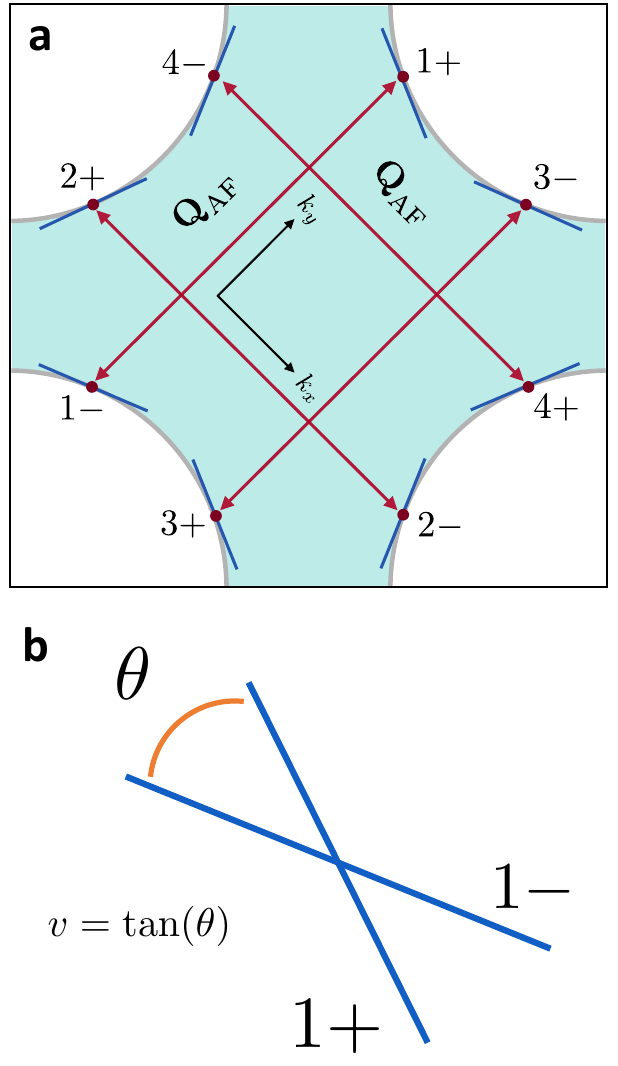}
    \caption{\textbf{The Fermi surface.} (a) The first Brillouin zone of the model in Eq. (\ref{eq:hotspot theory}). Note the axes rotated by $45 \degree$. The shaded region corresponds to the occupied states. The SDW ordering wavevector $\textbf{Q}_{\text{AF}}$ is denoted by red arrows. The hot spots are the red points connected by $\textbf{Q}_{\text{AF}}$. At each hot spot, the linearized Fermi surface is shown with a blue line. (b) A close-up of linearized dispersions of the paired hot spots $(1,\pm)$. One has been shifted by $\textbf{Q}_{\text{AF}}$, to show their crossing. The angle between them $\theta$ is called the `nesting angle', and its tangent $v = \tan(\theta)$ is called the `nesting parameter'.}
    \label{fig:hot spots and nesting angle}
\end{figure}

Our starting point is the theory that describes the low-energy degrees of freedom near a metallic SDW quantum critical point in two spatial dimensions. These are the order parameter for the transition, which is the collective spin excitation, and the electrons near points on the Fermi surface called `hotspots' that are connected by the AF ordering wavevector. For concreteness, we consider a single band with $\mathrm{C}_4$ symmetry and an ordering wavevector equal to $(\pi,\pi)$, as shown in Fig. \ref{fig:hot spots and nesting angle}. Generically for such a Fermi surface there are four pairs of coupled hotspots. The Euclidean-time action for this low-energy theory is then given by
\begin{equation}
\begin{split}
\mathcal{S} = & 
\sum_{n=1}^4 \sum_{m=\pm} \sum_{\sigma=\uparrow,\downarrow} 
\sum_{j=1}^{N_f}
T \sum_{\omega_k}
\int \frac{d\textbf k}{(2\pi)^2} 
\\ & \hspace{20mm}
{\psi}^{(m)*}_{n,\s,j}(k)
\left[ i \omega_k + e^{m}_n(\textbf k;v)  \right] 
\psi^{(m)}_{n,\sigma,j}(k) 
\\ & 
+ T \sum_{\omega_q} \int \frac{d\textbf q}{(2\pi)^2} \,
\f12
\Bigl[
\frac{\omega_q^2}{c^2} + (q_x^2 + q_y^2) + r
\Bigr]
\boldsymbol \phi(q) \cdot \boldsymbol \phi(-q)
\\ & 
+ \frac{g}{\sqrt{N_f}} \sum_{n=1}^4 
\sum_{\s,\s'=\uparrow,\downarrow}
\sum_{j=1}^{N_f}
\int d \tau \sum_{\textbf r}
\\ &
\Bigl[
\boldsymbol{\phi} \cdot 
{\psi}^{(+)*}_{n,\s,j} \boldsymbol{\tau}_{\s,\s'}  
\psi^{(-)}_{n,\s',j}
+ \mathrm{h.c.} \Bigr]
+ \frac{u}{4} \int d \tau \sum_{\textbf r} 
(\boldsymbol{\phi} \cdot \boldsymbol{\phi})^2.
\end{split}
\label{eq:hotspot theory}
\end{equation}
In Eq. (\ref{eq:hotspot theory}), $k = (\omega_k, \textbf{k})$ consists of the fermionic Matsubara frequency and the two-dimensional momentum $\textbf{k} = (k_x, k_y)$. $T$ is the temperature. The $\psi_{n,\s}^{(m)}$ correspond to electrons at the hot spots labeled by $n\in \{1,2,3,4 \}$ and $m \in \{+, - \}$ (cf. Fig. \ref{fig:hot spots and nesting angle}), and spin $\s \in \{ \uparrow, \downarrow \}$. 
The axes are chosen as follows: $\hat{k}_x$ is in the direction from hot spot $(2,+)$ to $(2,-)$ and $\hat{k}_y$ is in the direction from hot spot $(1,-)$ to $(1,+)$. Given this choice of axes, the ordering wave vector connecting the paired hot spots is $\textbf{Q}_{\text{AF}} = ( \pm \sqrt{2} \pi \hat{k}_x, \pm \sqrt{2} \pi \hat{k}_y )$ up to the reciprocal lattice vectors $\sqrt{2} \pi (\hat{k}_x \pm \hat{k}_y)$. See Fig. \ref{fig:hot spots and nesting angle} for details. The linearized electron dispersions are given by
$e^{\pm}_1(\textbf k;v) = -e^{\pm}_3(\textbf k;v) = v k_x \pm k_y$, 
$e^{\pm}_2(\textbf k;v) = -e^{\pm}_4(\textbf k;v) = \mp k_x + v k_y$, 
where the momentum $\textbf{k}$ is measured relative to each hot spot. 
For any $v \neq 0$, the curvature of the Fermi surface ($\mathcal{O}(k^2)$ terms) can be ignored, since the linearized dispersions at the coupled hostpots are not parallel to each other, and therefore the problem is fully two-dimensional. The component of the Fermi velocity along $\textbf{Q}_{\text{AF}}$ has been set to one by rescaling $\textbf{k}$. 
$v$ is the component of Fermi velocity that is perpendicular to $\textbf{Q}_{\text{AF}}$. It controls the degree of nesting between coupled hot spots and can be written as $v = \tan \th$, where $\th$ is the nesting angle (c.f. Fig. \ref{fig:hot spots and nesting angle}). $\boldsymbol{\phi}(q)$ is the three-component boson field
that describes the AFM collective mode in the fundamental representation of $\mathrm{O}(3)$,
with frequency $\omega_q$ and momentum $\textbf{q}$. Note that while the collective spin is centered at $\textbf{Q}_{\text{AF}}$, $\boldsymbol \phi$ is centered at zero, since $\textbf{Q}_{\text{AF}}$ is already incorporated into the hotspot label. $c$ is the boson velocity, and $u$ is its quartic interaction. $r$ is the squared mass of the boson, as is tuned drive the boson to criticality. $g$ is the Yukawa coupling between the boson and the electrons, which scatters the electrons between hot spot pairs via the spin-spin interaction. These couplings represent all the relevant and marginal terms obeying the symmetries of the problem. $\boldsymbol{\tau}$ consists of the three generators of the $SU(2)$ group. We have also generalized the action from one to $N_f$ fermion flavors. This generalization was used in the large $N_f$ expansion in previous renormalization group studies \cite{PhysRevLett.84.5608,PhysRevB.82.075128}, and we use this general form in subsequent sections.

\subsection{Sign-problem-free UV completion}

\begin{figure}[!ht]
    \centering
    \includegraphics[width=0.45\textwidth]{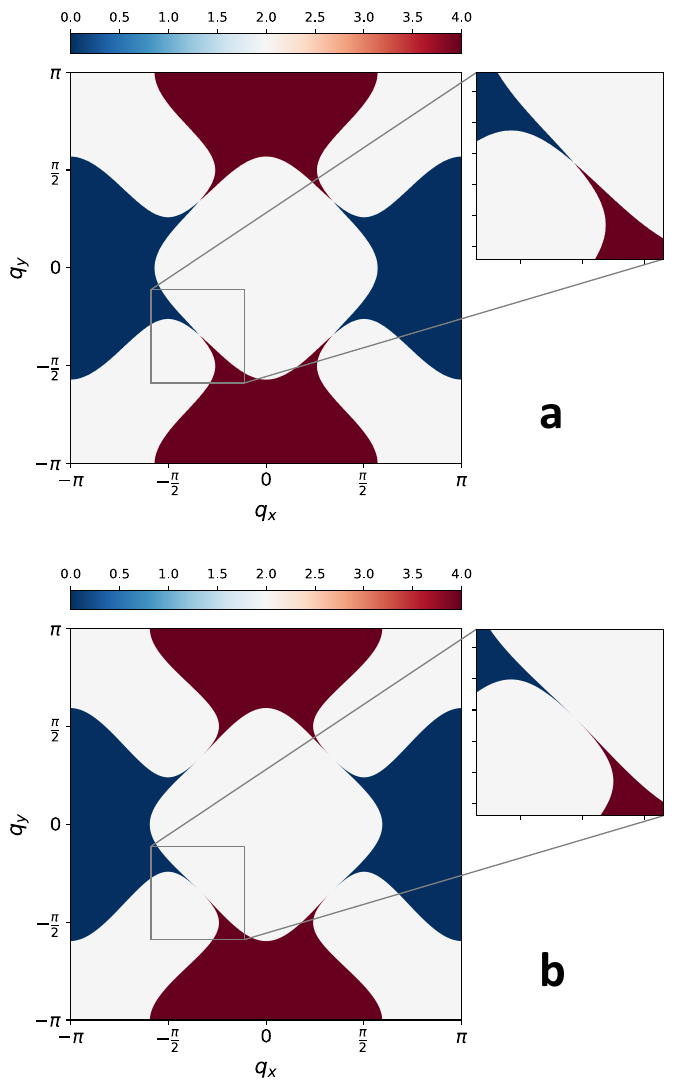}
    \caption{\textbf{Occupation number.} The total occupation number $n_{\boldsymbol q} = \sum_{\a,\s} \langle \psi^{\dagger}_{\a,\s,\boldsymbol q} \psi_{\a,\s,\boldsymbol q} \rangle$ of one flavor of fermions for the free theory, summed over spin and band degrees of freedom. Here, we have incorporated the phase shift $e^{i \boldsymbol{Q}_{\text{AF}} \cdot \boldsymbol r}$ into the $\a = y$ band. The nesting values shown are (a) $v = v_1$ and (b) $v = v_5$, which are the two extremes we study. Note that the Fermi surface at the hot-spots is actually connected, but due to the very small nesting parameter, that is hard to show graphically.}
    \label{fig:occupation_function v_1 and v_5}
\end{figure}

The action in Eq. (\ref{eq:hotspot theory}) captures the universal low-energy properties of the system associated with the divergent correlation length near the phase transition. Since the electrons far away from the hotspots do not enter into the theory, the precise shape of the Fermi surface beyond the momentum cutoffs does not play a role in the critical phenomena of this theory. Therefore, we can change the band structure while ensuring that the action of Eq. (\ref{eq:hotspot theory}) is not modified. One such UV completion was given in Ref. \cite{Berg21122012} and has the real-space action of
\begin{equation}
\begin{split}
\mathcal{S} = & 
\int d \tau
\sum_{\sigma=\uparrow,\downarrow} 
\sum_{\alpha=x,y} 
\sum_{j=1}^{N_f}
\sum_{\boldsymbol r, \boldsymbol r'} 
\\ &
\hspace{20mm} 
\psi^{*}_{\a,\s,j,\boldsymbol r}
\left[ (\partial_{\tau} - \mu) \delta_{\boldsymbol r, \boldsymbol r'} 
- t_{\a, \boldsymbol r, \boldsymbol r'} \right] 
\psi_{\a,\s,j,\boldsymbol r'}
\\ &
\int d \tau \sum_{\boldsymbol r}
\Bigl[
\frac{1}{c^2} (\partial_\tau \boldsymbol{\phi}_{\boldsymbol r})^2 + \f12 (\nabla \boldsymbol{\phi}_{\boldsymbol r})^2 + \frac{r}{2} (\boldsymbol{\phi}_{\boldsymbol r})^2
+ \frac{u}{4} (\boldsymbol{\phi}_{\boldsymbol r})^4
\Bigr]
\\ & 
+ \frac{g}{\sqrt{N_f}}
\sum_{\s,\s'=\uparrow,\downarrow}
\sum_{j=1}^{N_f}
\int d \tau 
\sum_{\boldsymbol r}
e^{i \boldsymbol{Q}_{\text{AF}} \cdot \boldsymbol r}
\;
\\ & \hspace{30mm}
\boldsymbol{\phi}_{\boldsymbol r} \cdot 
\Bigl[
\psi^{*}_{x,\s, \boldsymbol r} \;
\boldsymbol{\tau}_{\s,\s'}  \;
\psi_{y, \s', \boldsymbol r}
+ \text{h.c.} \Bigr].
\end{split}
\label{eq:sign-problem-free action}.
\end{equation}
Here, the number of bands has been doubled ($\alpha = x,y$ represents a band index). The boson now scatters electrons between the two bands. The band-dependent hopping amplitude $t_{\a, \boldsymbol r, \boldsymbol r'}$ still respects the $\mathrm{C}_4$ symmetry of the lattice, provided that the bands are also interchanged. The location of the hotspots and the dispersion linearized about them is unchanged from the one-band model. Note that the axes of this model are rotated $45 \degree$ relative to the model of Eq. (\ref{eq:hotspot theory}). The advantage of this model is that it now contains an inter-band anti-unitary symmetry that guarantees the positivity of the fermionic determinant \cite{Berg21122012} and makes it amenable to sign-problem-free Monte Carlo simulations \cite{PhysRevResearch.2.023008,PhysRevB.95.174520,doi:10.1146/annurev-conmatphys-031218-013339,PhysRevLett.120.247002,PhysRevB.95.035124,PhysRevLett.117.097002,doi:10.1146/annurev-conmatphys-033117-054307}. 
\\ \indent
We study the lattice model defined in Eq. (\ref{eq:sign-problem-free action}) using HMC. Our method is described in detail in Sec. \ref{sec:method}. As explained there, in order for our algorithm to function, we must work with an even $N_f$, and we choose $N_f = 2$. Unless stated otherwise, we fix the parameter values to be $u = 0, g = 0.7 \, \sqrt 2 \approx 1, c = 3$. We study five different Fermi surfaces, with the following values of the nesting parameter: $v_1 \approx 0.149, v_2 \approx 0.072, v_3 \approx 0.036, v_4 \approx 0.018, v_5 \approx 0.0092$, corresponding to nesting angles of $\theta_1 \approx 8.5 \degree, \theta_2 \approx 4.13 \degree, \theta_3 \approx 2.05 \degree, \theta_4 \approx 1.03 \degree, \theta_5 \approx 0.53 \degree$, respectively. We illustrate the Fermi surface corresponding to $v = v_1$ and $v = v_5$ in Fig. \ref{fig:occupation_function v_1 and v_5}.
The non-zero hopping amplitudes and chemical potentials corresponding to each are $t_{h,x} = t_{v,y} = 1$ and 
\begin{align}
\td t_{v,x}^{(1)} &= - \td t_{h,y}^{(1)} = 0.45, 
\quad 
\mu^{(1)}_x = - \mu^{(1)}_y = -0.47
\\ 
\td t_{v,x}^{(2)} &= - \td t_{h,y}^{(2)} = 0.48, 
\quad 
\mu^{(2)}_x = - \mu^{(2)}_y = -0.46
\\ 
\td t_{v,x}^{(3)} &= - \td t_{h,y}^{(3)} = 0.498, 
\quad 
\mu^{(3)}_x = - \mu^{(3)}_y = -0.44
\\ 
\td t_{v,x}^{(4)} &= - \td t_{h,y}^{(4)} = 0.505, 
\quad 
\mu^{(4)}_x = - \mu^{(4)}_y = -0.44
\\ 
\td t_{v,x}^{(5)} &= - \td t_{h,y}^{(5)} = 0.5085, 
\quad 
\mu^{(5)}_x = - \mu^{(5)}_y = -0.44.
\end{align}
Here, $t_{h,\a}, t_{v,\a}$ denote the nearest-neighbor hopping amplitudes in the $x$ and $y$ directions, respectively, which are the same for all nesting parameter values; $\td t_{h,\a}^{(i)}, \td t_{v,\a}^{(i)}$ denote the next-nearest-neighbor hopping amplitudes in the $x$ and $y$ directions, respectively, for nesting parameter value $i$; $\mu^{(i)}_{\a}$ are the chemical potentials for the nesting parameter value $i$. 
\\ \indent 
The reason for choosing these specific parameter values is to make contact with Ref. \cite{PhysRevResearch.2.023008}, where the authors in turn tried to make contact with Ref. \cite{PhysRevX.7.021010}. Of course, it is important to scan the values of $u,g,c$ to check the stability of our results and look for new behavior. However, such a detailed study is beyond the scope of the present work, and is left for future studies.

\subsection{Theoretical analysis near perfect nesting}
\label{sec:theory_nesting_fixed_point}

\begin{figure}[!ht]
    \centering
    \includegraphics[width=0.3\textwidth]{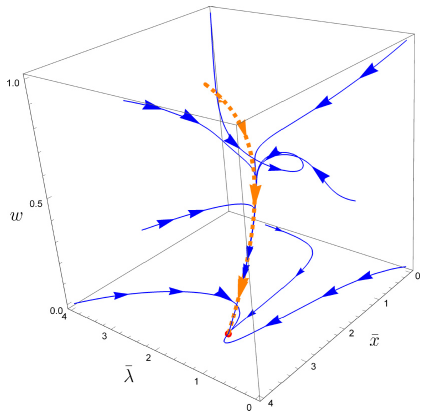}
    \caption{\textbf{Renormalization group flow.} RG flow of the theory in Eq. (\ref{eq:hotspot theory}), computed using the epsilon expansion, reproduced from Ref. \cite{PhysRevB.95.245109}. The axes are $w$, $\bar x = x/10$, $\bar \lambda = 10 \lambda$. Initially, there is a fast flow (blue lines) to a fixed one-dimensional manifold (dashed orange line). The flow along this manifold is slow, and towards $w=0$. The flow of $\kappa$ is excluded, since its flow towards zero is the fastest.}
    \label{fig:RG flow}
\end{figure}

Although in general the theory in Eq. (\ref{eq:hotspot theory}) cannot be understood analytically using a controlled approach, there exists a parameter regime where a controlled solution in the IR can be obtained \cite{PhysRevX.7.021010}. This parameter regime is primarily characterized by a small nesting parameter $v \ll 1$ and an effective coupling of intermediate strength, leading to strong correlations. We start by reviewing this IR fixed point of Refs. \cite{PhysRevX.7.021010, PhysRevB.95.245109, PhysRevB.98.075140}. 
\\ \indent 
At criticality, the UV the theory is best described in terms of the ratios $\lambda \equiv \frac{g^2 c^2}{v}, x \equiv \frac{g^2}{c}$, $\kappa \equiv u \, c^2$, and $w = \frac{v}{c}$. (We note that compared to Refs. \cite{PhysRevX.7.021010, PhysRevB.95.245109, PhysRevB.98.075140}, the action of Eq. (\ref{eq:hotspot theory}) has $\boldsymbol \phi \rightarrow \boldsymbol \phi /c$). The renormalization group flow of the UV theory to the IR fixed point is two-fold and is shown in Fig. \ref{fig:RG flow}, which is reproduced from Ref. \cite{PhysRevB.95.245109}. 
Initially, there is a fast (algebraically in the running energy scale $\mu$) flow of $\lambda$ and $x$ to $w$-dependent $\mathcal{O}(1)$ values and $\kappa$ to zero. Once the first step of the flow is complete, the couplings will keep flowing along the one-dimensional manifold defined by the fixed point values $\lambda, x, \kappa$. This one-dimensional manifold defines the only remaining free coupling of the theory. We note that since $\lambda \sim \mathcal{O}(1)$ is the effective coupling of the UV theory, this manifold represents the theory in the strongly correlated regime. Crucially, due to all the coupling inter-dependencies, along the manifold $w$ becomes a function of $v$ only. We can choose to parametrize the manifold by $w(v)$. This choice is useful, since it turns out that theory on the manifold has a perturbation theory that is organized in powers of $w(v)$ (along with powers of $\log(w)$). However, since $w(v)$ is only a function of $v$, such that $w(v) \rightarrow 0$ with $v \rightarrow 0$, perturbation theory can also be done in $v$ itself.
\\ \indent 
Finally, once the theory sits on the one-dimensional manifold, the remaining flow is towards $w(v) \rightarrow 0$. However, unlike the first part of the flow, the flow of $w(v)$ in this second part happens at a rate that is sub-logarithmic in $\mu$. Therefore, for all practical purposes we can take $w(v)$, and therefore $v$ to be fixed (scale independent) on this manifold. An important point is that due to the initial fast flow, the value of $v$ on the manifold is different than the bare (UV) value $v_B$. However, given that the bare parameters of the theory are tuned to minimize the length of this initial flow, we assume that $v \approx v_B$ in the analysis of Sec. \ref{sec:results}.  
\\ \indent 
If the value of $w(v)$ is small enough, the theory can be studied using perturbation theory in $w(v)$ to a finite order. Focusing on the critical spin susceptibility $\chi(\o, \boldsymbol q)$, its leading order in $w(v)$ behavior is given by (here we use the axes of Eq. (\ref{eq:hotspot theory}))
\begin{equation}
    \chi^{-1}(\o, \boldsymbol q) = \abs{\o} + c(w(v)) \, (\abs{q_x} + \abs{q_y}).
    \label{eq: chi tree level}
\end{equation}
Here, $c(w(v))$ is a new emergent boson velocity, not related to the bare value $c$, and dependent on the only coupling in the theory. It has its own expansion in powers of $w(v)$ \cite{PhysRevX.7.021010}. This form breaks the $\mathrm{O}(2)$ spatial symmetry (a $\abs{\boldsymbol q}$ dependence) of the spin susceptibility down to the $\mathrm{C}_4$ symmetry of the Fermi surface. This is a key prediction of the small $v$ fixed point theory \cite{PhysRevX.7.021010}. Eq. (\ref{eq: chi tree level}) can be thought of as the tree-level susceptibility at the fixed point, and perturbations to it are computed in powers of $w(v)$. The deviations of the scaling are encoded in the dynamical critical exponent $z$ and the anomalous boson dimension $\eta_{\phi}$, which were computed to leading order in Ref. \cite{PhysRevX.7.021010},
\begin{align}
z &= 1 + \frac{3}{4\pi N_f} w(v)
\label{eq:z leading corr}
\\
\eta_{\phi} &= \frac{1}{2\pi N_f} w(v) \log\left(\frac{1}{w(v)}\right).
\label{eq:eta_phi leading corr}
\end{align}
Ref. \cite{PhysRevX.7.021010} also computed the dependence of $w(v)$ on $v$ to lowest order in $v$,
\begin{equation}
    w = 4 \sqrt{N_f} \sqrt{\frac{v}{\log(1/v)}}.
    \label{eq: w of v}
\end{equation}
Together these relations give the leading contributions to $z,\eta_{\phi}$ for small $v$.
\\ \indent 
In order to make a connection between the leading-order pertubation theory in $w(v)$ results of Ref. \cite{PhysRevX.7.021010} and the theory of Eq. (\ref{eq:sign-problem-free action}) at finite $v$, we need to use the non-perturbative renormalization group (RG) equation. This equation for $\chi(\o, \boldsymbol q)$ at criticality ($r = r_c$) is given by
\begin{equation}
\begin{split}
&\left[\Delta(\eta_\phi, z)
+ \o \frac{\partial}{\partial \o} 
+ \frac{1}{z} \boldsymbol q \cdot \frac{\partial}{\partial \boldsymbol q}
+ T \frac{\partial}{\partial T} 
\right.
\\ &
\left.
+ \frac{1}{z} \left(\frac{1}{L}\right) \frac{\partial}{\partial \left(\frac{1}{L}\right)}
\right] \chi(\omega, \boldsymbol q; T, L) = 0,
\end{split}
\label{eq:RG equation for D}
\end{equation}
where $\Delta(\eta_\phi, z) = \frac{1 - 2 \eta_\phi - (z-1)}{z}$. Here, we have used the fact that $v \approx v_B$ is approximated as being constant as a function of scale (otherwise its beta function would enter in Eq. (\ref{eq:RG equation for D})). Since at criticality the correlation length is absent from Eq. (\ref{eq:RG equation for D}), we have included $T, L^{-1}$ as relevant energy scales. We note that for a scaling theory different from that of Ref. \cite{PhysRevX.7.021010},
the RG equation would be the same, with the only modification being the form of $\Delta$. The only requirement is that the flow of the couplings can be ignored, which is a good approximation when the couplings flow at most logarithmically in the running scale.
\\ \indent
We study systems of fixed size and temperature.
We focus on two limits of Eq. (\ref{eq:RG equation for D}). In the first limit, $\boldsymbol q = 0$, the solution is given by 
\begin{equation}
\chi(\omega) = \abs{\o}^{- \Delta(\eta_\phi, z)} f\left(\o/T, \o L^z \right),
\label{eq:RG equation for D with q = 0}
\end{equation}
where $f$ is some non-universal crossover function. Here, $T$ and $1/L$ act as IR cutoffs to the critical scaling. In other words, there is some effective IR frequency cutoff given roughly by $\o_{\textrm{IR}} \sim \max\left(\frac{2\pi}{\beta}, (\frac{2\pi}{L})^{z} \right)$, and the function $f\left(\o/T, \o L^z \right)$ is a finite constant for $\o \gg \o_{\textrm{IR}}$. On the other hand, the scaling equation itself only holds below some non-universal UV cutoff $\Lambda_\o$. Therefore, the pure algebraic scaling of $\chi(\omega, \boldsymbol q = 0) \sim \abs{\o} ^{- \Delta(\eta_\phi, z)}$ can be observed in an intermediate frequency window, $\o_{\textrm{IR}} \ll \o \ll \L_{\o}$.
\\ \indent 
Now we turn to the case of $\o = 0$. For simplicity, we first also set $q_y = 0$. Analogously to $\chi(\o)$, $\chi(q_x)$ is given by
\begin{equation}
\chi(q_x) = \abs{q_x}^{- z \, \Delta(\eta_\phi, z)} \; g(q_x/T^{1/z}, q_x L).
\label{eq:RG equation for D with omega = 0 and q_y = 0 with IR cutoff}
\end{equation}
Here, $g$ is another non-universal crossover function. The critical scaling $\chi(q_x) \sim \abs{q_x}^{- z \Delta(\eta_\phi, z)}$ holds in the intermediate region of $q_{\textrm{IR}} \ll q_x \ll \L_{q}$, where $q_{\textrm{IR}} = \max \left((\frac{2\pi}{\beta})^{1/z}, \frac{2\pi}{L} \right) $ and $\L_{q}$ are IR and UV momentum cutoffs, respectively. Once the intermediate-scale algebraic scaling of both $\chi(\omega)$ and $\chi(q_x)$ are known, $z$ is computed directly from their ratio: $z \Delta(\eta_\phi, z)/\Delta(\eta_\phi, z)$. 
\\ \indent 
When we turn on $q_y$, the RG equation does not say anything about the functional form of the dependence of $\chi(\boldsymbol q)$ on $q_x$ and $q_y$, which must be deduced by other means. All that is known from the tree-level form of Eq. (\ref{eq: chi tree level}) is that the spatial dependence must be $\mathrm{C}_4$-symmetric. Jumping ahead to Sec. \ref{sec:results}, the Monte Carlo data suggests the form
\begin{equation}
\begin{split}
\chi^{-1}(\boldsymbol q) = & (\abs{q_x}^{z \, \Delta(\eta_\phi, z)} + \abs{q_y}^{z \, \Delta(\eta_\phi, z)}) \; 
\\ &
g((\abs{q_x} + \abs{q_y})/T^{1/z},(\abs{q_x} + \abs{q_y}) \, L),
\end{split}
\label{eq:RG equation for D with omega = 0 with IR cutoff}
\end{equation}
which we take as our conjecture.
\\ \indent
As as aside, we note that this way of extracting critical exponents from a system at fixed $L,\beta$ is not the usual one of finite size scaling. Although the latter is more systematic, it would require the simultaneous scaling of $L$ and $\beta \propto L^z$. Since we do not know what $z$ is a priori, this requires an additional scan over its potential values. Also, the $z$ value we find later on is always $z > 1.6$, which implies the need for a large $\beta$ in the finite size scaling. Both of these are very computationally expensive, and we therefore opt for the present method.

\subsection{Numerical results}
\label{sec:num results}

\begin{figure}[!ht]
    \centering
    \includegraphics[width=0.45\textwidth]{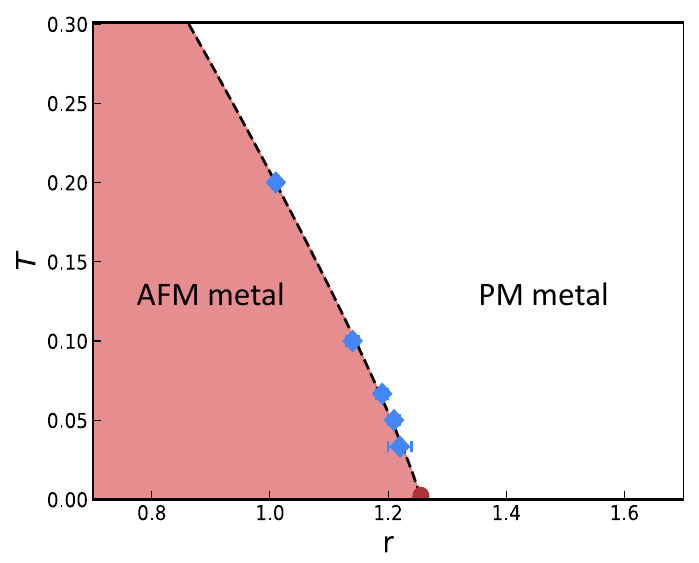}
    \caption{\textbf{Phase diagram in $T$ and $r$.} The red (white) region is a metal with AFM (no magnetic) order. The blue points denote the intersection of curves of the Binder cumulant for different system sizes at fixed temperatures. The error bars denote the finite resolution in $r$ that we have in those intersection points. The phase diagram for all nesting parameters studied is the same up to our resolution. Superconductivity is absent for all the parameters we study.}
    \label{fig:phase diagram}
\end{figure}

\begin{figure}[!ht]
    \centering
    \includegraphics[width=0.45\textwidth]{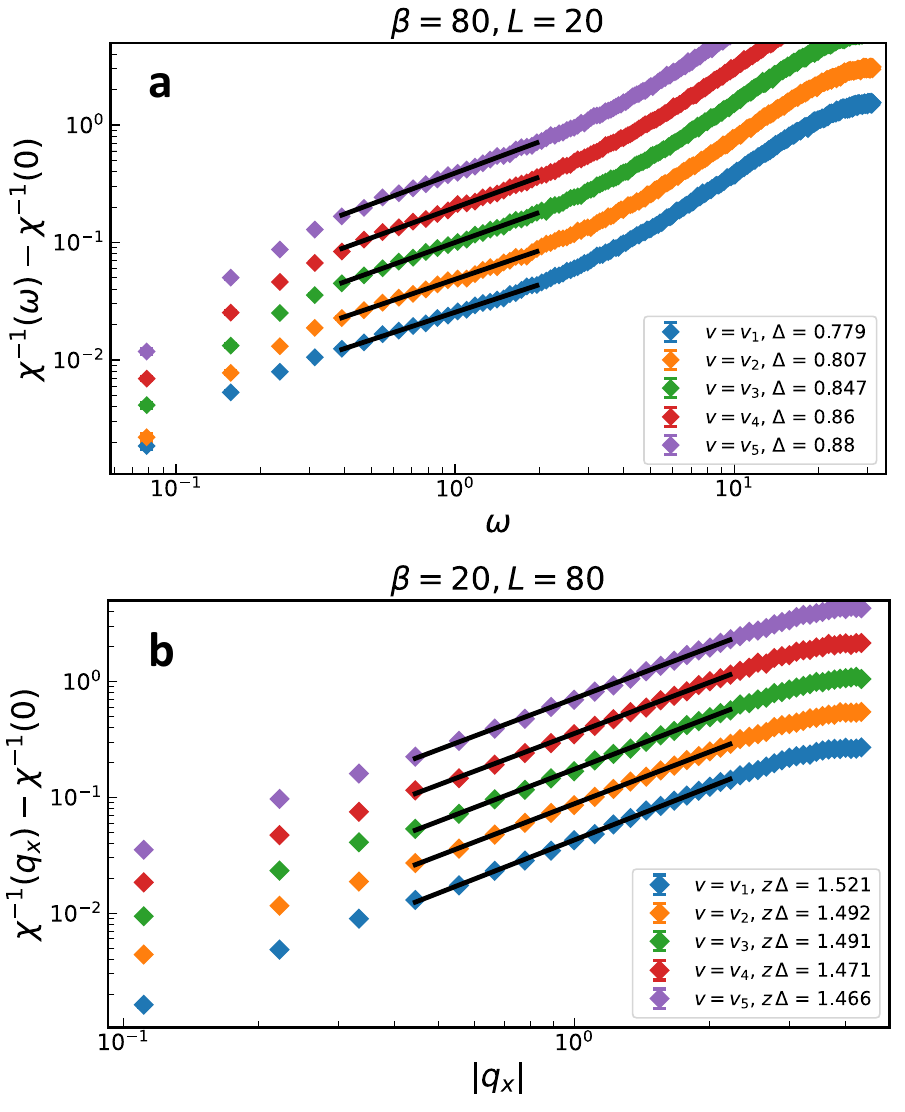}
    \caption{\textbf{Scaling of critical spin susceptibility.} The (a) dynamic and (b) static spin susceptibility, shown on log-log plots for all nesting parameters $v_i$ studied. The error bars are given by the one sigma statistical uncertainties from the stochastic calculation. The different curves are shifted relative to each other for visual clarity. We show only the largest (a) $\beta$ and (b) $L$ values that we simulated in order to extract the intermediate $\o$ and $q_x$ regimes of power-law scaling. For each curve, we choose the regions that look the most straight by eye (if there are two such regions, as in $\chi^{-1}(\o)$, we choose the one with smaller $\o$ or $q_x$). Choosing slightly different boundaries leads to slightly different exponents, but the trend with $v$ is always the same.}
    \label{fig:chi inv vs omega and q_x all nesting largest sizes}
\end{figure}

\begin{figure}[!ht]
    \centering
    \includegraphics[width=0.4\textwidth]{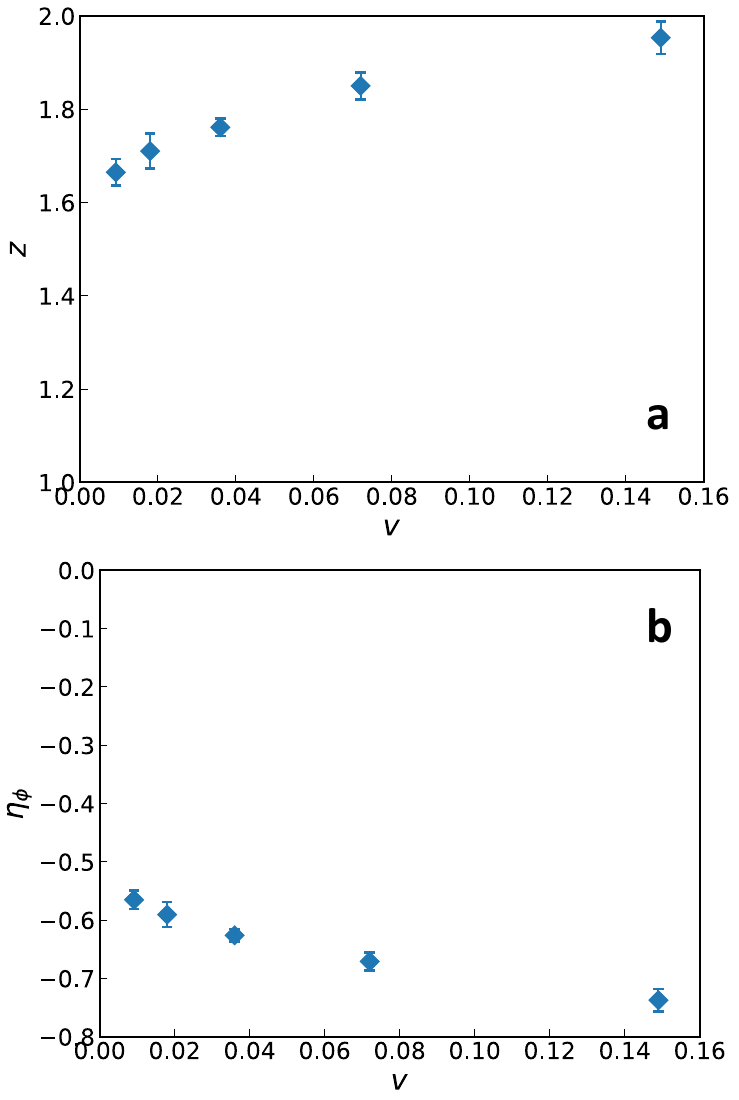}
    \caption{\textbf{Critical exponents.} The (a) dynamical critical exponent $z$ and (b) the anomalous boson dimension $\eta_{\phi}$ as a function of $v$. The error bars are computed using a bootstrap method on the fitted data, allowing for duplicates (c.f. Supplementary Note 1).}
    \label{fig:z and eta_phi vs v}
\end{figure}

\begin{figure*}[!ht]
    \centering
    \includegraphics[width=1.0\textwidth]{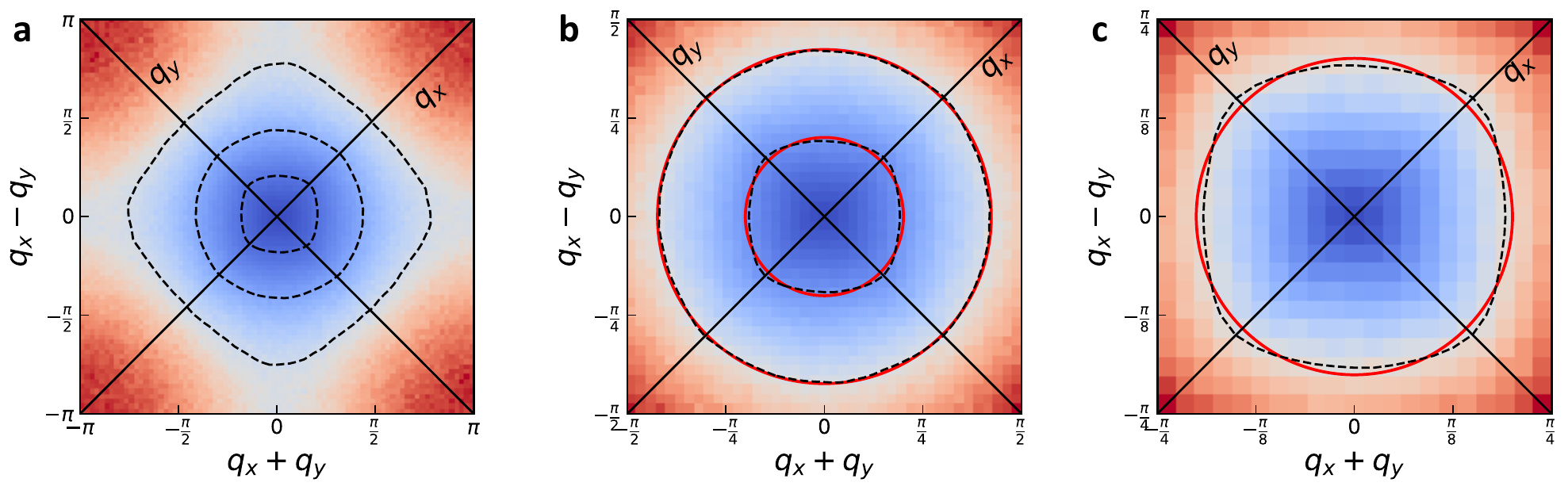}
    \caption{\textbf{Density plots static critical spin susceptibility.} All are for $v=v_5$, $\beta = 20$ and $L=80$. We are interested in showing the equal-density contours (we do not show a color bar as it is irrelevant for this purpose). The three plots are different levels of zooming in to $\boldsymbol q = 0$. In (a) we show the entire Brillouin zone, along with three dashed-line equal-density contours (passed through a Gaussian filter for smoothness), at large, intermediate, and small momenta. We can see that the largest contour has a $\mathrm{C}_4$-symmetric form, due to the lattice. In (b) we zoom in to focus on the inner two contours, which we overlay with two red circles at radii $0.63, 1.33$, in order to illustrate their symmetry. The larger circle overlays with the contour extremely well, indicating an $\mathrm{O}(2)$ symmetry at those momenta. The smaller circle overlays with its contour poorly, since that contour is much more square-like, indicating that at smaller momenta the symmetry is again $\mathrm{C}_4$. To show this more convincingly, in (c) we zoom in further to enlarge this smallest contour.}
    \label{fig:chi inv vs vec q density plots v_5}
\end{figure*}

The phase diagram of the theory in Eq. (\ref{eq:sign-problem-free action}) as a function of $r$ and $T$ is shown in Fig. \ref{fig:phase diagram}.
At large values of $r$ the system is a paramagnetic (PM) metal. As a function of decreasing $r$, for any finite $T$ (below some large temperature) there is a sharp crossover to a SDW metal. This crossover will not become a true second-order phase transition in the thermodynamic limit, since the dimensionality and symmetry of the order parameter prevent the system from ordering due to the Mermin-Wagner theorem. However, if we infinitesimally couple stacks of this two-dimensional system (as is the situation in many relevant experiments), order will be stabilized and the transition will become a true one. We can identify the sharp location of this potential second-order transition, $r_c(T)$, by studying finite-size scaling diagnostics, such as the Binder cumulant (c.f. Supplementary Note 1 for details). The line of these transition points terminates at $T = 0$ and $r = r_c$, which is the quantum critical point. This is the familiar picture of quantum criticality. 
\\ \indent 
For all nesting parameters we see no superconductivity down to the lowest measured temperatures. This is consistent with the arguments outlined in Ref. \cite{PhysRevResearch.2.023008} that the superconducting transition temperature $T_c$ of the spin-fermion model is suppressed with decreasing $v$ as $T_c \sim g^2 \sin(\arctan(v))$.
\\ \indent 
To study the critical scaling we tune $r = r_c$ and vary $\beta \equiv 1/T$. Following Sec. \ref{sec:theory_nesting_fixed_point}, we look at the dependence of the spin susceptibility on frequency and momentum, $\chi(\o)$, $\chi(q_x)$, $\chi(\boldsymbol q)$. Here, we use the axes of Eq. (\ref{eq:hotspot theory}), in order to compare to Sec. \ref{sec:theory_nesting_fixed_point}. From the scaling of $\chi(\o)$ and $\chi(q_x)$ in the intermediate scaling regions we obtain the exponents $\Delta(\eta_\phi, z)$ and $z \, \Delta(\eta_\phi, z)$, as illustrated in Fig. \ref{fig:chi inv vs omega and q_x all nesting largest sizes}. 
From $\Delta(\eta_\phi, z)$ and $z \, \Delta(\eta_\phi, z)$ we determine $z$ and $\eta_{\phi}$, which are plotted in Fig. \ref{fig:z and eta_phi vs v} as functions of $v$. 
An important point is that the intermediate regime of power-law scaling in Fig. \ref{fig:chi inv vs omega and q_x all nesting largest sizes} is chosen by eye, and different choices yield slightly different forms of $z(v)$ and $\eta_{\phi}(v)$. The specific choice we use in Fig. \ref{fig:chi inv vs omega and q_x all nesting largest sizes} is described in Supplementary Note 1, along with how we compute the corresponding error bars. However, crucially, for all (reasonable) choices of the scaling region boundaries, $z(v)$ monotonically decreases with decreasing $v$, starting from a value slightly less than $z=2$ (which is the Hertz-Millis prediction) for $v_1$. Similarly, $\eta_{\phi}(v)$ monotonically increases from $\eta_{\phi} \approx -0.75$ with decreasing $v$.
\\ \indent 
In addition to the dependence of $z(v)$ and $\eta_{\phi}(v)$ on $v$, we determine the spatial symmetry of $\chi(\boldsymbol q)$ and its functional form. In Fig. \ref{fig:chi inv vs vec q density plots v_5} we show density plots of $\chi^{-1}(\boldsymbol q)$ for $v = v_5$.
\begin{figure*}[!ht]
    \centering
    \includegraphics[width=1.0\textwidth]{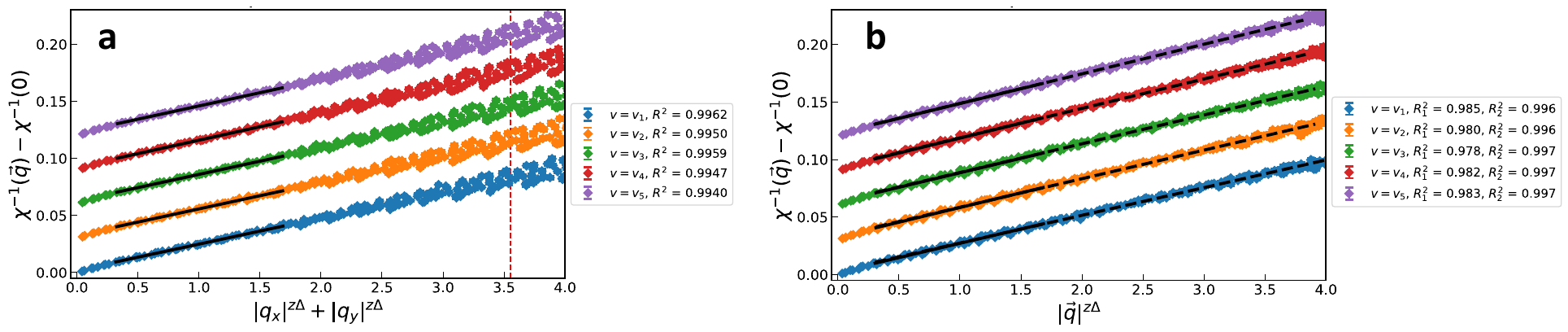}
    \caption{\textbf{Scaling form of static critical spin susceptibility.} $\chi^{-1}(\boldsymbol q) - \chi^{-1}(0)$ plotted against (a) $\abs{q_x}^{z \Delta} + \abs{q_y}^{z \Delta}$ and (b) $\abs{\boldsymbol q}^{z \Delta}$.
    Both plots are for $\beta = 20$ and $L=80$. The error bars are again the one sigma uncertainties.
    The curves for different $v_i$ are shifted to separate them for visual clarity. The initial region, up to $q^{z\Delta} \sim 1.7$ is fit with both forms. The $\mathrm{C}_4$ is much better in this region; see the solid lines in (a) and (b), and compare the coefficients of determination $R^2$ from (a) with $R_1^2$ from (b). However, over larger momenta (up to $q^{z \Delta} \sim 3.6$) the $\mathrm{O}(2)$ fit it much better; see the dashed line as well as $R_2^2$ in (b).}
    \label{fig:chi inv  vs qx^zDel + qy^zDel and vec q^zDel v_5}
\end{figure*}
We plot contours at various values. We can see that at large momenta, the symmetry of the contours is that of the lattice, $\mathrm{C}_4$. At intermediate momenta, the contours are circles, indicating an $\mathrm{O}_2$ symmetry, indicating that those momenta are small enough that $\cos(k_x) + \cos(k_y) \approx 2 - \boldsymbol{k}^2/2$ is a good approximation. However, at smaller momenta the contours transform again to a $\mathrm{C}_4$-symmetric form, with the maxima/minima now rotated by $45 \degree$. In Supplementary Note 1, we show more density plots at smaller $L$, in order to illustrate that this effect gets more pronounced as $L$ increases and is therefore a true long-wavelength effect present in the thermodynamic limit.
\\ \indent 
It is not possible to provably determine the exact functional form, as there are many forms that obey the $\mathrm{C}_4$ symmetry. However, $\abs{q_x}^{z \Delta} + \abs{q_y}^{z \Delta}$ seems to fit very well. We illustrate this in Fig. \ref{fig:chi inv  vs qx^zDel + qy^zDel and vec q^zDel v_5}, along with a comparison to $\abs{\boldsymbol q}^{z \Delta}$. We therefore conclude that $\chi^{-1}(\boldsymbol q) \sim \abs{q_x}^{z \Delta} + \abs{q_y}^{z \Delta}$ is the correct long-wavelength form.
\\ \indent 
These two features of the boson susceptibility---i.e., the monotonic forms of $z(v), \eta_{\phi}(v)$ and the symmetry reduction of $\mathrm{O}(2) \rightarrow \mathrm{C}_4$ in the long-wavelength limit---provide strong numerical evidence in favor of the critical scaling at the SDW QCP being governed by the theory of Ref. \cite{PhysRevX.7.021010} for all the nesting values we study. 
\begin{figure*}[!ht]
    \centering
    \includegraphics[width=1.0\textwidth]{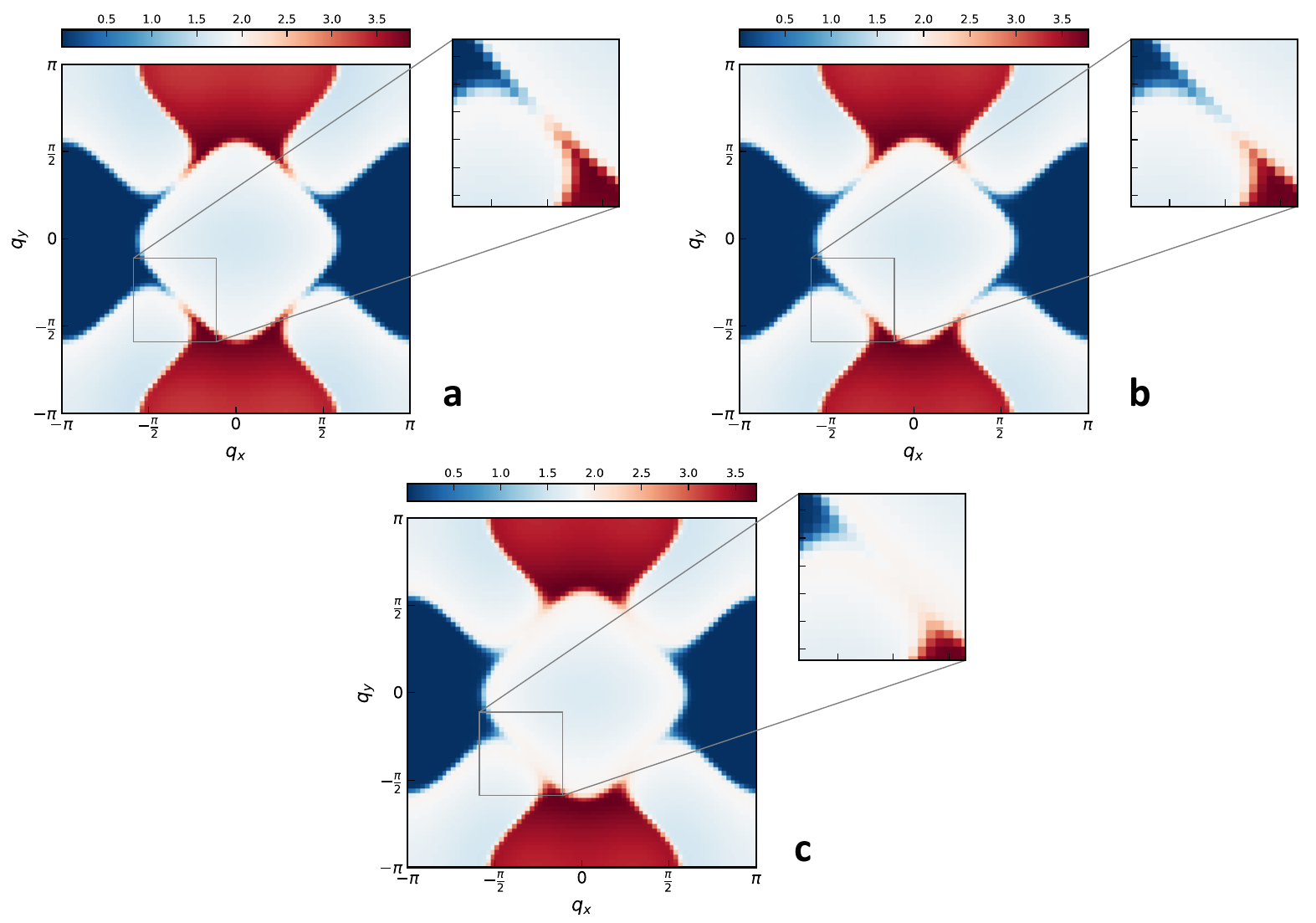}
    \caption{\textbf{Measured occupation number.} The total occupation number $n_{\boldsymbol q} = \sum_{\a,\s} \langle \psi^{\dagger}_{\a,\s,\boldsymbol q} \psi_{\a,\s,\boldsymbol q} \rangle$ for the (a) free theory, and the interacting theory (b) at criticality ($r = r_c \approx 1.26$) and (c) slightly in the ordered phase ($r = 1.15$). This explicitly shows the renormalization of the Fermi surface near the hot spots due to fluctuations of the order parameter. All plots are for $v = v_1$ and $\beta = 20, L = 80$.}
    \label{fig:occupation functions}    
\end{figure*}
\\ \indent 
Despite the similarities, there are two parts of the data that are seemingly in contrast to the predictions of Ref. \cite{PhysRevX.7.021010}. The first is the fact that $\eta_{\phi}$ is negative, whereas Eq. (\ref{eq:eta_phi leading corr}) predicts that as $\eta_{\phi} \rightarrow 0^+$ as $v \rightarrow 0$. However, Eq. (\ref{eq:eta_phi leading corr}) is only the leading order correction, and it particular, in obtaining it, terms of $\mathcal{O}(w(v))$ were ignored in Ref. \cite{PhysRevX.7.021010}. Therefore, it is perfectly possible that the actual leading-order coefficient behaves as $\eta_{\phi} \sim w(v) (\log(1/w(v)) - a)$, where $a>0$ is a constant, which would lead to a sign-change in the leading-order behavior when $1 \gg w(v) > e^{-a}$. On the other hand, the leading term of $z$ from Eq. \ref{eq:z leading corr} does not suffer from a similar problem, and its sign can be trusted at small $w(v)$.
\\ \indent 
The second part of the data that is seemingly in contrast to the predictions of Ref. \cite{PhysRevX.7.021010} is that $z(v)$ and $\eta_{\phi}(v)$ do not convincingly exhibit the limiting behaviors $z \rightarrow 1$ and $\eta_{\phi} \rightarrow 0$ as $v \rightarrow 0$. We conjecture the following explanation. As noted in Sec. \ref{sec:theory_nesting_fixed_point}, we assume that the renormalized value of $v$ is close to the bare one. However, as we decrease $v$ we are not tuning $g$, so the bare value of $\lambda = g^2 c^2/v$ is increasing, not staying fixed. This increases the RG-time needed to flow to the one-dimensional manifold discussed in Sec. \ref{sec:theory_nesting_fixed_point}. The large bare value of the effective coupling $\lambda$ might be renormalizing $v$ to larger values, before the one-dimensional manifold it reached. Remedying this requires decreasing $g$ with decreasing $v$ such that $\lambda$ remains fixed. This task is beyond the scope of this paper, and we leave it for future work.
\\ \indent 
Finally, we show the fermion occupation function at criticality in Fig. \ref{fig:occupation functions} (as well as Supplementary Note 1), where we can see the renormalization of the hot-spots by the Yukawa interaction. Unfortunately, within our spatial resolution it is not possible to measure the renormalized nesting parameter $v$, and therefore confirm the conjecture from the above paragraph.

\section{Discussion}
\label{sec:discussion}

In this work, we study the critical theory of the $\mathrm{O}(3)$ spin-fermion model as a model of the antiferromagnetic transition in two-dimensional metals. We use a novel HMC method that we extensively develop. Below we discuss some consequences of our work and outlook.

\subsection{Implications of theoretical results}

Our study of the critical spin susceptibility $\chi(\o, \boldsymbol q)$ provides one of the few controlled numerical studies of a model where there is an observed violation of Hertz-Millis scaling. The critical exponents and $\mathrm{C}_4$-symmetric form of $\chi(\o, \boldsymbol q)$ can be directly tested experimentally with neutron scattering in materials where the bare hot-spot nesting value $v_B$ is small. 
\\ \indent 
Additionally, we provide strong evidence that the physics of the quantum critical metal is governed by the fixed point of Ref. \cite{PhysRevX.7.021010}, even for appreciable nesting parameter values $v$. If one believes this evidence, the implication is that the theory of Ref. \cite{PhysRevX.7.021010}, valid at small $v$, augmented with numerical calculations at larger $v$, can be used to generate a fully controlled quantitative solution to the problem. For example, non-equilibrium properties such as conductivity, which are generally very difficult to resolve with imaginary time methods, but are the most common observable used to experimentally identify the quantum critical fan, can be computed using small $v$ perturbation theory about the fixed point of Ref. \cite{PhysRevX.7.021010} and supplemented with non-perturbative numerical calculation of the critical exponents. If valid, this approach would provide an unprecedented advance in the understanding of experimentally observed strange metals proximate to itinerant antiferromagnetism, characterized by, e.g., linear-in-temperature DC resistivity and present in many materials displaying unconventional superconductivity.

\subsection{Future directions with HMC}

As noted in the main text, Fig. \ref{fig:z and eta_phi vs v} does not convincingly show that $z \rightarrow 1$ as $v \rightarrow 0$, and the most likely reason for this is the increasing value of $\lambda = g^2 c^2 /v$ as we decrease $v$. It is therefore desirable to revisit this problem, while keeping $\lambda$ fixed by tuning $g$. We intend to address this in a future work.
\\ \indent
In this work, we focus on only two critical exponents of the spin-fermion model, $z$ and $\eta_{\phi}$, for the most straightforward comparison to the nearly-nested fixed point theory of Ref. \cite{PhysRevX.7.021010}. There are several other pieces of critical data that remain to be examined for a more comprehensive comparison. The two most important ones are the critical exponent $\nu$ that relates the correlation length to the deviation from the critical point and the scaling of the fermion Green's function. Measuring the critical scaling of thermodynamic quantities like specific heat is also an important task. One can also tune $g$ to a large enough value to see the onset of superconductivity in order to track the dependence of $T_c$ on $v$ and compare to the prediction of $T_c(v)$ made in Ref. \cite{PhysRevX.7.021010}. Finally, here we only study the parameter values $u = 0, c = 3$, and five small values of $v$. An obvious question is how the conclusions of this work change with varying parameters $u,c$, as well as larger values of $v$.
\\ \indent 
The HMC algorithm we use in this work is quite general and can be applied to many other theories of interest. The main restrictions are that the configuration (boson) field be continuous and that the number of fermions can be doubled without qualitatively changing the physics of the model. Some immediate candidates are other sign-problem-free `designer' models of various flavors of quantum criticality, such as easy-plane $XY$ \cite{PhysRevB.95.035124} and $Z_2$ \cite{doi:10.1073/pnas.1901751116} antiferromagnetic transitions in metals, the Ising nematic transition in metals \cite{PhysRevX.6.031028}, fermions coupled to emergent gauge fields \cite{Gazit_nature_phys, PhysRevX.6.041049, doi:10.1073/pnas.1806338115, PhysRevLett.121.086601, PhysRevX.9.021022, Chen_2020, PhysRevX.10.041057}, Kondo lattice physics \cite{PhysRevB.100.035118, PhysRevLett.120.107201, PhysRevB.104.L161105}, disorder-averaged criticality \cite{https://doi.org/10.48550/arxiv.2203.04990}, Gross-Neveu-Yukawa criticality \cite{PhysRevLett.123.137602}, and models of flat band physics \cite{https://doi.org/10.48550/arxiv.2204.02994, Zhang_2021, PhysRevB.100.115135, PhysRevX.12.011061}. Other classes of models are Hubbard-type models and electron-phonon models, which have been recently studied \cite{PhysRevB.36.8632,PhysRevB.38.12023,PhysRevB.97.085144,PhysRevLett.102.026802,PhysRevB.78.165423,PhysRevLett.111.056801,PhysRevB.98.235129,brower2012hybrid,PhysRevB.100.075141,PhysRevB.89.195429,KRIEG201915,PhysRevB.102.245105,PhysRevB.104.155142,ULYBYSHEV2019118,arxiv.2203.01291,arxiv.2203.07380} using various HMC and DQMC algorithms. We note that for theories with dynamical gauge fields of continuous gauge groups or non-linear-sigma models, HMC is less efficient due to the trouble it has in sampling the various topological sectors via molecular dynamics. In these cases, perhaps an HMC augmented with generative model sampling might make for a very efficient sampling \cite{PhysRevLett.125.121601}. This issue is not present in the present model, since the magnitude of the $\mathrm{O}(3)$ bosonic field is not restricted.

\subsection{Potential algorithmic improvements}
\label{sec:alg-improvements}

The implementation could benefit from more sophisticated approaches to solving the linear system arising from each integration step within the HMC algorithm. Currently, the computational cost of our preconditioned CG approach enjoys ideal scaling in the limit of large $L$. However, the scaling with respect to $N_\tau$ for $\Delta \tau$ fixed, i.e., the zero-temperature limit, is superlinear with respect to $\Delta \tau$. Several approaches to similar linear solves have been introduced in the LQCD and condensed matter literatures that could pay dividends here. These include Hasenbuch preconditioning\cite{Hasenbusch:2001xh, KRIEG201915}, multigrid preconditioners and solvers \cite{clark-multigrid2016}, and non-iterative ideas based on Schur complements \cite{ULYBYSHEV2019118}. These approaches should also help for larger values of the Yukawa coupling, when the preconditioner used in this work would be less effective.
Moreover, a mixed-precision implementation of CG on the GPU could provide significant speedup \cite{Clark2010}.
\\ \indent 
There is also room for further investigation into the HMC hyperparameter auto-tuning procedure, as there exist alternative criteria for tuning $\eps$ and  $n_{\text{leap}}$. For example, one could instead tune $\eps$ based on the average acceptance rate or based on the stability of the numerical integrator. Meanwhile, the tuning of $n_{\text{leap}}$ is by far the most expensive part of the warmup phase. The current state-of-the-art approach in the statistics literature for tuning $\nleap$ is called the No U-Turn Sampler (NUTS) \cite{JMLR:v15:hoffman14a}. Implementing NUTS in our setting might offer a similar advantage.

\section{Methods}
\label{sec:method}

\subsection{QMC path integral with fermions}
\label{sec:target-theory}
 
In order to simulate the action of Eq. (\ref{eq:sign-problem-free action}) we put the system on a finite square lattice of size $L$ with periodic boundary conditions. The lattice spacing is set to one. The imaginary time direction is also discretized into $N_{\tau}$ slices, with the imaginary time spacing, or Trotter step, labeled as $\Delta \tau$. We set $\Delta \tau = 0.1$. Fermions and boson obey anti-periodicity and periodicity, respectively, in the time direction. As in any path integral quantum Monte Carlo method involving fermions, we start by writing out the partition function and performing the Grassmann integral over the fermionic fields:
\begin{equation}
    \begin{split}
    Z & = \int d \boldsymbol{\phi} \; e^{-S_B(\boldsymbol \phi)} \int d \psi^{*} d \psi \; 
    e^{-S_F(\psi,\psi^*) - S_{FB}(\boldsymbol \phi, \psi, \psi^*)}
    \\ &
    = \int d \boldsymbol{\phi} \; e^{-S_B(\boldsymbol \phi)} \det(D(\boldsymbol{\phi}))^{N_f}.
    \end{split}
    \label{eq:Z with determinant Nf general}
\end{equation}
Here, $D(\boldsymbol \phi)$ is the fermion matrix of size $2 \cdot  2 \cdot  N_\tau \cdot L \cdot L = 4 N_\tau L^2 $, where the two factors of $2$ correspond to the band and spin indices. Since the fermionic action is diagonal in the flavor index $j$, $N_f$ enters trivially as a power of the determinant. Component-wise $D(\boldsymbol \phi)$ is given by
\begin{equation}
    \begin{split}
    & D(\boldsymbol \phi)_{(\a, s, \tau, x, y), (\a', s', \tau', x', y')} = 
    \\ &
    \delta_{\a,\a'} \delta_{s,s'} 
    \{
    \delta_{x,x'} \delta_{y,y'} 
    (\delta_{\tau,\tau'}[-1 - \Delta \tau \, \mu_{\a}] 
    \\ &
    + \delta_{\tau+1,\tau'} [1 - 2 \, \delta_{\tau,N_{\tau} - 1}])
    - \delta_{\tau,\tau'} \, \Delta \tau \, t_{\a,(x,y),(x',y')} \}
    \\ &
    + \delta_{x,x'} \delta_{y,y'} \delta_{\tau,\tau'} \; \Delta \tau \; g \; e^{i \boldsymbol{Q}_{AF} \cdot (x,y)} \; \boldsymbol{\phi}(\tau, x, y) \cdot \boldsymbol{\tau}_{\s,\s'} \, \sigma^{(x)}_{\a, \a'},
    \end{split}
    \label{eq:D_formula}
\end{equation}
which is read directly from the action in Eq. (\ref{eq:sign-problem-free action}). Here, $\tau$ is in the range $\{0, \dots, N_{\tau} - 1\}$, and $x,y$ are in the range $\{0, \dots, L - 1\}$. $\delta_{a,b}$ is the Kronecker delta function. $\sigma^{(x)}_{\a,\a'}$ is the usual Pauli matrix, but acting on the band indices instead of the spin indices.

\subsection{Stochastic formulation of the determinant}
\label{sec:pseudofermions}

In the rest of this section, we will use the notation $\phi$ (not bold) to denote the bosonic field, to emphasize its interpretation alternatively (dependent on context) as either a vector of length $3 N_\tau L^2$ or an array of size $3 \times N_\tau \times L \times L$, depending on context, as opposed to its equivalent representation as 3-component field $\boldsymbol{\phi} = \boldsymbol{\phi}(\tau,x,y)$.
\\ \indent 
In the usual BSS algorithm \cite{PhysRevD.24.2278}, the determinant in Eq. (\ref{eq:Z with determinant Nf general}) is rewritten as
\begin{equation}
    \det(D(\phi)) = \det(\mathbf{1} + \prod_{l = 0}^{N_\tau-1} B_{l}(\phi)),
\end{equation}
where $\mathbf{1}$ and the $B_{l}(\phi)$ are square matrices of size $2 \cdot 2 \cdot L \cdot  L = 4L^2$. This reduces the cost of exact computation of the determinant from the naive scaling of $\mathcal{O}((4 N_\tau L^2)^3)$ to the improved scaling of $\mathcal{O}(N_\tau (4 L^2)^3)$ for every configuration $\phi$. Although this is better than the direct computation of the larger determinant, the scaling with respect to $L$ is nonetheless quite severe. 
\\ \indent
As an alternative to the exact computation of the determinant, we use the pseudofermion method \cite{GattLang, Fucito:1980fh} to evaluate it stochastically. This method is based on the expression
\begin{equation}
\det(A) \propto \int \mathcal D \varphi \mathcal D \varphi^* \; e^{-\varphi^* A^{-1} \varphi},
\label{eq:pf-integral}
\end{equation}
where $\varphi$ is an auxiliary complex bosonic field (called a `pseudofermion') and $A$ is any {Hermitian positive definite} matrix.
\\ \indent 
Although the anti-unitary symmetry of Ref. \cite{Berg21122012} guarantees that $\det(D(\phi))$ is always non-negative, $D(\phi)$ itself is not necessarily positive definite. However, we can write
\begin{equation}
\begin{split}
\det(D)^{N_f} & = (\det(D) \, \det(D)^*)^{N_f/2} 
\\ &  
= (\det(D) \, \det(D^{\dagger}))^{N_f/2} 
= \det(D D^{\dagger})^{N_f/2}.
\end{split}
\end{equation}
The matrix $D(\phi) D(\phi)^{\dagger}$ is guaranteed to be positive definite. Note that when $N_f$ is odd, the integral in Eq. (\ref{eq:pf-integral}) comes with a fractional power, which limits us to the case of even $N_f$. From here on, we set $N_f = 2$, and moreover this value is used in our simulations.
\\ \indent 
Rewriting the determinant in this way, the partition function becomes
\begin{equation}
    Z = \int d \phi \, d \varphi^* d\varphi \; e^{-\left(\mathcal{S}_B(\phi) + \mathcal{S}_{PF}(\phi,\varphi) \right)},
    \label{eq:Z with pfs}
\end{equation}
where $\mathcal{S}_{PF}(\phi,\varphi) = \varphi^* (D(\phi) D(\phi)^{\dagger})^{-1} \varphi$. Note that the dimension of $\varphi$ is the same as $D(\phi)$, i.e., $4 N_{\tau} L^2$. This new partition function defines a joint distribution
$p(\phi, \varphi)$, which can be sampled with Markov chain Monte Carlo (MCMC).

\subsection{Solving the linear system}
\label{sec:precond}

While Eq. (\ref{eq:Z with pfs}) introduces a matrix inverse to the action, the expression can be evaluated efficiently by treating the application of $(D D^{\dagger})^{-1}$ to $\varphi$ as the solution to a linear system:
\begin{equation}
\label{eq:linsys}
(DD^\dagger) \eta = \varphi, 
\end{equation}
where $\eta$ is the unknown. A wide class of iterative solvers are available to tackle this problem efficiently. Additionally, other non-iterative solvers have recently been employed in the application of HMC to the Hubbard model \cite{ULYBYSHEV2019118}. Here, we use the conjugate gradient (CG) method \cite{fletcher1964}, a commonly used technique with practical advantages for Hermitian positive definite systems.

\subsubsection{Choice of preconditioner}
\label{sec:precond}
Iterative solvers can often be preconditioned with a transformation that improves the conditioning of the linear system. For any preconditioner $P$, the equation $P^{-1} (DD^\dagger)\eta = P^{-1} \phi$ has the same solution as the original system but may be better conditioned. Within the CG algorithm, we choose a preconditioner as follows.

Note that $D(\phi)$ as defined in Eq. \eqref{eq:D_formula} can be split into two terms,
\begin{equation}
    \label{eq:Dsplit}
    D(\phi) = K + V(\phi),
\end{equation}
where the first term, $K$, i.e. the fermionic kinetic energy, consists of all terms in Eq. \eqref{eq:D_formula} that are independent of $\phi$. Since $K$ is a translation-invariant operator (modulo fermionic imaginary-time antiperiodicity), it can be diagonalized and inverted efficiently using fast Fourier operations and then used to construct an efficient preconditioner.

To wit, let $\mathcal{F}$ denote the discrete Fourier transform with respect to the spacetime lattice indices $(\tau, x, y)$, and let $\mathcal{T}$ be the unitary diagonal `twist' operator defined by 
\begin{equation}
\label{eq:twist}
\mathcal{T}_{(\alpha,s,\tau,x,y), (\alpha',s',\tau',x',y')} = \delta_{(\alpha,s,\tau,x,y), (\alpha',s',\tau',x',y') } \, e^{\pi i \tau}.
\end{equation}
Note that $\mathcal{T}$ transforms imaginary-time-periodic fields to imaginary-time-antiperiodic fields, and as such we can write 
\begin{equation}
    \label{eq:Kdiag}
    K = \widetilde{\mathcal{F}}\, \widehat{K} \, \widetilde{\mathcal{F}}^*,
\end{equation}
where $ \widetilde{\mathcal{F}} := \mathcal{T} \mathcal{F} $ is the `twisted' Fourier transform and $\widehat{K}$ is diagonal.

Our preconditioner is then defined by
\begin{equation}
    \label{eq:precon}
    P := (KK^{\dagger})^{-1} = \widetilde{\mathcal{F}}\, \widehat{P} \, \widetilde{\mathcal{F}}^*,
\end{equation}
where $\widehat{P} = \widehat{K} \widehat{K}^{\dagger}$ is diagonal. As such, $P$ can be applied with computational cost that is linear (up to log factors) in the spacetime volume $N_\tau L^2 $. In the limit where the coupling constant $g \rightarrow 0$, this is the perfect preconditioner, $P = (DD^{\dagger})^{-1}$, achieving convergence in a single CG iteration. More generally, the preconditioner is effective in addressing poor conditioning due to the unboundedness of the differential operator term of $D$ as the imaginary time step $\Delta \tau$ is refined, in analogy with, e.g., Laplacian preconditioners \cite{doi:10.1137/18M1212458} in the numerical PDE literature.

See also \cite{PhysRevB.36.8632} for a similar approach to preconditioning based on a noninteracting model. Note that such a preconditioner may not be satisfactory in general in the strong coupling limit, and we discuss possible future improvements in Section~\ref{sec:alg-improvements}. For the couplings studied in this paper, we do nonetheless observe improvement due to the preconditioner, as detailed in Supplementary Note 5.

\subsection{HMC}
Hybrid/Hamiltonian Monte Carlo (HMC) is a Markov chain Monte Carlo (MCMC) method in which numerical integration of Hamilton's equations in a fictitious phase space produces sequential updates in a Markov chain \cite{DUANE1987216,NealHMC}. Historically used as tool for the study of lattice quantum chromodynamics (LQCD), HMC is currently the state-of-the-art method in the field, where it has yielded sub-$1\%$ errors on over $\sim 10^{10}$ degrees of freedom \cite{luscher-master, clark-multigrid2016}. 
\\ \indent 
We now give a brief overview of HMC as applied to theories with fermions. As explained above, Eq. (\ref{eq:Z with pfs}) defines a joint probability distribution,
\begin{equation}
    p(\phi, \varphi) = \frac{1}{Z} e^{-\mathcal{S}_B(\phi) - \mathcal{S}_{PF}(\phi,\varphi)},
    \label{eq:effective-target-density}
\end{equation}
that needs to be sampled to estimate observables. We adopt a Gibbs sampler approach to sampling the two fields $\phi, \varphi$, wherein we update each one conditioned on the other in alternating fashion. 
\\ \indent 
The $\varphi$ sample can be drawn exactly from the conditional distribution for fixed $\phi$, which is the complex normal distribution $\mathcal N(0, D(\phi) D(\phi)^\dagger)$. This is achieved by sampling a standard complex Gaussian vector $\chi$ and setting $\varphi = D(\phi) \chi $.
\\ \indent 
Because the action is not quadratic in $\phi$, sampling of this field remains non-trivial but can be handled with MCMC. We employ the HMC method for this purpose. The first step is to introduce a field $\pi$ that acts as an artificial conjugate momentum to the field $\phi$. Together, $\phi$ and $\pi$ define phase space coordinates for a fictitious Hamiltonian system
\begin{equation}
\td H (\phi, \varphi, \pi) = K(\pi ; M) + \mathcal{S}(\phi, \varphi) = \frac{1}{2} \pi^\top M^{-1} \pi + \mathcal{S}(\phi, \varphi),
\label{eq:hmc Hamiltonian}
\end{equation}
consisting of `kinetic energy' and `potential energy' terms. The kinetic energy specifically is defined with respect to a choice of metric $M$, also called a mass matrix, which is given to be positive definite. The choice of $M$ will be discussed later. This Hamiltonian allows us to define a new joint distribution $p(\phi, \varphi,  \pi) \sim e^{-\td H (\phi, \varphi, \pi)}$, compatible with our target distribution after marginalization of $\pi$.
\\ \indent 
 The new joint distribution can now be efficiently sampled by drawing $\pi$ directly as a normal random vector with covariance $M$ and then integrating the equations of motion of $\td H$,
\begin{align}
\frac{d \phi}{d t}= M^{-1} \pi \quad \text{and} \quad \frac{d \pi}{d t}=-\frac{\partial \mathcal S (\phi, \varphi)}{\partial \phi}.
\label{eq:hamiltonseq}
\end{align}
Here, $t$ is a fictitious `time' variable. The trajectory defined by Eqs. (\ref{eq:hamiltonseq}) is also known as the molecular dynamics trajectory. Once a new sample of $(\phi,\pi)$ is obtained, its $\phi$ component is taken as the new sample of $p(\phi, \varphi)$. 
\\ \indent 
With these tools, we can construct the Markov chain update for $\phi$ as follows: 
\begin{enumerate}
\itemsep0em 
    \item Generate a momentum sample from $\pi \sim \mathcal{N}(0, M)$.
    \item Approximately integrate the molecular dynamics (\ref{eq:hamiltonseq}) with initial condition $(\phi$, $\pi)$ for some time to produce new configuration $(\phi', \pi')$.
    \item Accept the new configuration ($\phi', \pi'$) with probability $\alpha = \mathrm{min}\left(1, e^{-\big\{\td H(\phi', \pi') - \td H(\phi, \pi)\big\}}\right)$. 
\end{enumerate}
Exact integration of Eqs. (\ref{eq:hamiltonseq}) is not feasible and must therefore be performed numerically in discrete time. A suitable choice of numerical integration scheme is leapfrog integration \cite{hockney2021}, which preserves phase space volume and ensures that the final Metropolis accept/reject step preserves detailed balance.

\subsection{Scale-invariant sampling}

In high dimensions, local MCMC methods encounter difficulties in the setting of `poorly scaled,' i.e., severely anisotropic, distributions. In order to guarantee a nonvanishing acceptance rate without correcting for this difficulty, the local moves in the MCMC sampler are constrained by the smallest scale present in the problem, and as such the autocorrelation time with respect to the large scales must grow accordingly. For a critical lattice model, the smallest scale present in the distribution is not bounded away from zero in the thermodynamic limit, and this difficulty must be addressed to maintain constant autocorrelation time in this limit. Existing approaches based on affine invariance are unsuitable in this context due either to a curse of dimensionality \cite{GoodmanWeare:2010:EnsembleMCMC} or to unacceptable (i.e., at least quadratic) scaling \cite{doi:10.1137/19M1304891} with respect to the spacetime volume $N_\tau L^2$. Relative to these approaches, we are able to exploit the known \emph{a priori} structure of our model, i.e., translation invariance, to achieve a fast scale-invariant sampler. Specifically, within the context of HMC, we learn an optimal metric $M$ `online.' In contrast with \cite{arxiv.2203.01291}, beyond translation invariance we do not assume any \emph{a priori} functional form for the metric, as any \emph{a priori} choice may struggle near criticality. Since $M$ is updated adaptively, it is then important to tune the HMC hyperparameters $\varepsilon$ and $n_{\mathrm{nleap}}$ adaptively in a scale-invariant fashion. For this task we adapt best practices from the statistics literature (as implemented, e.g., in \cite{osti_1430202}) to our setting, as we shall discuss below. These quantities (i.e., $M$, $\eps$, and $\nleap$) are all determined during a warm-up phase, after which they are fixed for the duration of the run.

\subsubsection{Online metric estimation}

HMC as described above allows for an arbitrary choice of metric or mass matrix $M$. When the underlying probability distribution to be sampled is poorly scaled, the metric must be adapted to the scaling of the distribution to maintain a constant autocorrelation time. An idealized choice for $M$ is given by the inverse of the exact covariance matrix ${\Sigma}_{\mathrm{true}} := \langle \phi \phi^\top \rangle$ of the underlying distribution $p$, i.e., by $M = {\Sigma}_{\mathrm{true}}^{-1}$. In the case where $p$ is the Gaussian $\mathcal{N}(0,\Sigma_{\mathrm{true}})$, such a choice is equivalent to rescaling $p$ to the standard (isotropic) Gaussian distribution $\mathcal{N}(0,\mathbf{1})$. More generally, such a choice can be viewed as rescaling $p$ to a distribution with unit covariance matrix.

There are two obstacles to achieving this idealized choice. First, the exact covariance ${\Sigma}_{\mathrm{true}}$ is not known \emph{a priori}. Second, even if it were known or estimated as $\Sigma \approx \Sigma_{\mathrm{true}}$, the cost of even storing this matrix scales as $\sim (V N_\tau)^2$, which already imposes a computational bottleneck. Worse still, within the HMC algorithm, we must generate samples from $\mathcal{N}(0,\Sigma^{-1})$, which in general scales as $\sim (VN_\tau)^2$, i.e., as the cost of a Cholesky factorization of $\Sigma$.

The first of these obstacles can be tackled via online estimation of the covariance matrix ${\Sigma}_{\mathrm{true}}$ from samples. To wit, we can choose some initialization for $\Sigma$ (typically $\Sigma = \mathbf{1})$ and run the HMC algorithm to produce a batch of $S$ samples $\phi^{(1)},\ldots ,\phi^{(S)}$ of the bosonic field, then reset $\Sigma \leftarrow \Sigma_{\mathrm{est}} := \frac{1}{S} \sum_{s=1}^S \phi^{(s)} \phi^{(s),T}$ and repeat the procedure until $\Sigma$ converges.

There is an appearance of circular reasoning to such an online estimation scheme in that it assumes the ability to draw samples $\phi \sim p$ via our MCMC sampler (which may be bottlenecked by a long autocorrelation time) in order to improve the autocorrelation of the MCMC method itself! However, it is important to realize that in the initial iteration, the autocorrelation time for the smallest eigenmodes of $\Sigma_{\mathrm{true}}$ is not long. Indeed, to get a nonvanishing acceptance probability, the step size in HMC must be tuned to be on the order of the smallest eigenvalue of the covariance. Due to anisotropy, this may result in very slow MCMC mixing in the highest eigendirections, but this does not interfere with our ability to estimate the smallest eigenmodes. After the first iteration, having corrected for the smallest eigenmodes, the procedure proceeds to correct the next smallest, etc.

Despite its theoretical appeal, there are two problems with this approach, from the point of view of computational scaling. First, to get an estimate $\Sigma_{\mathrm{est}} \approx \Sigma_{\mathrm{true}}$ of fixed accuracy requires a size-extensive number $S$ of samples, due to the size-extensive numerical rank of $\Sigma_{\mathrm{true}}$. Second, as mentioned earlier, the cost of storing and factorizing $\Sigma_{\mathrm{est}}$ is prohibitive. In fact, both problems can be addressed by exploiting translation invariance.

Indeed, since $\Sigma_{\mathrm{true}}$ is translation-invariant, it is equivalently diagonal in Fourier space---or more precisely, $(3 \times 3)$-block-diagonal with respect to the vector index $l = 1,2,3$ of $\boldsymbol \phi$. Therefore $\Sigma_{\mathrm{true}} = \mathcal{F} \,\widehat{\Sigma}_{\mathrm{true}}\, \mathcal{F}^*$, where $\mathcal{F}$ represents the discrete Fourier transform with respect to the spacetime lattice indices and $\widehat{\Sigma}_{\mathrm{true}}$ is block-diagonal. As such the block-diagonal can be obtained as the expectation $\mathrm{diag} \left( [\widehat{\Sigma}_{\mathrm{true}}]_{l l'} \right) = \left\langle \widehat{\phi}_l \odot \overline{\widehat{\phi}_{l'}} \right\rangle$, where samples $\widehat{\phi}$ are obtained as $\widehat{\phi} = \mathcal{F} \phi$ from samples $\phi$ and `$\odot$' indicates the entrywise product. In practice, for simplicity we consider only the true diagonal of $\widehat{\Sigma}$, which is sufficient to address the increasing anisotropy due to large volume. Concretely, we therefore form
\begin{equation}
\label{eq:covest}
\widehat{\sigma} \leftarrow \widehat{\sigma}_{\mathrm{est}} := \frac{1}{S} \sum_{s=1}^S \widehat{\phi} \odot \overline{\widehat{\phi}}
\end{equation}
and let $\Sigma = \mathcal{F} \, \widehat{\Sigma} \, \mathcal{F}^*$, where $\widehat{\Sigma}$ is the diagonal matrix with diagonal $\widehat{\sigma}$.

In order to draw sample momenta $\pi \sim \mathcal{N}(0, M)$ within the HMC algorithm, observe that $M = \Sigma^{-1} = (\mathcal{F} \, \widehat{\Sigma}^{-1/2} ) (\mathcal{F} \, \widehat{\Sigma}^{-1/2})^*$. Therefore we can simply draw $z \sim \mathcal{N}(0,\mathbf{1})$ and let $\pi = \mathcal{F} (\widehat{\sigma}^{-1/2} \odot z)$, where $\widehat{\sigma}^{-1/2}$ indicates the entrywise inverse square root of $\widehat{\sigma}$. The total cost of drawing such a sample is linear in the spacetime volume $L^2 N_\tau$, up to a log factor, as is the cost of evaluating the HMC energy of Eq. (\ref{eq:hmc Hamiltonian}).

Since the off-diagonal elements of $\widehat{\Sigma}$ are known \emph{a priori} to be zero (and implicitly set to zero automatically), we only need $O(1)$ effective samples to estimate $\widehat{\sigma}$ to fixed accuracy. Hence the total cost of each iterative update of $\widehat{\Sigma}$ scales linearly in the spacetime volume $L^2 N_\tau$, up to log factors.

\subsubsection{Online HMC hyperparameter tuning}

Within the HMC algorithm, once the metric $M$ is fixed, it remains to choose (subordinate to this choice) appropriate values for the integration step size $\varepsilon > 0$ and the number $\nleap$ of leapfrog integration steps per proposal.

First we consider the choice of $\eps$. Roughly speaking, one wants to choose $\eps$ as large as possible without sabotaging the acceptance rate. Quantitatively, let $\alpha(\eps)$ denote the expected acceptance probability for a single leapfrog integration step of size $\eps$. More precisely, recalling that $\alpha = \alpha(\phi,\pi, \phi',\pi')$ denotes the acceptance probability of a proposed move $(\phi,\pi) \rightarrow (\phi',\pi')$, we may define $\alpha(\eps) = \langle \alpha(\phi,\phi') \rangle_{\eps, \nleap = 1}$, where the statistical expectation is computed by sampling $(\phi,\pi) \sim p$ and then sampling $(\phi',\pi')$ as a proposal from $\phi$ according to the HMC algorithm with step size $\eps$ and $\nleap = 1$. Then one wants to maximize $\eps$ subject to the inequality 
\begin{equation}
\label{eq:acc_ineq}
(1 - \alpha(\eps/2))^2 \leq 2(1 - \alpha(\eps)).
\end{equation}
The left-hand side estimates the probability of accepting two steps of size $\eps/2$, whereas the right-hand side estimates the probability of accepting one step of size $\eps$. Since the former strategy costs twice as many linear solves $(DD^{\dagger})^{-1} \varphi$ as the latter, the latter is computationally preferable as long as the inequality holds.
\\ \indent 
In practice, we maintain an estimate for $\alpha(\eps)$ and $\alpha(\eps/2)$ based on an empirical average over a recent history of samples, and we increase or decrease $\eps$ when Eq. \eqref{eq:acc_ineq} is satisfied or violated, respectively.
\\ \indent 
Now we turn to the choice of $\nleap$. Within the actual HMC algorithm, we choose $\nleap$ uniformly at random from the set $\{1,\ldots, n_{\mathrm{max}} \}$, independently for each iteration, where $n_{\mathrm{max}}$ is a hyperparameter to be determined adaptively. This `jittering' procedure is standard in the statistics community \cite{osti_1430202}. The hyperparameter $n_{\mathrm{max}}$ is determined as the maximizer of the expected squared jump distance (ESJD) \cite{10.2307/24308995} 
\begin{equation}
\label{eq:ESJD}
\mathbf{ESJD} (n) := \left\langle\, (\phi - \phi')^\top M \,  (\phi - \phi') \, \alpha(\phi,\phi') \, \right\rangle_{\eps, \nleap = n}.
\end{equation}
Importantly, the metric $M$ is needed here to correctly define the squared distance $\left\Vert \phi - \phi'\right\Vert^2_{M} = (\phi - \phi')^\top M \,  (\phi - \phi')$.
\\ \indent 
In practice, similarly to $\alpha(\eps)$, we maintain an estimate for $\mathbf{ESJD} (n)$ based on an empirical average over recent history, and we choose $n_{\mathrm{max}}$ to maximize it.
\\ \indent 
During the warmup phase, the three components of $\eps, n_{\mathrm{max}}$ and $M$ are each tuned several times, until they stop changing appreciably.  

Note that since the acceptance probability in HMC is directly determined by the energy conservation error in the integration of Hamilton's equations~\eqref{eq:hamiltonseq}, alternative higher-order integration schemes besides leapfrog can be considered to improve acceptance rates at the price of more expensive integration steps, cf. for example~\cite{KENNEDY2001456,CALVO2021110333}.

\subsection{Estimation of observables}

Here we describe the methods we use to compute physical observables directly in Fourier space, while keeping the computational cost near-linear in $\beta V$. 

\subsubsection{Bosonic observables}
Bosonic observables can be computed directly as a sample average over the bosonic field variable $\phi$. Specifically of interest is the SDW susceptibility
\begin{equation}
\label{eq:SDWsus}
\chi(\omega, \boldsymbol q) = \int_0^\beta \sum_{\boldsymbol r} \langle \boldsymbol{\phi}(\tau, \boldsymbol{r}) \cdot \boldsymbol{\phi}(0,\boldsymbol{0}) \rangle  e^{i \omega \tau -i \boldsymbol q \cdot \boldsymbol r} \,d \tau.
\end{equation}
The  SDW susceptibility can be estimated at linear cost (up to log factors) as 
\begin{equation}
\label{eq:SDWsus}
\chi(\omega, \boldsymbol q) = \left\langle  \vert \widehat{\phi}(\omega ,\boldsymbol{q}) \vert^2 \right\rangle \,  \Delta \tau,
\end{equation}
where we recall that $\widehat{\phi} = \mathcal{F} \phi$ is the spacetime discrete Fourier transform of $\phi$.

\subsubsection{Fermionic observables}
\label{sec:fermion observables main text}
Fermionic observables require more care to estimate while maintaining almost-linear scaling. Of particular interest are the two-point correlator (in Fourier space) 
\begin{equation}
\label{eq:2pt}
G_{\alpha,s} (\omega, \boldsymbol q) = \int_0^\beta \sum_{\boldsymbol r} \langle  \psi_{\alpha, s} (0, \boldsymbol 0) \psi_{\alpha, s} (\tau, \boldsymbol r)^* \rangle  e^{i \omega \tau - i \boldsymbol q \cdot \boldsymbol r} \,d \tau.
\end{equation}
and the superfluid (SF) density, which is defined in Supplementary Note 3.
\\ \indent 
The key observation is that all fermionic observables of interest can be phrased in terms of the expectation value of the diagonal of a matrix for which it is possible to perform efficient matrix-vector multiplications, i.e., as
\begin{equation}
\langle \mathrm{diag}(O_\phi) \rangle,
\label{eq:fermi_obs}
\end{equation}
where the statistical average is taken with respect to the bosonic density for $\phi$ and where $O_\phi$ is an operator that depends on the bosonic field. Rather than compute the diagonal entries individually, it is more efficient to recover them simultaneously via the identity
\begin{equation}
\mathrm{diag}(O_\phi) = \mathbb{E}[v \odot (O_\phi v)],
\end{equation}
where $v$ is a random vector with independent entries that take values $\pm 1$ each with probability $1/2$ and `$\mathbb{E}$' indicates the expectation with respect to this distribution over $v$. As before, `$\odot$' indicates the entrywise product of vectors. This identity defines a randomized matrix-free algorithm for recovering a matrix diagonal from only $O(1)$ matrix-vector multiplications, which has appeared before in various works \cite{EstimatorDiagonalSaad, https://doi.org/10.1002/nla.779}. Moreover, after taking traces of both sides, one recovers the famous Hutchinson trace estimator \cite{doi:10.1080/03610919008812866}. During the preparation of this work, this randomized diagonal estimator has also appeared in \cite{arxiv.2203.01291} for the same purpose of computing fermionic observables, though their approach to computing quartic fermionic observables such as the SF density differ somewhat from ours.

Since we must average $\mathrm{diag}(O_\phi)$ over the bosonic distribution for $\phi$, we can in fact obtain a consistent estimator by independently drawing a single vector $v^{(s)}$ for each bosonic sample $\phi^{(s)}$, $s=1,\ldots, S$, where $S$ is the sample size of our empirical average and estimating
\begin{equation}
\label{eq:fermi_est}
\mathrm{diag}(O_\phi) \approx \frac{1}{S} \sum_{s=1}^S v^{(s)} \odot O_\phi v^{(s)}.
\end{equation}
Further details of the implementation of the approach to computing fermionic observables (i.e., the specification of $O_\phi$ for observables of interest) can be found in Supplementary Note 3.

\subsection{Numerical Performance}
\label{sec:performance}

Here we summarize the numerical performance of our algorithm, with more details provided in Supplementary Note 6. As noted in the Introduction, in the presence of a critical slowing down, the HMC algorithm can require a number of integration steps per effective sample of $O(\beta^{1/4 + z_1}V^{1/4 + z_2})$, where $z_1, z_2$ are to be determined empirically. To extract these exponents, we benchmark our algorithm across lattice sizes at the critical parameters of the theory, scaling with respect to $V=L^2$ and $N_{\tau}$ separately. We track the growth of the integrated autocorrelation time $\tau_{\mathrm{int}}$ of the total SDW susceptibility $\chi \equiv \chi(0, 0)$ at criticality, and use it to quantify the wallclock time and number of HMC integration steps per effective sample: $\tau_{\mathrm{int}} \times n_{\mathrm{leap}}$.

We find that our algorithm exhibits constant scaling of $\tau_{\mathrm{int}}$ with respect to both lattice volume and inverse temperature. For the exponents, we find $z_1 \approx 0.5 , z_2 \approx 0$. This implies an absence of critical slowing down with respect to the lattice volume $V$ for the auto-tuned HMC algorithm presented in this work.

The actual wall clock time in turn depends on the cost of the linear solves of the form Eq. \eqref{eq:linsys}. With the preconditioned CG approach described in \Cref{sec:precond}, we observe near linear scaling in wall clock time per effective sample with respect to $V$, while the scaling is superlinear with respect to $N_\tau$. The linear solve can be approached with more advanced techniques, as well as more refined GPU parallelism. These directions are discussed in \Cref{sec:alg-improvements}.

\section{Data availability}
The data analyzed in the current manuscript is available from the corresponding author upon reasonable request.

\section{Code availability}
All the custom codes used in this study can be requested from the corresponding author. 


\section{Acknowledgements}
We are extremely indebted to Bob Carpenter for teaching us the auto-tuning procedure used in industry applications of HMC and employed in this paper. We thank Phiala Shanahan for insightful discussions. P.L. thanks Subir Sachdev, the attendees of the Aspen Center for Physics conference `New Directions in Strong Correlation Physics: From Strange Metals to Topological Superconductivity,' and in particular Sung-Sik Lee for useful discussions. We thank Sung-Sik Lee for comments on the manuscript. We thank Nick Carriero and the other members of the Scientific Computing Core at Flatiron Institute for their assistance in running the codes. M.A. and M.L. are grateful for the hospitality of the Center for Computational Quantum Physics at the Flatiron Institute. The Flatiron Institute is a division of the Simons Foundation. This work was partially completed at the Aspen Center for Physics, which is supported by National Science Foundation grant PHY-1607611. P.L. is supported by the Simons Foundation. M.A. is supported by the National Science Foundation under the award PHY-2141336. M.L. is supported by the National Science Foundation under Award No. 1903031.

\section{Author contributions}
P.L. conceived and directed the project, wrote the majority of the code, and collected and analyzed the data. M.A. provided the initial HMC code design, conceived and led the GPU implementation, wrote the code for several observables, and helped with the data analysis. M.L. conceived the ideas of using translation invariance in the metric, the stochastic estimators for the fermionic observables and the choice of preconditioner, worked out the expression for the superfluid density, and wrote the code for several observables. All authors extensively discussed the project and co-wrote the paper.

\section{Competing interests}
The authors declare no competing interests.

\onecolumngrid
\appendix

\section*{Supplementary Note 1. Details of results}
\label{sec:results details}

Here we provide more details and supplementary plots for our data analysis, the results of which are presented in the Results Section.

\subsection{Phase diagram}

\begin{figure}[!ht]
    \centering
    \includegraphics[width=0.45\textwidth]{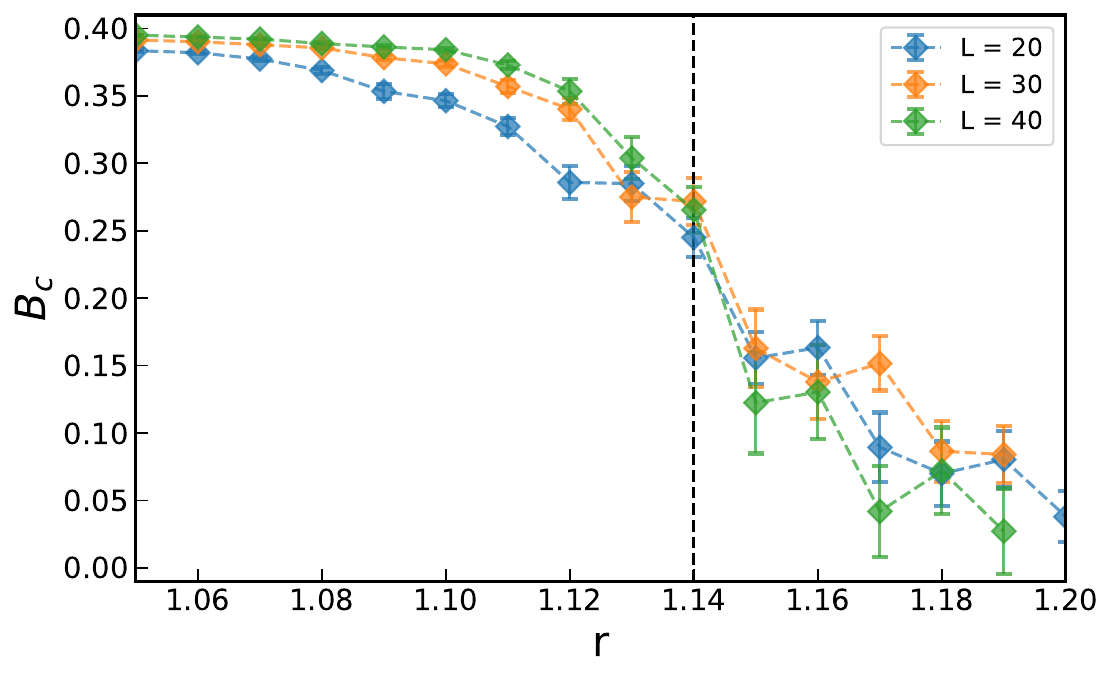}
    \caption{\textbf{Crossing of the Binder cumulant for different $L$.} Binder cumulant $B_c$ as a function of $r$ for $v = v_1$, $\beta = 10$. It goes from $B_c = 0$ in the disordered phase to $B_c = 0.4$ in the ordered phase. The `critical point' is identified as the crossing point of the different $L$ curves, here taken to be $r_c(T) = 1.14 \pm 0.01$.}
    \label{fig:Binder_c v_2 beta_10}
\end{figure}

The phase diagram of Fig. 4 is obtained by first identifying the location of the thermal `phase transition,' $r_c(T)$, at various temperatures. Due to the Mermin-Wagner theorem, this transition is actually a crossover. However, a sharp $r_c(T)$ can still be extracted from the finite-size behavior of the Binder cumulant. The Binder cumulant of the order parameter $\boldsymbol \phi$ is defined as $B_c \equiv 1 - \frac{3\langle \boldsymbol{\Phi}^4\rangle}{5 \langle \boldsymbol{\Phi}^2\rangle^2}$, where $\boldsymbol{\Phi} \equiv \sum_{\tau, x, y} \boldsymbol{\phi} (\tau, x, y)$. We show an example plot in Fig. \ref{fig:Binder_c v_2 beta_10} for $v = v_1$ and $\beta = 10$. The transition point $r_c(T)$ is defined as the value of $r$ at which all the different $L$ curves cross (at least for large enough $L$). In Supplementary Note 4 we show that the `would-be' transition is indeed second-order. Examining the curves in Supplementary Fig. \ref{fig:Binder_c v_2 beta_10} by eye gives us a value $r_c(T)$ with an error $\Delta r_c(T)$. 
\\ \indent 
Once $r_c(T)$ is obtained for several $T$ values, we extrapolate them to $T = 0$ using a simple polynomial fit, $r_c(T) = r_c - a \, T^{b}$, which gives us our estimate of the quantum critical point $r_c \equiv r_c(T = 0)$. The values of $r_c$ that we obtain for $v_i$, $i = 1, \dots, 5$ are all given by $r_c \approx 1.2554$. Putting everything together gives the phase diagram in Fig. 4. The error bars denote the one sigma confidence interval. 
\\ \indent 
To probe for superconductivity we use the method of Ref. \cite{PhysRevB.47.7995}. We observe no superconductivity at $r_c$ down to the lowest temperature that we study, $T = 1/80$. We give more details on this computation in Supplementary Note 2.
\\ \indent 
In Supplementary Fig. \ref{fig:occupation functions v = v_5} we show plots of the occupation function for more nesting parameter values, namely $v = v_5$.
\begin{figure}[!ht]
    \centering
    \includegraphics[width=1.0\textwidth]{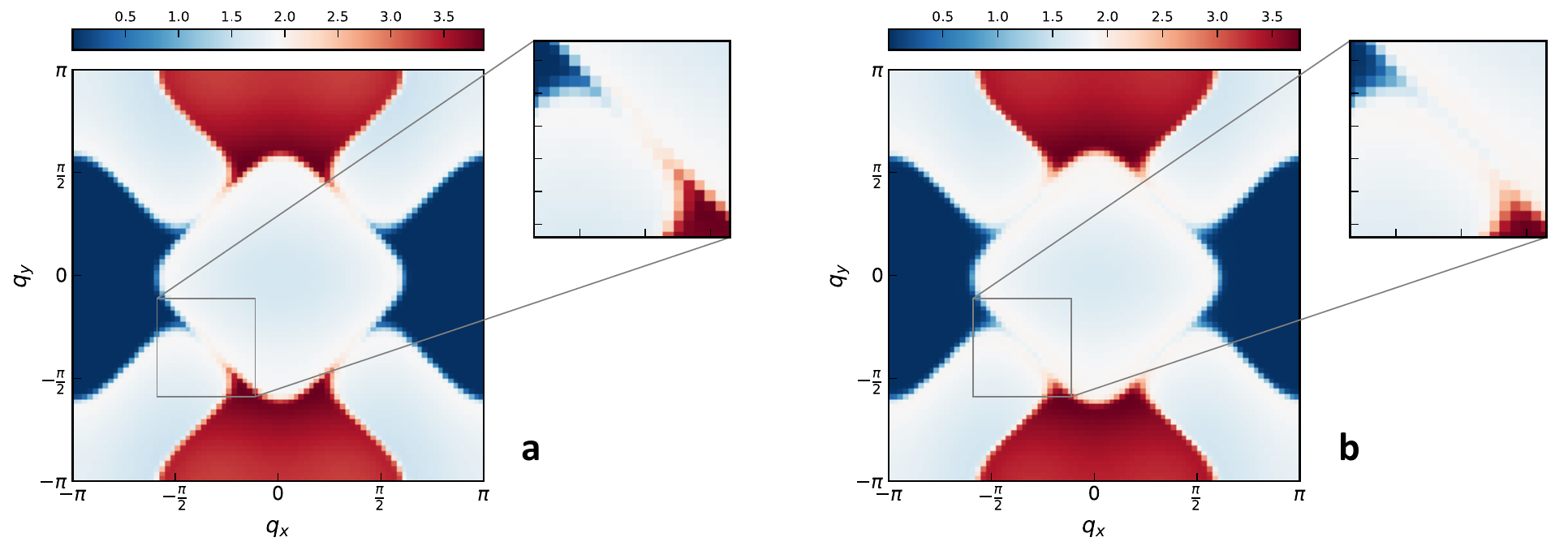}
    \caption{\textbf{Measured occupation number.} The total occupation number $n_{\boldsymbol q} = \sum_{\a,\s} \langle \psi^{\dagger}_{\a,\s,\boldsymbol q} \psi_{\a,\s,\boldsymbol q} \rangle$ for the (a) free theory and the (b) interacting theory at criticality ($r = r_c$). Both plots are at $v = v_5, \beta = 20, L = 80$.}
    \label{fig:occupation functions v = v_5}    
\end{figure}

\subsection{Scaling of spin susceptibility at criticality}
Having established the phase diagram and located the QCP, we examine the boson correlation function at criticality. We set $r = 1.26 \gtrsim r_c$, which is very slightly further into the disordered phase than the extrapolated value $r_c$. This is done to stay on the safe side of the transition, i.e. within the critical fan. 
\\ \indent 
The log-log plots of $\chi^{-1}(\o) - \chi^{-1}(0)$ for $\beta = 80, L = 20$ and of $\chi^{-1}(q_x) - \chi^{-1}(0)$ for $\beta = 20, L = 80$ are shown in Fig. 5a and Fig. 5b, respectively. As noted there, the regions of intermediate algebraic scaling are chosen by eye. For all nesting parameter values we define the regions as $ 10\pi/\beta \leq \o < 2.0$ and $ 8\pi/L \leq q_x < 2.3$. The slope of the linear fits in these regions determine the critical exponents $z$ and $\eta_{\phi}$, shown in Fig. 6. To compute the error bars, we used a bootstrap method, where the selected data points were re-sampled $10000$ times, allowing for duplicates.

The symmetry of the full spatial dependence is seen from Fig. 7 for $v = v_5$. In Supplementary Fig. \ref{fig:chi inv vs vec q density plots v_i i not 5} we provide similar zoomed-in plots for the other $v_i$, showing the $\mathrm{C}_4$ symmetry at small momenta.
\begin{figure*}[!ht]
    \centering
    \includegraphics[width=1.0\textwidth]{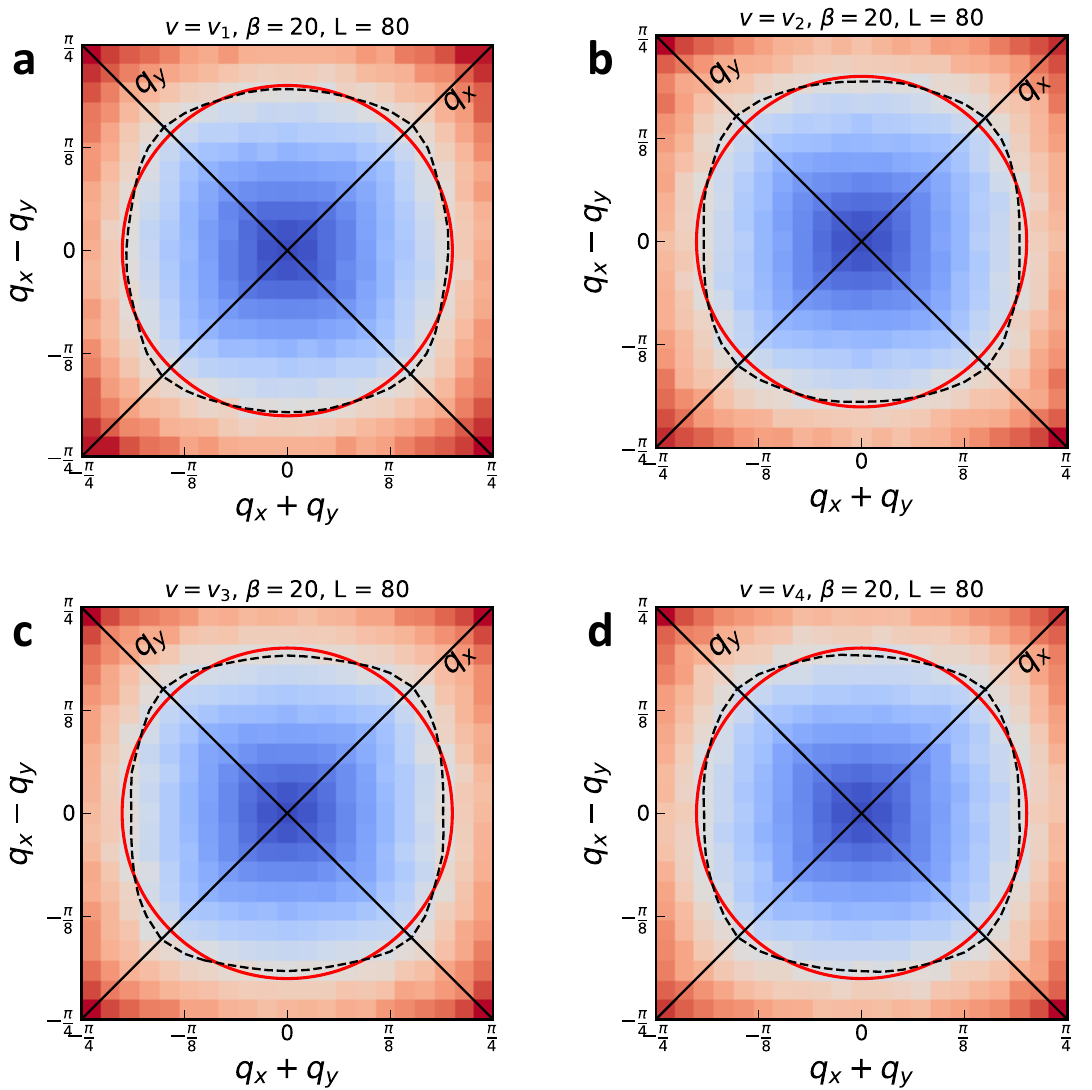}
    \caption{\textbf{Density plots of static spin susceptibility at criticality.} Zoomed-in plot of $\chi^{-1}(\boldsymbol q)$ for $v_i$, $i = 1,2,3,4$ at $\beta = 20, L = 80$, $r = r_c$. The color bar is absent for the reason explained in the caption of Fig. 7.}
    \label{fig:chi inv vs vec q density plots v_i i not 5}
\end{figure*}
In Supplementary Fig. \ref{fig:chi inv vs vec q density plots v_1 smaller L}, we also show that the contour at the same small momenta only get more `square-like' with increasing $L$, indicating that the emergent $\mathrm{C}_4$ symmetry is not a finite-size effect but a true long-wavelength feature of the theory in the thermodynamic limit.
\begin{figure*}[!htb]
    \centering
    \includegraphics[width=1.0\textwidth]{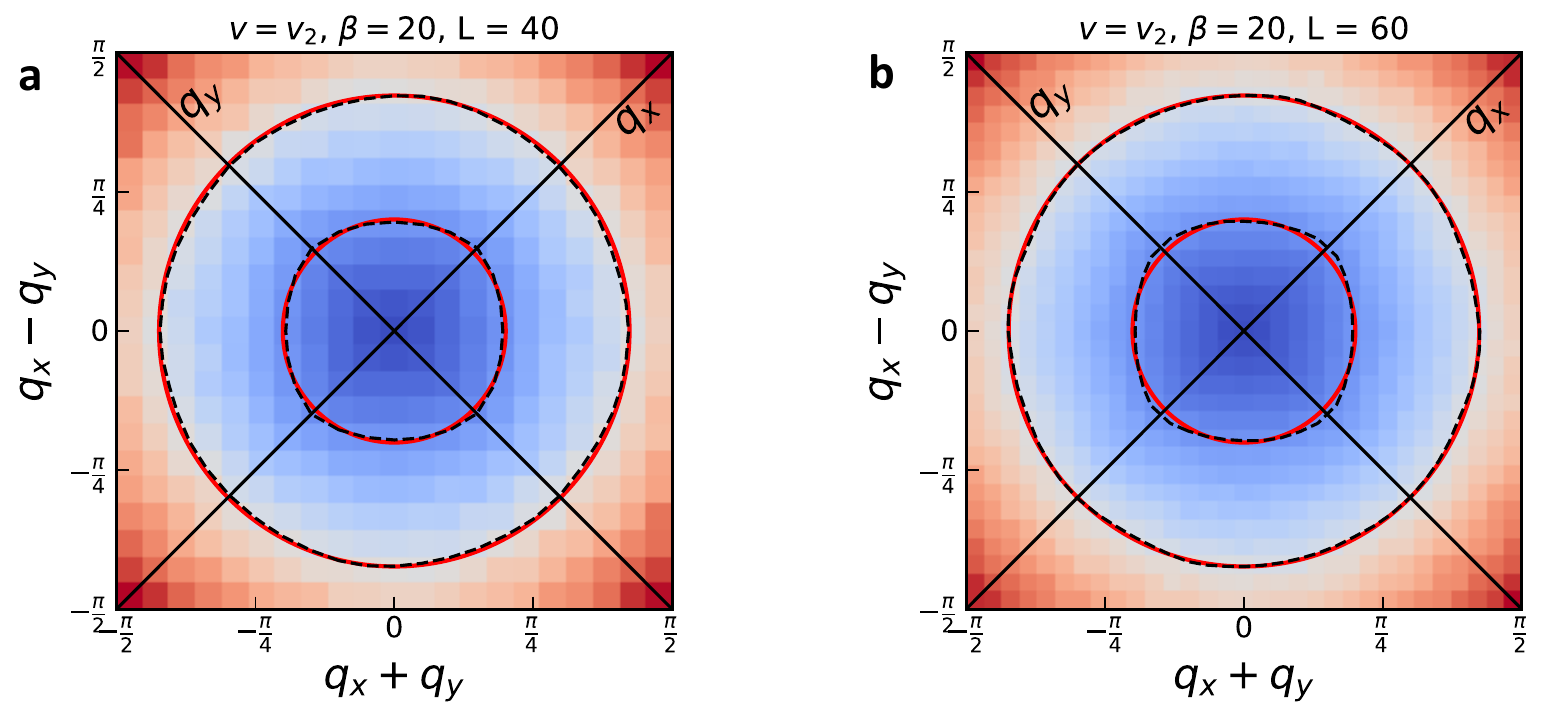}
    \caption{\textbf{Density plots of static spin susceptibility at criticality at smaller $L$.} Partially zoomed-in plot of $\chi^{-1}(\boldsymbol q)$ for $v = v_2, r = r_c, \beta = 20$ and (a) $L = 40$ and (b) $L = 60$. The red circles are drawn at the same radii for all system sizes. Compare to Fig. 7b. The color bar is absent for the reason explained in the caption of Fig. 7.}
    \label{fig:chi inv vs vec q density plots v_1 smaller L}
\end{figure*}


\section*{Supplementary Note 2. Superfluid density}
\label{sec:superfluid density}

In order to test for superconductivity in our system, we apply the method of Ref. \cite{PhysRevB.47.7995} to compute the superfluid density $\rho_s$. The system is a superconductor if $\rho_s$ surpasses the universal BKT value of $\Delta \rho_s = 2 T /\pi$.  

The current density operator in the $\hat x$ direction is given by
\begin{equation}
j^{(\mathrm{\hat x})}_{\tau,x,y} = \sum_{\alpha, \s, x'} i \, t_{\a,(x,y),(x',y)} \;  \psi^*_{\alpha,\s,\tau,x,y} \psi_{\alpha,\s,\tau,x',y} + \mathrm{h.c.}.
\label{eq:x-current density}
\end{equation}
The static current-current correlator in momentum space is defined as
\begin{equation}
\Lambda^{(\mathrm{xx})} (\boldsymbol q) = \Delta \tau \sum_{\tau, x,y} e^{-i \boldsymbol{q} \cdot \boldsymbol{r}}\,  \left \langle j^{(\mathrm{\hat x})}_{\tau,x,y} j^{(\mathrm{\hat x})}_{0,0,0} \right\rangle,
\label{eq:currentcurrentfourier}
\end{equation}
where we understand $\boldsymbol r = (x,y)$ as usual.

The superfluid density in the thermodynamic limit is then given by
\begin{equation}
    \rho_{\mathrm{s}} = \frac{1}{4} \left[ \lim_{q_x \rightarrow 0} \Lambda^{(\mathrm{xx})} (q_x,0) - 
    \lim_{q_y \rightarrow 0}
    \Lambda^{(\mathrm{xx})} (0,q_y) \right].
    \label{eq:sfdensity therm limit}    
\end{equation}
For a finite $L$, we use
\begin{equation}
    \rho_{\mathrm{s}}^{(L)} = \frac{1}{4} \left[ \Lambda^{(\mathrm{xx})} \left(\frac{2\pi}{L}, 0\right) - \Lambda^{(\mathrm{xx})} \left(0, \frac{2\pi}{L}\right) \right],
    \label{eq:sfdensity finite L}
\end{equation}
as a finite-size estimate of $\rho_{\mathrm{s}}$. In practice, $\rho_{\mathrm{s}}^{(L)}$ depends weakly on $L$, as is seen by others \cite{PhysRevLett.117.097002}. For all nesting parameters, at the lowest temperature ($\beta = 80$), we measure $\rho_{\mathrm{s}}^{(L)} \approx 0$, indicating the lack of superconductivity at the parameters values we study.

\section*{Supplementary Note 3. Calculation of fermionic observables}
\label{sec:fermionic obs}

Here we expound upon the computation of the fermion Green's function of Eq. (35) and the superfluid density of Supplementary Eq. (\ref{eq:currentcurrentfourier}) using the techniques outlined in the Methods Section.
\\ \indent 
In order to recover the Green's function as an expected diagonal of the form of Eq. (36) we simply take
\begin{equation}
    O_\phi = \widetilde{\mathcal{F}}^\dagger  \, D(\phi)^{-1} \,  \widetilde{\mathcal{F}},
\end{equation}
where $\widetilde{\mathcal{F}}$ is the twisted discrete Fourier transform defined in the Methods Section. Note that by rewriting 
\begin{equation}
    D(\phi)^{-1} = D(\phi)^\dagger \left[ D(\phi) D(\phi)^\dagger \right]^{-1}
\end{equation}
we can perform a matrix-vector multiplications by $O_\phi$ by performing linear solves of the form of Eq. (22), together with fast Fourier operations and sparse matrix-vector multiplications.

It can be shown via Wick's theorem together with laborious algebraic manipulations that 
\begin{widetext}
\begin{eqnarray}
\label{eq:bigcheese}
\Lambda^{(\mathrm{xx})} (\boldsymbol q) & \ \  = \ \ &  \Big\langle \left[\mathrm{diag}(A G^0_\phi BA G^0_\phi )\cdot\mathbf{1}\right]-\left[\mathrm{diag}(A G^0_\phi )\cdot\mathrm{diag}(B)\right]\left[\mathrm{diag}(A G^0_\phi )\cdot\mathbf{1}\right] \nonumber \\
 &  & \quad + \ \ [\mathrm{diag}(G^0_\phi A^{\dagger}B G^0_\phi A^{\dagger})\cdot\mathbf{1}]-\left[\mathrm{diag}(G^0_\phi A^{\dagger})\cdot\mathrm{diag}(B)\right]\left[\mathrm{diag}(G^0_\phi A^{\dagger})\cdot\mathbf{1}\right]\nonumber \\
 &  & \quad - \ \ \left[\mathrm{diag}(A G^0_\phi A^{\dagger}B G^0_\phi )\cdot\mathbf{1}\right]+\left[\mathrm{diag}(G^0_\phi A^{\dagger})\cdot\mathrm{diag}(B)\right]\left[\mathrm{diag}(A G^0_\phi )\cdot\mathbf{1}\right] \nonumber \\
 &  & \quad - \ \ \left[\mathrm{diag}(G^0_\phi BA G^0_\phi A^{\dagger})\cdot\mathbf{1}\right]+\left[\mathrm{diag}(A G^0_\phi )\cdot\mathrm{diag}(B)\right]\left[\mathrm{diag}(G^0_\phi A^{\dagger})\cdot\mathbf{1}\right] \  \Big\rangle,
\end{eqnarray}
\end{widetext}
where $G^0_\phi := D(\phi)^{-1}$, and the diagonal matrices $A$ and $B$ are defined by 
\begin{equation}
    A_{(\a,\s,\tau,x,y),(\a',\s',\tau',x',y')}=\delta_{\a\a'}\delta_{\s\s'}\delta_{\tau\tau'}\delta_{yy'}\ t_{\a,(x,y),(x',y')}
\end{equation}

and 
\begin{equation}
    \begin{split}
    & B_{(\a,\s,\tau,x,y),(\a',\s',\tau',x',y')} =
    \\ &
    \frac{\Delta \tau}{N_\tau L^2}\  \delta_{\a\a'}\delta_{\s\s'}\delta_{\tau\tau'}\delta_{xx'}\delta_{yy'} \ e^{-i\boldsymbol q \cdot (x,y)}.
    \end{split}
\end{equation}
Observe that $B = B_{\boldsymbol q}$ implicitly depends on the choice $\boldsymbol q$, but for any specific choice of $\boldsymbol q$, the angle-bracketed expression in Supplementary Eq.  \eqref{eq:bigcheese} can be obtained in computational time dominated by the cost of a constant number of linear solves of the form of Eq. (22). Note that for any term in which diagonals appear within a product, independent diagonal estimators of the form of Eq. (38) must be used for each in order to obtain a consistent estimator for the expression in Supplementary Eq. \eqref{eq:bigcheese}.


\section*{Supplementary Note 4. Order of the transition}
\label{sec:order of transition}

Although the effective action of Eq. (1) is modeling a continuous phase transition, the transition can be first-order, especially for $u=0$ \cite{PhysRevB.50.14048}. In order to check that the transition is indeed second-order for all the parameter values we use in the main text, we plot the inverse spin susceptibility $\chi^{-1} = \chi^{-1}(0,\boldsymbol 0)$ as a function of $r$ for $v = v_1$, $\beta = 40, L=20$. We can see that even at low temperatures, the susceptibility turns on in a continuous fashion. 
\begin{figure}[!htb]
    \centering
    \vspace{6mm}
    \includegraphics[width=0.45\textwidth]{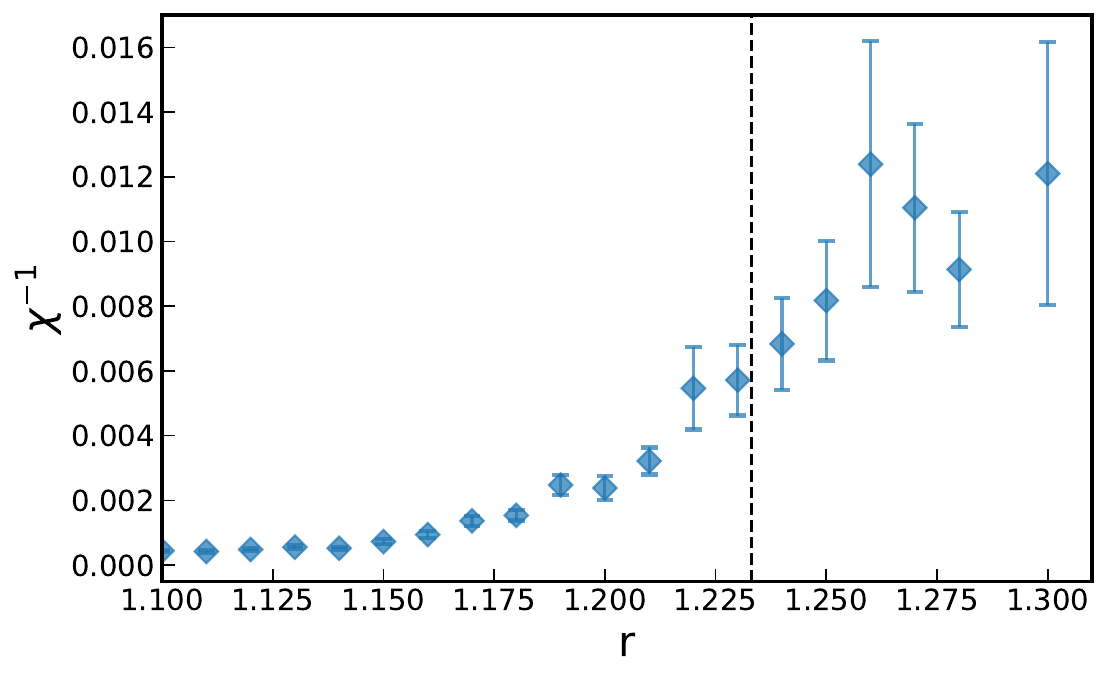}
    \caption{\textbf{The (unnormalized) inverse susceptibility.} $\chi^{-1}$ for $v = v_1$, $\beta = 40, L=20$ as a function of $r$. The error bars denote the one sigma confidence interval.}
    \label{fig:chi inv nesting 2 beta 40 L 20}
\end{figure}
This is in contrast to the first-order behavior observed in Ref. \cite{PhysRevB.98.045102} for certain parameter values, where $\chi^{-1}$ develops a pronounced jump at lower temperatures.

\section*{Supplementary Note 5. Preconditioner performance}
\label{app:preconditioner}
In this section we numerically validate our choice of preconditioner as described in the Methods Section. All experiments are performed for our standard choices of model parameters
$u = 0.0$, $g=0.7 \,\sqrt{2}$, $c = 3.0$, as well as the choice $v=v_2\approx 0.072$ for the nesting parameter and the critical parameter value $r=1.26$.

In Supplementary Fig. \ref{fig:preconSpeedup}, we fix $N_\tau = 100$ and plot the average speedup provided by the preconditioner over an HMC run, measured both in terms of wall clock time and CG iterations.

\begin{figure*}[!htb]
    \centering
    \includegraphics[width=0.9\textwidth]{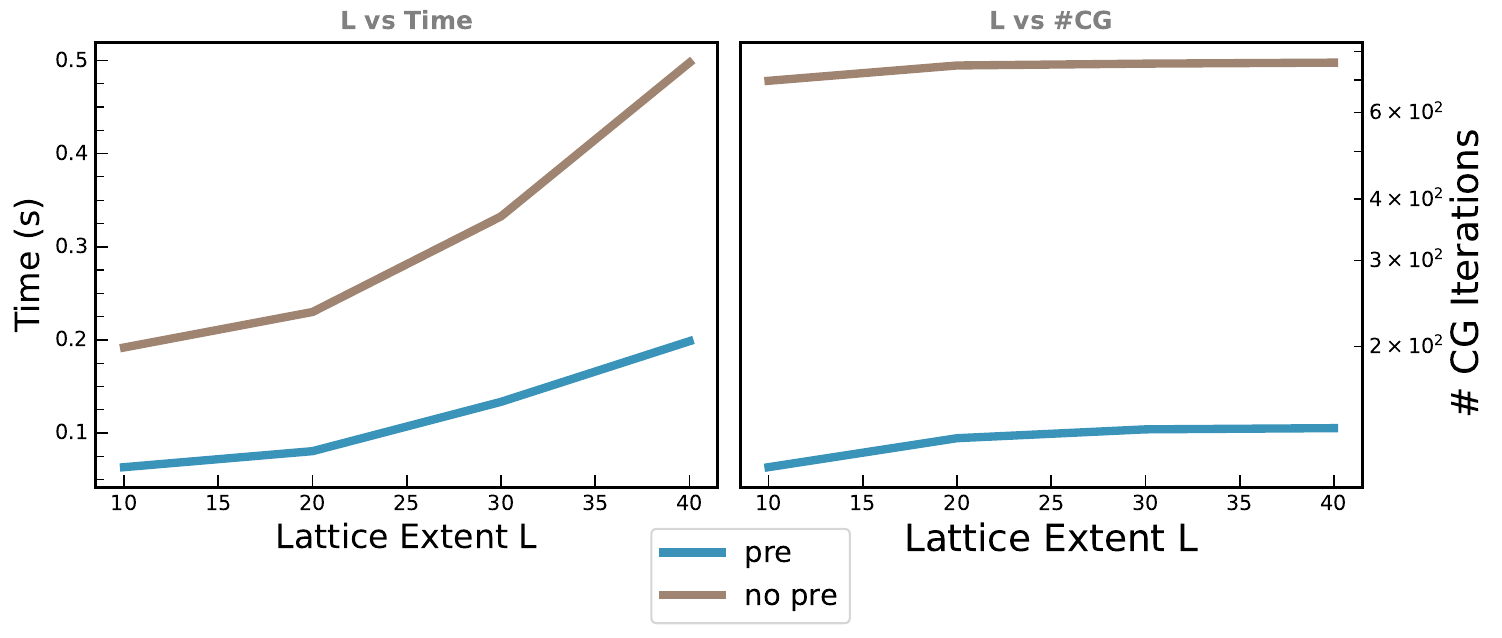}
    \caption{\textbf{Preconditioner performance.} Average preconditioner speedup over an HMC run as a function of lattice extent $L$, where $N_\tau = 100$ is fixed, measured both in terms of wall clock time and number of CG iterations.}
    \label{fig:preconSpeedup}
\end{figure*}

To bolster the practical conclusions of Supplementary Fig. \ref{fig:preconSpeedup} and see more concretely how the preconditioner affects the conditioning of the linear solve, we plot in Supplementary Fig. \ref{fig:cond} a histogram of the condition numbers of the matrices $M=DD^\dagger$ and $M=PDD^\dagger$, where $D = D(\phi)$ is the fermion matrix and $P$ is the preconditioner of Eq. (26), collected over configurations $\phi$ obtained from an HMC run, for two alternative choices $N_\tau= 50, 100$.

\begin{figure*}[!ht]
    \centering
    \includegraphics[width=1.0\textwidth]{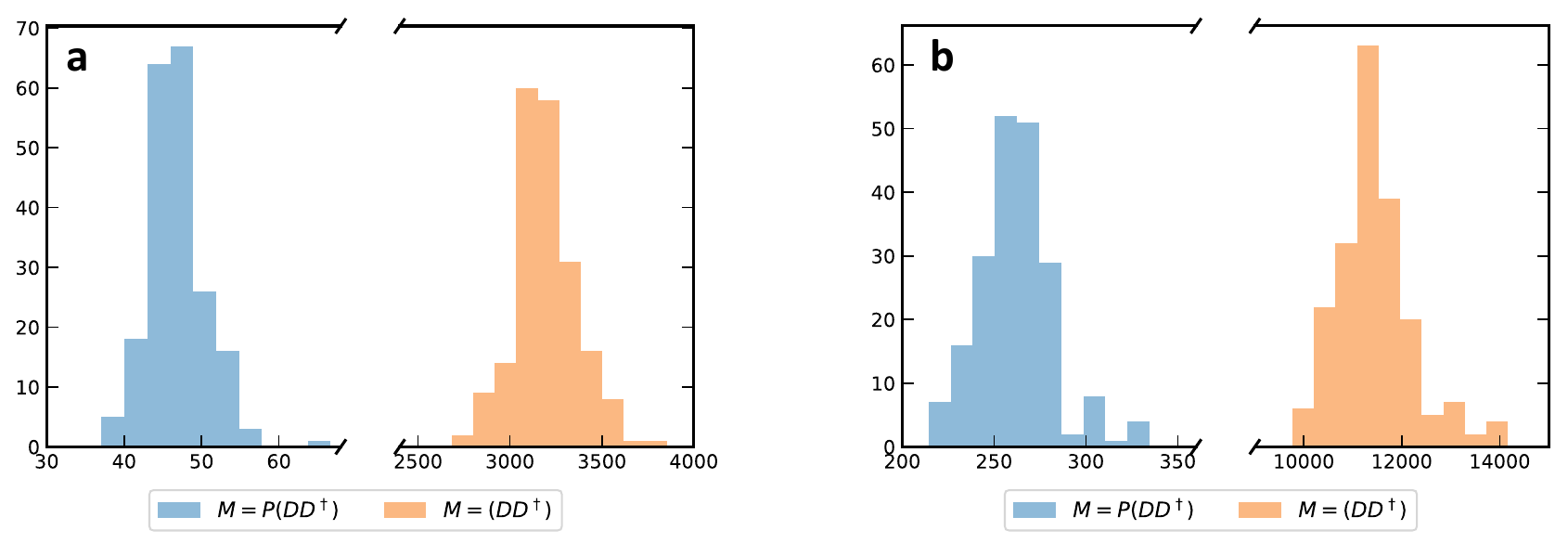}
    \caption{\textbf{Effect of preconditioner on condition numbers.} Histogram of condition numbers of $M=DD^\dagger, PDD^\dagger$, where $D=D(\phi)$ is the fermion matrix and $P$ is the preconditioner, based on samples $\phi$ collected from an HMC run. (a) $N_\tau = 50$. (b) $N_\tau= 100$.}
    \label{fig:cond}
\end{figure*}

Finally, in Supplementary Fig. \ref{fig:trotter} we validate the claim from Methods Section that the advantage of the preconditioner becomes more significant in the limit of small Trotter step $\Delta_\tau$. Although in the rest of this work we only consider $\Delta_\tau = 0.1$, this limit nonetheless provides one way of understanding the advantage of the preconditioner besides the motivation via exactness for a the choice of zero coupling $g=0$.

\begin{figure*}[!htb]
    \centering
    \includegraphics[width=0.9\textwidth]{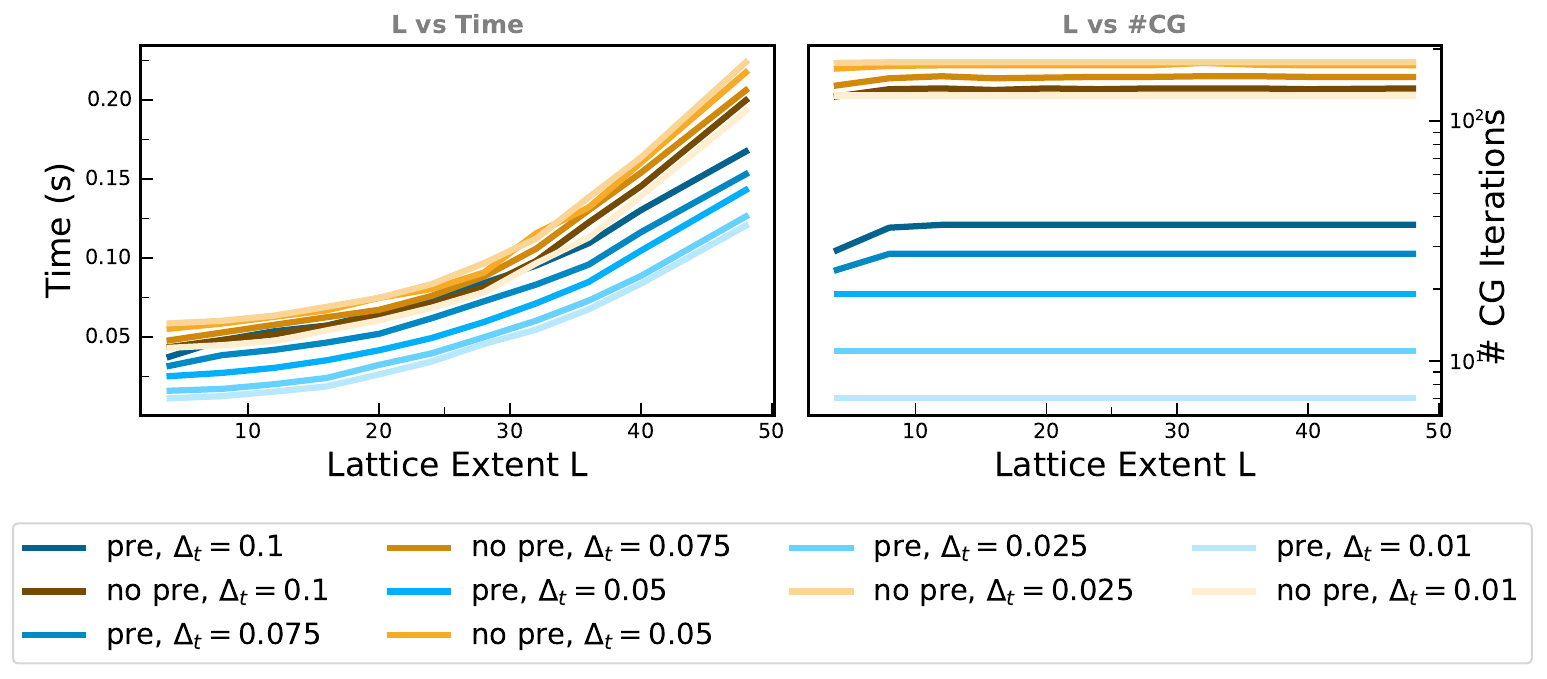}
    \caption{\textbf{Trotter steps and preconditioner performance.} Average preconditioner speedup over an HMC run as a function of lattice extent $L$, where $N_\tau = 50$ is fixed, measured both in terms of wall clock time and number of CG iterations. Lines correspond to various choices of Trotter step size $\Delta_\tau$ with and without precondtioning.}
    \label{fig:trotter}
\end{figure*}


\section*{Supplementary Note 6. Numerical Performance Details}
\label{app:performance details}

Here we provide more details on the numerical performance of our algorithm, expounding on the summary main in the the Methods Section. As a concrete observable, we track the growth of the integrated autocorrelation time $\tau_{\mathrm{int}}$ of the total SDW susceptibility $\chi \equiv \chi(0, \textbf 0)$ at criticality.

\begin{figure*}[!ht]
    \centering
    \includegraphics[width=1.0\textwidth]{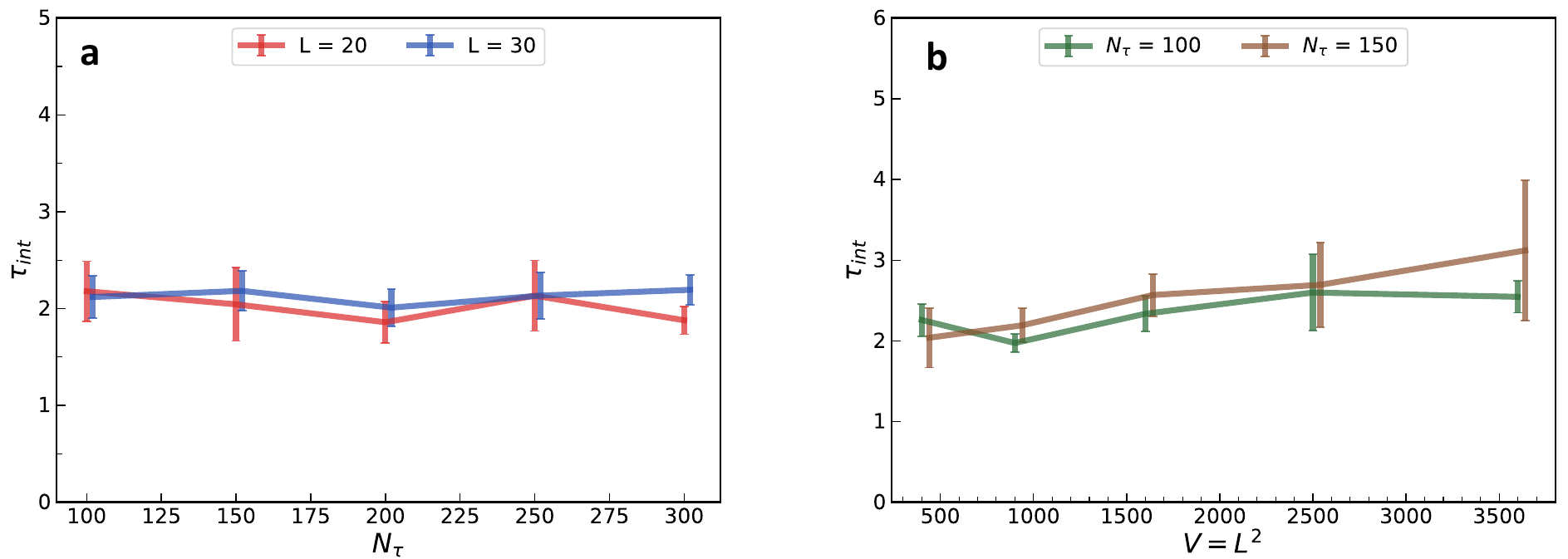}
    \caption{\textbf{Scaling of autocorrelation time.}
    Integrated autocorrelation time $\tau_{\mathrm{int}}$ for the SDW susceptibility $\chi$ under the auto-tuned HMC algorithm. Each value is computed from the average of five Markov chains, and the error bars are the corresponding one sigma deviation. (a) Plot against $N_{\tau}$ for $L=20$ and $L=30$. (b) Plot against $L$ for $N_{\tau}=100$ and $N_{\tau}=150$. In both cases, we observe evidence of constant scaling of $\tau_{\mathrm{int}}$ with respect to system size.}

    \label{fig:iats}
\end{figure*}

To test the performance, we ran the HMC algorithm in two studies: one in which the lattice volume is fixed to sizes $L=20, 30$ and inverse temperature scaled from $N_{\tau} = 100$ to $N_{\tau} = 300$, and one in which the number of imaginary time steps is fixed to $N_{\tau} = 100, 150$ while lattice extent $L$ is scaled from 20 to 60. For each scenario, empirical averages and error bars were computed using 5 Markov chains. In Supplementary Fig. \ref{fig:iats}, we see that our algorithm exhibits constant scaling of $\tau_{\mathrm{int}}$ with respect to both lattice volume and inverse temperature.

The number of leapfrog integration steps per effective sample is plotted in Supplementary Fig. \ref{fig:algcost}. This number roughly tracks the total algorithmic cost in terms of the number of linear solves of the form Eq. (22). A power law fit $y = a x^b$ is applied to the measurements to extract the scaling exponents $z_1, z_2$. With respect to $N_{\tau}$, we observe that $z_1 = b - 1/4 \approx 0.5 $. With respect to $V$, we observe that $z_2 = b - 1/4 \approx 0$. As noted in the main text, this implies an absence of critical slowing down with respect to the lattice volume $V$ for the auto-tuned HMC algorithm presented in this work. 

Finally, in Supplementary Fig. \ref{fig:wallclocks}, we show the wall clock time per effective sample with respect to $V$ and $N_\tau$, where we can see that the scaling is roughly linear with respect to $V$ and superlinear with respect to $N_\tau$.
\begin{figure}[!ht]
    \centering
    \includegraphics[width=1.0\textwidth]{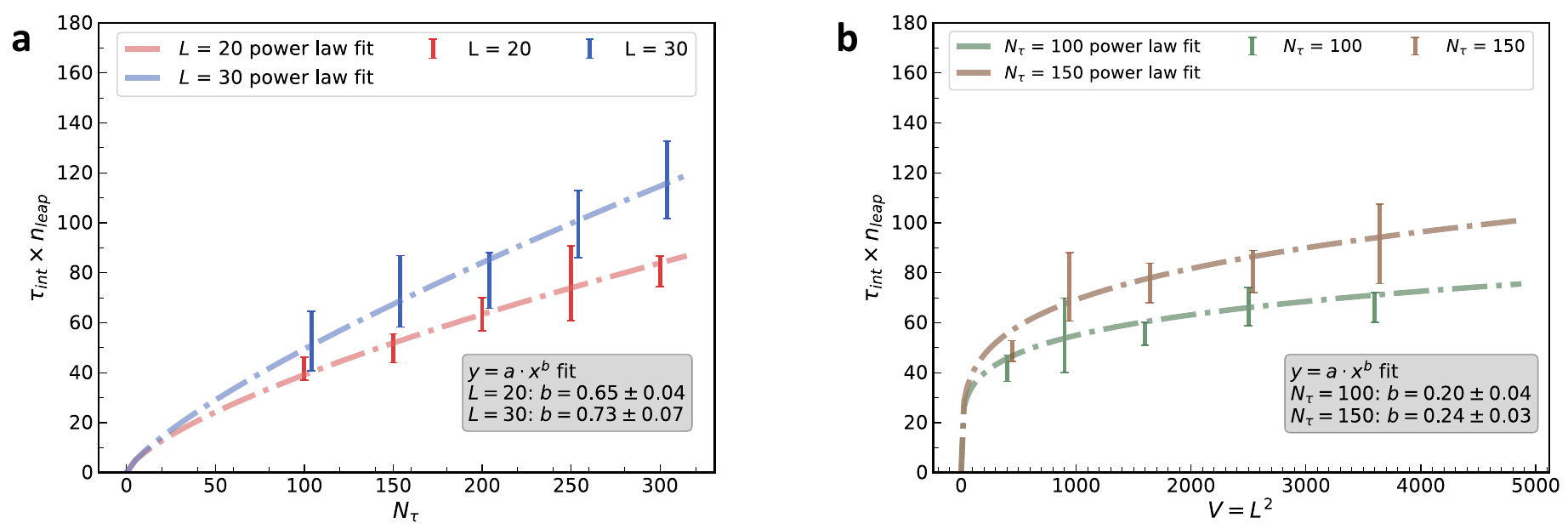}
    \caption{\textbf{Scaling of computational cost.} Benchmarks of integration steps per effective sample, $\tau_{\mathrm{int}} \times n_{\mathrm{leap}}$, for the SDW susceptibility $\chi$. Each value is computed from the average of five Markov chains, and the error bars are the corresponding one sigma deviation. (a) Plot against $N_{\tau}$ for $L=20$ and $L=30$. (b) Plot against $V=L^2$ for $N_{\tau}=100$ and $N_{\tau}=150$. In both cases, a power-law fit is shown.}
    \label{fig:algcost}
\end{figure}

\begin{figure*}[!ht]
    \centering
    \includegraphics[width=1.0\textwidth]{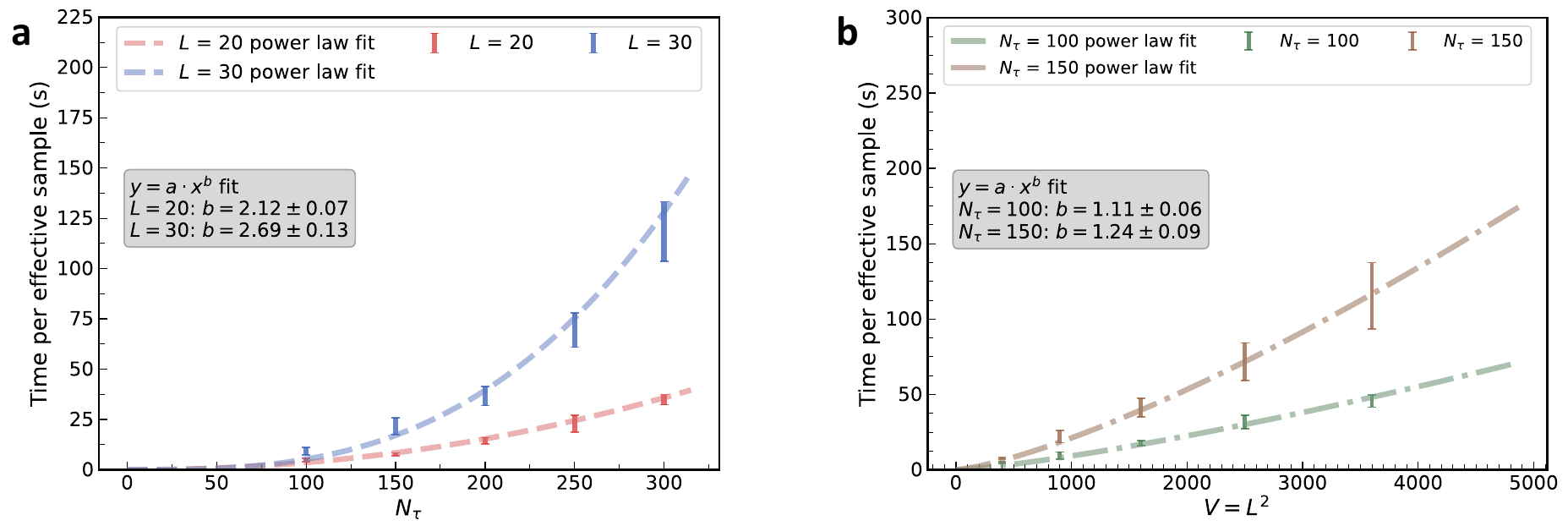}
    \caption{\textbf{Scaling of wallclock time.} Benchmarks of wallclock time per effective sample under the auto-tuned HMC algorithm. Each value is computed from the average of five Markov chains, and the error bars are the corresponding one sigma deviation. (a) Times scaled across $N_{\tau}$ for $L$ values of $L=20$ and $L=30$. (b) Times scaled across $V=L^2$ for $N_{\tau}$ values of $N_{\tau}=100$ and $N_{\tau}=150$. In both cases, a power-law fit is shown.}
    \label{fig:wallclocks}
\end{figure*}



\begin{thebibliography}{10}
\expandafter\ifx\csname url\endcsname\relax
  \def\url#1{\texttt{#1}}\fi
\expandafter\ifx\csname urlprefix\endcsname\relax\def\urlprefix{URL }\fi
\providecommand{\bibinfo}[2]{#2}
\providecommand{\eprint}[2][]{\url{#2}}

\bibitem{sachdev_2011}
\bibinfo{author}{Sachdev, S.}
\newblock \emph{\bibinfo{title}{Quantum Phase Transitions}}
  (\bibinfo{publisher}{Cambridge University Press}, \bibinfo{year}{2011}).

\bibitem{doi:10.1146/annurev-conmatphys-031016-025531}
\bibinfo{author}{Lee, S.-S.}
\newblock \bibinfo{title}{Recent developments in non-fermi liquid theory}.
\newblock \emph{\bibinfo{journal}{Annual Review of Condensed Matter Physics}}
  \textbf{\bibinfo{volume}{9}}, \bibinfo{pages}{227--244}
  (\bibinfo{year}{2018}).
\newblock
  \urlprefix\url{https://doi.org/10.1146/annurev-conmatphys-031016-025531}.

\bibitem{doi:10.1146/annurev-conmatphys-031119-050558}
\bibinfo{author}{Greene, R.~L.}, \bibinfo{author}{Mandal, P.~R.},
  \bibinfo{author}{Poniatowski, N.~R.} \& \bibinfo{author}{Sarkar, T.}
\newblock \bibinfo{title}{The strange metal state of the electron-doped
  cuprates}.
\newblock \emph{\bibinfo{journal}{Annual Review of Condensed Matter Physics}}
  \textbf{\bibinfo{volume}{11}}, \bibinfo{pages}{213--229}
  (\bibinfo{year}{2020}).
\newblock
  \urlprefix\url{https://doi.org/10.1146/annurev-conmatphys-031119-050558}.
\newblock \eprint{https://doi.org/10.1146/annurev-conmatphys-031119-050558}.

\bibitem{Green_Fe_review}
\bibinfo{author}{Paglione, J.} \& \bibinfo{author}{Greene, R.~L.}
\newblock \bibinfo{title}{High-temperature superconductivity in iron-based
  materials}.
\newblock \emph{\bibinfo{journal}{Nature Physics}}
  \textbf{\bibinfo{volume}{6}}, \bibinfo{pages}{645--658}
  (\bibinfo{year}{2010}).
\newblock \urlprefix\url{https://doi.org/10.1038/nphys1759}.

\bibitem{doi:10.1146/annurev-conmatphys-062910-140546}
\bibinfo{author}{Stockert, O.} \& \bibinfo{author}{Steglich, F.}
\newblock \bibinfo{title}{Unconventional quantum criticality in heavy-fermion
  compounds}.
\newblock \emph{\bibinfo{journal}{Annual Review of Condensed Matter Physics}}
  \textbf{\bibinfo{volume}{2}}, \bibinfo{pages}{79--99} (\bibinfo{year}{2011}).
\newblock
  \urlprefix\url{https://doi.org/10.1146/annurev-conmatphys-062910-140546}.
\newblock \eprint{https://doi.org/10.1146/annurev-conmatphys-062910-140546}.

\bibitem{doi:10.1080/0001873021000057123}
\bibinfo{author}{Abanov, A.}, \bibinfo{author}{Chubukov, A.~V.} \&
  \bibinfo{author}{Schmalian, J.}
\newblock \bibinfo{title}{Quantum-critical theory of the spin-fermion model and
  its application to cuprates: Normal state analysis}.
\newblock \emph{\bibinfo{journal}{Advances in Physics}}
  \textbf{\bibinfo{volume}{52}}, \bibinfo{pages}{119--218}
  (\bibinfo{year}{2003}).
\newblock
  \urlprefix\url{http://www.tandfonline.com/doi/abs/10.1080/0001873021000057123}.

\bibitem{PhysRevLett.84.5608}
\bibinfo{author}{Abanov, A.} \& \bibinfo{author}{Chubukov, A.~V.}
\newblock \bibinfo{title}{Spin-fermion model near the quantum critical point:
  One-loop renormalization group results}.
\newblock \emph{\bibinfo{journal}{Phys. Rev. Lett.}}
  \textbf{\bibinfo{volume}{84}}, \bibinfo{pages}{5608--5611}
  (\bibinfo{year}{2000}).
\newblock \urlprefix\url{http://link.aps.org/doi/10.1103/PhysRevLett.84.5608}.

\bibitem{PhysRevLett.93.255702}
\bibinfo{author}{Abanov, A.} \& \bibinfo{author}{Chubukov, A.}
\newblock \bibinfo{title}{Anomalous scaling at the quantum critical point in
  itinerant antiferromagnets}.
\newblock \emph{\bibinfo{journal}{Phys. Rev. Lett.}}
  \textbf{\bibinfo{volume}{93}}, \bibinfo{pages}{255702}
  (\bibinfo{year}{2004}).
\newblock
  \urlprefix\url{http://link.aps.org/doi/10.1103/PhysRevLett.93.255702}.

\bibitem{PhysRevB.82.075128}
\bibinfo{author}{Metlitski, M.~A.} \& \bibinfo{author}{Sachdev, S.}
\newblock \bibinfo{title}{Quantum phase transitions of metals in two spatial
  dimensions. ii. spin density wave order}.
\newblock \emph{\bibinfo{journal}{Phys. Rev. B}} \textbf{\bibinfo{volume}{82}},
  \bibinfo{pages}{075128} (\bibinfo{year}{2010}).
\newblock \urlprefix\url{http://link.aps.org/doi/10.1103/PhysRevB.82.075128}.

\bibitem{PhysRevB.91.125136}
\bibinfo{author}{Sur, S.} \& \bibinfo{author}{Lee, S.-S.}
\newblock \bibinfo{title}{Quasilocal strange metal}.
\newblock \emph{\bibinfo{journal}{Phys. Rev. B}} \textbf{\bibinfo{volume}{91}},
  \bibinfo{pages}{125136} (\bibinfo{year}{2015}).
\newblock \urlprefix\url{http://link.aps.org/doi/10.1103/PhysRevB.91.125136}.

\bibitem{PhysRevB.95.245109}
\bibinfo{author}{Lunts, P.}, \bibinfo{author}{Schlief, A.} \&
  \bibinfo{author}{Lee, S.-S.}
\newblock \bibinfo{title}{Emergence of a control parameter for the
  antiferromagnetic quantum critical metal}.
\newblock \emph{\bibinfo{journal}{Phys. Rev. B}} \textbf{\bibinfo{volume}{95}},
  \bibinfo{pages}{245109} (\bibinfo{year}{2017}).
\newblock \urlprefix\url{https://link.aps.org/doi/10.1103/PhysRevB.95.245109}.

\bibitem{PhysRevX.7.021010}
\bibinfo{author}{Schlief, A.}, \bibinfo{author}{Lunts, P.} \&
  \bibinfo{author}{Lee, S.-S.}
\newblock \bibinfo{title}{Exact critical exponents for the antiferromagnetic
  quantum critical metal in two dimensions}.
\newblock \emph{\bibinfo{journal}{Phys. Rev. X}} \textbf{\bibinfo{volume}{7}},
  \bibinfo{pages}{021010} (\bibinfo{year}{2017}).
\newblock \urlprefix\url{https://link.aps.org/doi/10.1103/PhysRevX.7.021010}.

\bibitem{PhysRevB.100.235104}
\bibinfo{author}{Lunts, P.} \& \bibinfo{author}{Patel, A.~A.}
\newblock \bibinfo{title}{Many-body chaos in the antiferromagnetic quantum
  critical metal}.
\newblock \emph{\bibinfo{journal}{Phys. Rev. B}}
  \textbf{\bibinfo{volume}{100}}, \bibinfo{pages}{235104}
  (\bibinfo{year}{2019}).
\newblock \urlprefix\url{https://link.aps.org/doi/10.1103/PhysRevB.100.235104}.

\bibitem{PhysRevB.14.1165}
\bibinfo{author}{Hertz, J.~A.}
\newblock \bibinfo{title}{Quantum critical phenomena}.
\newblock \emph{\bibinfo{journal}{Phys. Rev. B}} \textbf{\bibinfo{volume}{14}},
  \bibinfo{pages}{1165--1184} (\bibinfo{year}{1976}).
\newblock \urlprefix\url{http://link.aps.org/doi/10.1103/PhysRevB.14.1165}.

\bibitem{PhysRevB.48.7183}
\bibinfo{author}{Millis, A.~J.}
\newblock \bibinfo{title}{Effect of a nonzero temperature on quantum critical
  points in itinerant fermion systems}.
\newblock \emph{\bibinfo{journal}{Phys. Rev. B}} \textbf{\bibinfo{volume}{48}},
  \bibinfo{pages}{7183--7196} (\bibinfo{year}{1993}).
\newblock \urlprefix\url{http://link.aps.org/doi/10.1103/PhysRevB.48.7183}.

\bibitem{Berg21122012}
\bibinfo{author}{Berg, E.}, \bibinfo{author}{Metlitski, M.~A.} \&
  \bibinfo{author}{Sachdev, S.}
\newblock \bibinfo{title}{Sign-problem-free quantum monte carlo of the onset
  of antiferromagnetism in metals}.
\newblock \emph{\bibinfo{journal}{Science}} \textbf{\bibinfo{volume}{338}},
  \bibinfo{pages}{1606--1609} (\bibinfo{year}{2012}).
\newblock
  \urlprefix\url{http://www.sciencemag.org/content/338/6114/1606.abstract}.

\bibitem{PhysRevB.95.035124}
\bibinfo{author}{Gerlach, M.~H.}, \bibinfo{author}{Schattner, Y.},
  \bibinfo{author}{Berg, E.} \& \bibinfo{author}{Trebst, S.}
\newblock \bibinfo{title}{Quantum critical properties of a metallic
  spin-density-wave transition}.
\newblock \emph{\bibinfo{journal}{Phys. Rev. B}} \textbf{\bibinfo{volume}{95}},
  \bibinfo{pages}{035124} (\bibinfo{year}{2017}).
\newblock \urlprefix\url{https://link.aps.org/doi/10.1103/PhysRevB.95.035124}.

\bibitem{PhysRevB.95.174520}
\bibinfo{author}{Wang, X.}, \bibinfo{author}{Schattner, Y.},
  \bibinfo{author}{Berg, E.} \& \bibinfo{author}{Fernandes, R.~M.}
\newblock \bibinfo{title}{Superconductivity mediated by quantum critical
  antiferromagnetic fluctuations: The rise and fall of hot spots}.
\newblock \emph{\bibinfo{journal}{Phys. Rev. B}} \textbf{\bibinfo{volume}{95}},
  \bibinfo{pages}{174520} (\bibinfo{year}{2017}).
\newblock \urlprefix\url{https://link.aps.org/doi/10.1103/PhysRevB.95.174520}.

\bibitem{PhysRevLett.120.247002}
\bibinfo{author}{Wang, X.}, \bibinfo{author}{Wang, Y.},
  \bibinfo{author}{Schattner, Y.}, \bibinfo{author}{Berg, E.} \&
  \bibinfo{author}{Fernandes, R.~M.}
\newblock \bibinfo{title}{Fragility of charge order near an antiferromagnetic
  quantum critical point}.
\newblock \emph{\bibinfo{journal}{Phys. Rev. Lett.}}
  \textbf{\bibinfo{volume}{120}}, \bibinfo{pages}{247002}
  (\bibinfo{year}{2018}).
\newblock
  \urlprefix\url{https://link.aps.org/doi/10.1103/PhysRevLett.120.247002}.

\bibitem{PhysRevLett.117.097002}
\bibinfo{author}{Schattner, Y.}, \bibinfo{author}{Gerlach, M.~H.},
  \bibinfo{author}{Trebst, S.} \& \bibinfo{author}{Berg, E.}
\newblock \bibinfo{title}{Competing orders in a nearly antiferromagnetic
  metal}.
\newblock \emph{\bibinfo{journal}{Phys. Rev. Lett.}}
  \textbf{\bibinfo{volume}{117}}, \bibinfo{pages}{097002}
  (\bibinfo{year}{2016}).
\newblock
  \urlprefix\url{https://link.aps.org/doi/10.1103/PhysRevLett.117.097002}.

\bibitem{PhysRevResearch.2.023008}
\bibinfo{author}{Bauer, C.}, \bibinfo{author}{Schattner, Y.},
  \bibinfo{author}{Trebst, S.} \& \bibinfo{author}{Berg, E.}
\newblock \bibinfo{title}{Hierarchy of energy scales in an o(3) symmetric
  antiferromagnetic quantum critical metal: A monte carlo study}.
\newblock \emph{\bibinfo{journal}{Phys. Rev. Research}}
  \textbf{\bibinfo{volume}{2}}, \bibinfo{pages}{023008} (\bibinfo{year}{2020}).
\newblock
  \urlprefix\url{https://link.aps.org/doi/10.1103/PhysRevResearch.2.023008}.

\bibitem{Li2016925}
\bibinfo{author}{Li, Z.-X.}, \bibinfo{author}{Wang, F.}, \bibinfo{author}{Yao,
  H.} \& \bibinfo{author}{Lee, D.-H.}
\newblock \bibinfo{title}{What makes the tc of monolayer fese on srtio3 so
  high: a sign-problem-free quantum monte carlo study}.
\newblock \emph{\bibinfo{journal}{Science Bulletin}}
  \textbf{\bibinfo{volume}{61}}, \bibinfo{pages}{925 -- 930}
  (\bibinfo{year}{2016}).
\newblock
  \urlprefix\url{//www.sciencedirect.com/science/article/pii/S2095927316300962}.

\bibitem{PhysRevB.95.214505}
\bibinfo{author}{Li, Z.-X.}, \bibinfo{author}{Wang, F.}, \bibinfo{author}{Yao,
  H.} \& \bibinfo{author}{Lee, D.-H.}
\newblock \bibinfo{title}{Nature of the effective interaction in electron-doped
  cuprate superconductors: A sign-problem-free quantum monte carlo study}.
\newblock \emph{\bibinfo{journal}{Phys. Rev. B}} \textbf{\bibinfo{volume}{95}},
  \bibinfo{pages}{214505} (\bibinfo{year}{2017}).
\newblock \urlprefix\url{https://link.aps.org/doi/10.1103/PhysRevB.95.214505}.

\bibitem{doi:10.1146/annurev-conmatphys-031218-013339}
\bibinfo{author}{Berg, E.}, \bibinfo{author}{Lederer, S.},
  \bibinfo{author}{Schattner, Y.} \& \bibinfo{author}{Trebst, S.}
\newblock \bibinfo{title}{Monte carlo studies of quantum critical metals}.
\newblock \emph{\bibinfo{journal}{Annual Review of Condensed Matter Physics}}
  \textbf{\bibinfo{volume}{10}}, \bibinfo{pages}{63--84}
  (\bibinfo{year}{2019}).
\newblock
  \urlprefix\url{https://doi.org/10.1146/annurev-conmatphys-031218-013339}.
\newblock \eprint{https://doi.org/10.1146/annurev-conmatphys-031218-013339}.

\bibitem{doi:10.1146/annurev-conmatphys-033117-054307}
\bibinfo{author}{Li, Z.-X.} \& \bibinfo{author}{Yao, H.}
\newblock \bibinfo{title}{Sign-problem-free fermionic quantum monte carlo:
  Developments and applications}.
\newblock \emph{\bibinfo{journal}{Annual Review of Condensed Matter Physics}}
  \textbf{\bibinfo{volume}{10}}, \bibinfo{pages}{337--356}
  (\bibinfo{year}{2019}).
\newblock
  \urlprefix\url{https://doi.org/10.1146/annurev-conmatphys-033117-054307}.
\newblock \eprint{https://doi.org/10.1146/annurev-conmatphys-033117-054307}.

\bibitem{PhysRevX.6.031028}
\bibinfo{author}{Schattner, Y.}, \bibinfo{author}{Lederer, S.},
  \bibinfo{author}{Kivelson, S.~A.} \& \bibinfo{author}{Berg, E.}
\newblock \bibinfo{title}{Ising nematic quantum critical point in a metal: A
  monte carlo study}.
\newblock \emph{\bibinfo{journal}{Phys. Rev. X}} \textbf{\bibinfo{volume}{6}},
  \bibinfo{pages}{031028} (\bibinfo{year}{2016}).
\newblock \urlprefix\url{http://link.aps.org/doi/10.1103/PhysRevX.6.031028}.

\bibitem{doi:10.1073/pnas.1620651114}
\bibinfo{author}{Lederer, S.}, \bibinfo{author}{Schattner, Y.},
  \bibinfo{author}{Berg, E.} \& \bibinfo{author}{Kivelson, S.~A.}
\newblock \bibinfo{title}{Superconductivity and non-fermi liquid behavior near
  a nematic quantum critical point}.
\newblock \emph{\bibinfo{journal}{Proceedings of the National Academy of
  Sciences}} \textbf{\bibinfo{volume}{114}}, \bibinfo{pages}{4905--4910}
  (\bibinfo{year}{2017}).
\newblock \urlprefix\url{https://www.pnas.org/doi/abs/10.1073/pnas.1620651114}.
\newblock \eprint{https://www.pnas.org/doi/pdf/10.1073/pnas.1620651114}.

\bibitem{PhysRevX.10.031053}
\bibinfo{author}{Klein, A.}, \bibinfo{author}{Chubukov, A.~V.},
  \bibinfo{author}{Schattner, Y.} \& \bibinfo{author}{Berg, E.}
\newblock \bibinfo{title}{Normal state properties of quantum critical metals at
  finite temperature}.
\newblock \emph{\bibinfo{journal}{Phys. Rev. X}} \textbf{\bibinfo{volume}{10}},
  \bibinfo{pages}{031053} (\bibinfo{year}{2020}).
\newblock \urlprefix\url{https://link.aps.org/doi/10.1103/PhysRevX.10.031053}.

\bibitem{PhysRevX.7.031058}
\bibinfo{author}{Xu, X.~Y.}, \bibinfo{author}{Sun, K.},
  \bibinfo{author}{Schattner, Y.}, \bibinfo{author}{Berg, E.} \&
  \bibinfo{author}{Meng, Z.~Y.}
\newblock \bibinfo{title}{Non-fermi liquid at ($2+1$)$\mathrm{D}$ ferromagnetic
  quantum critical point}.
\newblock \emph{\bibinfo{journal}{Phys. Rev. X}} \textbf{\bibinfo{volume}{7}},
  \bibinfo{pages}{031058} (\bibinfo{year}{2017}).
\newblock \urlprefix\url{https://link.aps.org/doi/10.1103/PhysRevX.7.031058}.

\bibitem{QMC_nFL_Yang}
\bibinfo{author}{Xu, X.~Y.}, \bibinfo{author}{Klein, A.}, \bibinfo{author}{Sun,
  K.}, \bibinfo{author}{Chubukov, A.~V.} \& \bibinfo{author}{Meng, Z.~Y.}
\newblock \bibinfo{title}{Identification of non-fermi liquid fermionic
  self-energy from quantum monte carlo data}.
\newblock \emph{\bibinfo{journal}{npj Quantum Materials}}
  \textbf{\bibinfo{volume}{5}}, \bibinfo{pages}{65} (\bibinfo{year}{2020}).
\newblock \urlprefix\url{https://doi.org/10.1038/s41535-020-00266-6}.

\bibitem{doi:10.1073/pnas.1901751116}
\bibinfo{author}{Liu, Z.~H.}, \bibinfo{author}{Pan, G.}, \bibinfo{author}{Xu,
  X.~Y.}, \bibinfo{author}{Sun, K.} \& \bibinfo{author}{Meng, Z.~Y.}
\newblock \bibinfo{title}{Itinerant quantum critical point with fermion pockets
  and hotspots}.
\newblock \emph{\bibinfo{journal}{Proceedings of the National Academy of
  Sciences}} \textbf{\bibinfo{volume}{116}}, \bibinfo{pages}{16760--16767}
  (\bibinfo{year}{2019}).
\newblock \urlprefix\url{https://www.pnas.org/doi/abs/10.1073/pnas.1901751116}.
\newblock \eprint{https://www.pnas.org/doi/pdf/10.1073/pnas.1901751116}.

\bibitem{PhysRevB.105.L041111}
\bibinfo{author}{Liu, Y.} \emph{et~al.}
\newblock \bibinfo{title}{Dynamical exponent of a quantum critical itinerant
  ferromagnet: A monte carlo study}.
\newblock \emph{\bibinfo{journal}{Phys. Rev. B}}
  \textbf{\bibinfo{volume}{105}}, \bibinfo{pages}{L041111}
  (\bibinfo{year}{2022}).
\newblock
  \urlprefix\url{https://link.aps.org/doi/10.1103/PhysRevB.105.L041111}.

\bibitem{doi:10.1021/acs.jctc.8b00996}
\bibinfo{author}{Motta, M.}, \bibinfo{author}{Shee, J.},
  \bibinfo{author}{Zhang, S.} \& \bibinfo{author}{Chan, G. K.-L.}
\newblock \bibinfo{title}{Efficient ab initio auxiliary-field quantum monte
  carlo calculations in gaussian bases via low-rank tensor decomposition}.
\newblock \emph{\bibinfo{journal}{Journal of Chemical Theory and Computation}}
  \textbf{\bibinfo{volume}{15}}, \bibinfo{pages}{3510--3521}
  (\bibinfo{year}{2019}).
\newblock \urlprefix\url{https://doi.org/10.1021/acs.jctc.8b00996}.
\newblock \bibinfo{note}{PMID: 31091103},
  \eprint{https://doi.org/10.1021/acs.jctc.8b00996}.

\bibitem{PhysRevLett.123.136402}
\bibinfo{author}{He, Y.-Y.}, \bibinfo{author}{Shi, H.} \&
  \bibinfo{author}{Zhang, S.}
\newblock \bibinfo{title}{Reaching the continuum limit in finite-temperature ab
  initio field-theory computations in many-fermion systems}.
\newblock \emph{\bibinfo{journal}{Phys. Rev. Lett.}}
  \textbf{\bibinfo{volume}{123}}, \bibinfo{pages}{136402}
  (\bibinfo{year}{2019}).
\newblock
  \urlprefix\url{https://link.aps.org/doi/10.1103/PhysRevLett.123.136402}.

\bibitem{PhysRevB.36.8632}
\bibinfo{author}{Scalettar, R.~T.}, \bibinfo{author}{Scalapino, D.~J.},
  \bibinfo{author}{Sugar, R.~L.} \& \bibinfo{author}{Toussaint, D.}
\newblock \bibinfo{title}{Hybrid molecular-dynamics algorithm for the numerical
  simulation of many-electron systems}.
\newblock \emph{\bibinfo{journal}{Phys. Rev. B}} \textbf{\bibinfo{volume}{36}},
  \bibinfo{pages}{8632--8641} (\bibinfo{year}{1987}).
\newblock \urlprefix\url{https://link.aps.org/doi/10.1103/PhysRevB.36.8632}.

\bibitem{PhysRevB.38.12023}
\bibinfo{author}{Hirsch, J.~E.}
\newblock \bibinfo{title}{Stable monte carlo algorithm for fermion lattice
  systems at low temperatures}.
\newblock \emph{\bibinfo{journal}{Phys. Rev. B}} \textbf{\bibinfo{volume}{38}},
  \bibinfo{pages}{12023--12026} (\bibinfo{year}{1988}).
\newblock \urlprefix\url{https://link.aps.org/doi/10.1103/PhysRevB.38.12023}.

\bibitem{PhysRevB.97.085144}
\bibinfo{author}{Beyl, S.}, \bibinfo{author}{Goth, F.} \&
  \bibinfo{author}{Assaad, F.~F.}
\newblock \bibinfo{title}{Revisiting the hybrid quantum monte carlo method for
  hubbard and electron-phonon models}.
\newblock \emph{\bibinfo{journal}{Phys. Rev. B}} \textbf{\bibinfo{volume}{97}},
  \bibinfo{pages}{085144} (\bibinfo{year}{2018}).
\newblock \urlprefix\url{https://link.aps.org/doi/10.1103/PhysRevB.97.085144}.

\bibitem{PhysRevB.98.235129}
\bibinfo{author}{Buividovich, P.}, \bibinfo{author}{Smith, D.},
  \bibinfo{author}{Ulybyshev, M.} \& \bibinfo{author}{von Smekal, L.}
\newblock \bibinfo{title}{Hybrid monte carlo study of competing order in the
  extended fermionic hubbard model on the hexagonal lattice}.
\newblock \emph{\bibinfo{journal}{Phys. Rev. B}} \textbf{\bibinfo{volume}{98}},
  \bibinfo{pages}{235129} (\bibinfo{year}{2018}).
\newblock \urlprefix\url{https://link.aps.org/doi/10.1103/PhysRevB.98.235129}.

\bibitem{PhysRevB.100.075141}
\bibinfo{author}{Wynen, J.-L.}, \bibinfo{author}{Berkowitz, E.},
  \bibinfo{author}{K\"orber, C.}, \bibinfo{author}{L\"ahde, T.~A.} \&
  \bibinfo{author}{Luu, T.}
\newblock \bibinfo{title}{Avoiding ergodicity problems in lattice
  discretizations of the hubbard model}.
\newblock \emph{\bibinfo{journal}{Phys. Rev. B}}
  \textbf{\bibinfo{volume}{100}}, \bibinfo{pages}{075141}
  (\bibinfo{year}{2019}).
\newblock \urlprefix\url{https://link.aps.org/doi/10.1103/PhysRevB.100.075141}.

\bibitem{KRIEG201915}
\bibinfo{author}{Krieg, S.}, \bibinfo{author}{Luu, T.},
  \bibinfo{author}{Ostmeyer, J.}, \bibinfo{author}{Papaphilippou, P.} \&
  \bibinfo{author}{Urbach, C.}
\newblock \bibinfo{title}{Accelerating hybrid monte carlo simulations of the
  hubbard model on the hexagonal lattice}.
\newblock \emph{\bibinfo{journal}{Computer Physics Communications}}
  \textbf{\bibinfo{volume}{236}}, \bibinfo{pages}{15--25}
  (\bibinfo{year}{2019}).
\newblock
  \urlprefix\url{https://www.sciencedirect.com/science/article/pii/S0010465518303564}.

\bibitem{PhysRevB.102.245105}
\bibinfo{author}{Ostmeyer, J.} \emph{et~al.}
\newblock \bibinfo{title}{Semimetal--mott insulator quantum phase transition of
  the hubbard model on the honeycomb lattice}.
\newblock \emph{\bibinfo{journal}{Phys. Rev. B}}
  \textbf{\bibinfo{volume}{102}}, \bibinfo{pages}{245105}
  (\bibinfo{year}{2020}).
\newblock \urlprefix\url{https://link.aps.org/doi/10.1103/PhysRevB.102.245105}.

\bibitem{PhysRevB.104.155142}
\bibinfo{author}{Ostmeyer, J.} \emph{et~al.}
\newblock \bibinfo{title}{Antiferromagnetic character of the quantum phase
  transition in the hubbard model on the honeycomb lattice}.
\newblock \emph{\bibinfo{journal}{Phys. Rev. B}}
  \textbf{\bibinfo{volume}{104}}, \bibinfo{pages}{155142}
  (\bibinfo{year}{2021}).
\newblock \urlprefix\url{https://link.aps.org/doi/10.1103/PhysRevB.104.155142}.

\bibitem{ULYBYSHEV2019118}
\bibinfo{author}{Ulybyshev, M.}, \bibinfo{author}{Kintscher, N.},
  \bibinfo{author}{Kahl, K.} \& \bibinfo{author}{Buividovich, P.}
\newblock \bibinfo{title}{Schur complement solver for quantum monte-carlo
  simulations of strongly interacting fermions}.
\newblock \emph{\bibinfo{journal}{Computer Physics Communications}}
  \textbf{\bibinfo{volume}{236}}, \bibinfo{pages}{118--127}
  (\bibinfo{year}{2019}).
\newblock
  \urlprefix\url{https://www.sciencedirect.com/science/article/pii/S0010465518303710}.

\bibitem{arxiv.2203.01291}
\bibinfo{author}{Cohen-Stead, B.} \emph{et~al.}
\newblock \bibinfo{title}{Fast and scalable quantum monte carlo simulations of
  electron-phonon models} (\bibinfo{year}{2022}).
\newblock \urlprefix\url{https://arxiv.org/abs/2203.01291}.

\bibitem{PhysRevLett.102.026802}
\bibinfo{author}{Drut, J.~E.} \& \bibinfo{author}{L\"ahde, T.~A.}
\newblock \bibinfo{title}{Is graphene in vacuum an insulator?}
\newblock \emph{\bibinfo{journal}{Phys. Rev. Lett.}}
  \textbf{\bibinfo{volume}{102}}, \bibinfo{pages}{026802}
  (\bibinfo{year}{2009}).
\newblock
  \urlprefix\url{https://link.aps.org/doi/10.1103/PhysRevLett.102.026802}.

\bibitem{PhysRevB.78.165423}
\bibinfo{author}{Hands, S.} \& \bibinfo{author}{Strouthos, C.}
\newblock \bibinfo{title}{Quantum critical behavior in a graphenelike model}.
\newblock \emph{\bibinfo{journal}{Phys. Rev. B}} \textbf{\bibinfo{volume}{78}},
  \bibinfo{pages}{165423} (\bibinfo{year}{2008}).
\newblock \urlprefix\url{https://link.aps.org/doi/10.1103/PhysRevB.78.165423}.

\bibitem{PhysRevLett.111.056801}
\bibinfo{author}{Ulybyshev, M.~V.}, \bibinfo{author}{Buividovich, P.~V.},
  \bibinfo{author}{Katsnelson, M.~I.} \& \bibinfo{author}{Polikarpov, M.~I.}
\newblock \bibinfo{title}{Monte carlo study of the semimetal-insulator phase
  transition in monolayer graphene with a realistic interelectron interaction
  potential}.
\newblock \emph{\bibinfo{journal}{Phys. Rev. Lett.}}
  \textbf{\bibinfo{volume}{111}}, \bibinfo{pages}{056801}
  (\bibinfo{year}{2013}).
\newblock
  \urlprefix\url{https://link.aps.org/doi/10.1103/PhysRevLett.111.056801}.

\bibitem{arxiv.1204.5424}
\bibinfo{author}{Brower, R.}, \bibinfo{author}{Rebbi, C.} \&
  \bibinfo{author}{Schaich, D.}
\newblock \bibinfo{title}{Hybrid monte carlo simulation on the graphene
  hexagonal lattice}  (\bibinfo{year}{2012}).
\newblock \urlprefix\url{https://arxiv.org/abs/1204.5424}.

\bibitem{PhysRevB.89.195429}
\bibinfo{author}{Smith, D.} \& \bibinfo{author}{von Smekal, L.}
\newblock \bibinfo{title}{Monte carlo simulation of the tight-binding model of
  graphene with partially screened coulomb interactions}.
\newblock \emph{\bibinfo{journal}{Phys. Rev. B}} \textbf{\bibinfo{volume}{89}},
  \bibinfo{pages}{195429} (\bibinfo{year}{2014}).
\newblock \urlprefix\url{https://link.aps.org/doi/10.1103/PhysRevB.89.195429}.

\bibitem{NealHMC}
\bibinfo{author}{Neal, R.}
\newblock \bibinfo{title}{Mcmc using hamiltonian dynamics}.
\newblock In \bibinfo{editor}{Brooks, S.}, \bibinfo{editor}{Gelman, A.},
  \bibinfo{editor}{Jones, G.} \& \bibinfo{editor}{Meng, X.-L.} (eds.)
  \emph{\bibinfo{booktitle}{Handbook of Markov Chain Monte Carlo}},
  chap.~\bibinfo{chapter}{5} (\bibinfo{publisher}{Chapman \& Hall / CRC Press},
  \bibinfo{year}{2011}).

\bibitem{10.2307/24308995}
\bibinfo{author}{Pasarica, C.} \& \bibinfo{author}{Gelman, A.}
\newblock \bibinfo{title}{Adaptively scaling the metropolis algorithm using
  expected squared jumped distance}.
\newblock \emph{\bibinfo{journal}{Statistica Sinica}}
  \textbf{\bibinfo{volume}{20}}, \bibinfo{pages}{343--364}
  (\bibinfo{year}{2010}).
\newblock \urlprefix\url{http://www.jstor.org/stable/24308995}.

\bibitem{osti_1430202}
\bibinfo{author}{Carpenter, B.} \emph{et~al.}
\newblock \bibinfo{title}{Stan : A probabilistic programming language}.
\newblock \emph{\bibinfo{journal}{Journal of Statistical Software}}
  \textbf{\bibinfo{volume}{76}} (\bibinfo{year}{2017}).
\newblock \urlprefix\url{https://www.osti.gov/biblio/1430202}.

\bibitem{PhysRevB.98.075140}
\bibinfo{author}{Schlief, A.}, \bibinfo{author}{Lunts, P.} \&
  \bibinfo{author}{Lee, S.-S.}
\newblock \bibinfo{title}{Noncommutativity between the low-energy limit and
  integer dimension limits in the $\ensuremath{\epsilon}$ expansion: A case
  study of the antiferromagnetic quantum critical metal}.
\newblock \emph{\bibinfo{journal}{Phys. Rev. B}} \textbf{\bibinfo{volume}{98}},
  \bibinfo{pages}{075140} (\bibinfo{year}{2018}).
\newblock \urlprefix\url{https://link.aps.org/doi/10.1103/PhysRevB.98.075140}.

\bibitem{Gazit_nature_phys}
\bibinfo{author}{Gazit, S.}, \bibinfo{author}{Randeria, M.} \&
  \bibinfo{author}{Vishwanath, A.}
\newblock \bibinfo{title}{Emergent dirac fermions and broken symmetries in
  confined and deconfined phases of z2 gauge theories}.
\newblock \emph{\bibinfo{journal}{Nature Physics}}
  \textbf{\bibinfo{volume}{13}}, \bibinfo{pages}{484--490}
  (\bibinfo{year}{2017}).
\newblock \urlprefix\url{https://doi.org/10.1038/nphys4028}.

\bibitem{PhysRevX.6.041049}
\bibinfo{author}{Assaad, F.~F.} \& \bibinfo{author}{Grover, T.}
\newblock \bibinfo{title}{Simple fermionic model of deconfined phases and phase
  transitions}.
\newblock \emph{\bibinfo{journal}{Phys. Rev. X}} \textbf{\bibinfo{volume}{6}},
  \bibinfo{pages}{041049} (\bibinfo{year}{2016}).
\newblock \urlprefix\url{https://link.aps.org/doi/10.1103/PhysRevX.6.041049}.

\bibitem{doi:10.1073/pnas.1806338115}
\bibinfo{author}{Gazit, S.}, \bibinfo{author}{Assaad, F.~F.},
  \bibinfo{author}{Sachdev, S.}, \bibinfo{author}{Vishwanath, A.} \&
  \bibinfo{author}{Wang, C.}
\newblock \bibinfo{title}{Confinement transition of z2 gauge theories coupled
  to massless fermions: Emergent quantum chromodynamics and so(5) symmetry}.
\newblock \emph{\bibinfo{journal}{Proceedings of the National Academy of
  Sciences}} \textbf{\bibinfo{volume}{115}}, \bibinfo{pages}{E6987--E6995}
  (\bibinfo{year}{2018}).
\newblock \urlprefix\url{https://www.pnas.org/doi/abs/10.1073/pnas.1806338115}.
\newblock \eprint{https://www.pnas.org/doi/pdf/10.1073/pnas.1806338115}.

\bibitem{PhysRevLett.121.086601}
\bibinfo{author}{Hohenadler, M.} \& \bibinfo{author}{Assaad, F.~F.}
\newblock \bibinfo{title}{Fractionalized metal in a falicov-kimball model}.
\newblock \emph{\bibinfo{journal}{Phys. Rev. Lett.}}
  \textbf{\bibinfo{volume}{121}}, \bibinfo{pages}{086601}
  (\bibinfo{year}{2018}).
\newblock
  \urlprefix\url{https://link.aps.org/doi/10.1103/PhysRevLett.121.086601}.

\bibitem{PhysRevX.9.021022}
\bibinfo{author}{Xu, X.~Y.} \emph{et~al.}
\newblock \bibinfo{title}{Monte carlo study of lattice compact quantum
  electrodynamics with fermionic matter: The parent state of quantum phases}.
\newblock \emph{\bibinfo{journal}{Phys. Rev. X}} \textbf{\bibinfo{volume}{9}},
  \bibinfo{pages}{021022} (\bibinfo{year}{2019}).
\newblock \urlprefix\url{https://link.aps.org/doi/10.1103/PhysRevX.9.021022}.

\bibitem{Chen_2020}
\bibinfo{author}{Chen, C.}, \bibinfo{author}{Xu, X.~Y.}, \bibinfo{author}{Qi,
  Y.} \& \bibinfo{author}{Meng, Z.~Y.}
\newblock \bibinfo{title}{Metal to orthogonal metal transition}.
\newblock \emph{\bibinfo{journal}{Chinese Physics Letters}}
  \textbf{\bibinfo{volume}{37}}, \bibinfo{pages}{047103}
  (\bibinfo{year}{2020}).
\newblock \urlprefix\url{https://doi.org/10.1088/0256-307x/37/4/047103}.

\bibitem{PhysRevX.10.041057}
\bibinfo{author}{Gazit, S.}, \bibinfo{author}{Assaad, F.~F.} \&
  \bibinfo{author}{Sachdev, S.}
\newblock \bibinfo{title}{Fermi surface reconstruction without symmetry
  breaking}.
\newblock \emph{\bibinfo{journal}{Phys. Rev. X}} \textbf{\bibinfo{volume}{10}},
  \bibinfo{pages}{041057} (\bibinfo{year}{2020}).
\newblock \urlprefix\url{https://link.aps.org/doi/10.1103/PhysRevX.10.041057}.

\bibitem{PhysRevB.100.035118}
\bibinfo{author}{Hofmann, J.~S.}, \bibinfo{author}{Assaad, F.~F.} \&
  \bibinfo{author}{Grover, T.}
\newblock \bibinfo{title}{Fractionalized fermi liquid in a frustrated kondo
  lattice model}.
\newblock \emph{\bibinfo{journal}{Phys. Rev. B}}
  \textbf{\bibinfo{volume}{100}}, \bibinfo{pages}{035118}
  (\bibinfo{year}{2019}).
\newblock \urlprefix\url{https://link.aps.org/doi/10.1103/PhysRevB.100.035118}.

\bibitem{PhysRevLett.120.107201}
\bibinfo{author}{Sato, T.}, \bibinfo{author}{Assaad, F.~F.} \&
  \bibinfo{author}{Grover, T.}
\newblock \bibinfo{title}{Quantum monte carlo simulation of frustrated kondo
  lattice models}.
\newblock \emph{\bibinfo{journal}{Phys. Rev. Lett.}}
  \textbf{\bibinfo{volume}{120}}, \bibinfo{pages}{107201}
  (\bibinfo{year}{2018}).
\newblock
  \urlprefix\url{https://link.aps.org/doi/10.1103/PhysRevLett.120.107201}.

\bibitem{PhysRevB.104.L161105}
\bibinfo{author}{Sato, T.}, \bibinfo{author}{Hohenadler, M.},
  \bibinfo{author}{Grover, T.}, \bibinfo{author}{McGreevy, J.} \&
  \bibinfo{author}{Assaad, F.~F.}
\newblock \bibinfo{title}{Topological terms on topological defects: A quantum
  monte carlo study}.
\newblock \emph{\bibinfo{journal}{Phys. Rev. B}}
  \textbf{\bibinfo{volume}{104}}, \bibinfo{pages}{L161105}
  (\bibinfo{year}{2021}).
\newblock
  \urlprefix\url{https://link.aps.org/doi/10.1103/PhysRevB.104.L161105}.

\bibitem{https://doi.org/10.48550/arxiv.2203.04990}
\bibinfo{author}{Patel, A.~A.}, \bibinfo{author}{Guo, H.},
  \bibinfo{author}{Esterlis, I.} \& \bibinfo{author}{Sachdev, S.}
\newblock \bibinfo{title}{Universal $t$-linear resistivity in two-dimensional
  quantum-critical metals from spatially random interactions}
  (\bibinfo{year}{2022}).
\newblock \urlprefix\url{https://arxiv.org/abs/2203.04990}.

\bibitem{PhysRevLett.123.137602}
\bibinfo{author}{Lang, T.~C.} \& \bibinfo{author}{L\"auchli, A.~M.}
\newblock \bibinfo{title}{Quantum monte carlo simulation of the chiral
  heisenberg gross-neveu-yukawa phase transition with a single dirac cone}.
\newblock \emph{\bibinfo{journal}{Phys. Rev. Lett.}}
  \textbf{\bibinfo{volume}{123}}, \bibinfo{pages}{137602}
  (\bibinfo{year}{2019}).
\newblock
  \urlprefix\url{https://link.aps.org/doi/10.1103/PhysRevLett.123.137602}.

\bibitem{https://doi.org/10.48550/arxiv.2204.02994}
\bibinfo{author}{Hofmann, J.~S.}, \bibinfo{author}{Berg, E.} \&
  \bibinfo{author}{Chowdhury, D.}
\newblock \bibinfo{title}{Superconductivity, charge density wave, and
  supersolidity in flat bands with tunable quantum metric}
  (\bibinfo{year}{2022}).
\newblock \urlprefix\url{https://arxiv.org/abs/2204.02994}.

\bibitem{Zhang_2021}
\bibinfo{author}{Zhang, X.}, \bibinfo{author}{Pan, G.}, \bibinfo{author}{Zhang,
  Y.}, \bibinfo{author}{Kang, J.} \& \bibinfo{author}{Meng, Z.~Y.}
\newblock \bibinfo{title}{Momentum space quantum monte carlo on twisted bilayer
  graphene}.
\newblock \emph{\bibinfo{journal}{Chinese Physics Letters}}
  \textbf{\bibinfo{volume}{38}}, \bibinfo{pages}{077305}
  (\bibinfo{year}{2021}).
\newblock \urlprefix\url{https://doi.org/10.1088/0256-307x/38/7/077305}.

\bibitem{PhysRevB.100.115135}
\bibinfo{author}{Fang, S.-C.}, \bibinfo{author}{Liu, G.-K.},
  \bibinfo{author}{Lin, H.-Q.} \& \bibinfo{author}{Huang, Z.-B.}
\newblock \bibinfo{title}{Quantum monte carlo study of magnetic ordering and
  superconducting pairing symmetry in twisted bilayer graphene}.
\newblock \emph{\bibinfo{journal}{Phys. Rev. B}}
  \textbf{\bibinfo{volume}{100}}, \bibinfo{pages}{115135}
  (\bibinfo{year}{2019}).
\newblock \urlprefix\url{https://link.aps.org/doi/10.1103/PhysRevB.100.115135}.

\bibitem{PhysRevX.12.011061}
\bibinfo{author}{Hofmann, J.~S.}, \bibinfo{author}{Khalaf, E.},
  \bibinfo{author}{Vishwanath, A.}, \bibinfo{author}{Berg, E.} \&
  \bibinfo{author}{Lee, J.~Y.}
\newblock \bibinfo{title}{Fermionic monte carlo study of a realistic model of
  twisted bilayer graphene}.
\newblock \emph{\bibinfo{journal}{Phys. Rev. X}} \textbf{\bibinfo{volume}{12}},
  \bibinfo{pages}{011061} (\bibinfo{year}{2022}).
\newblock \urlprefix\url{https://link.aps.org/doi/10.1103/PhysRevX.12.011061}.

\bibitem{brower2012hybrid}
\bibinfo{author}{Brower, R.}, \bibinfo{author}{Rebbi, C.} \&
  \bibinfo{author}{Schaich, D.}
\newblock \bibinfo{title}{Hybrid monte carlo simulation on the graphene
  hexagonal lattice} (\bibinfo{year}{2012}).
\newblock \eprint{1204.5424}.

\bibitem{arxiv.2203.07380}
\bibinfo{author}{Zhang, C.} \emph{et~al.}
\newblock \bibinfo{title}{Bipolaronic high-temperature superconductivity}
  (\bibinfo{year}{2022}).
\newblock \urlprefix\url{https://arxiv.org/abs/2203.07380}.

\bibitem{PhysRevLett.125.121601}
\bibinfo{author}{Kanwar, G.} \emph{et~al.}
\newblock \bibinfo{title}{Equivariant flow-based sampling for lattice gauge
  theory}.
\newblock \emph{\bibinfo{journal}{Phys. Rev. Lett.}}
  \textbf{\bibinfo{volume}{125}}, \bibinfo{pages}{121601}
  (\bibinfo{year}{2020}).
\newblock
  \urlprefix\url{https://link.aps.org/doi/10.1103/PhysRevLett.125.121601}.

\bibitem{Hasenbusch:2001xh}
\bibinfo{author}{Hasenbusch, M.} \& \bibinfo{author}{Jansen, K.}
\newblock \bibinfo{title}{{Speeding up the Hybrid Monte Carlo algorithm for
  dynamical fermions}}.
\newblock \emph{\bibinfo{journal}{Nucl. Phys. B Proc. Suppl.}}
  \textbf{\bibinfo{volume}{106}}, \bibinfo{pages}{1076--1078}
  (\bibinfo{year}{2002}).
\newblock \eprint{hep-lat/0110180}.

\bibitem{clark-multigrid2016}
\bibinfo{author}{Clark, M.~A.} \emph{et~al.}
\newblock \bibinfo{title}{Accelerating lattice qcd multigrid on gpus using
  fine-grained parallelization}.
\newblock In \emph{\bibinfo{booktitle}{Proceedings of the International
  Conference for High Performance Computing, Networking, Storage and
  Analysis}}, SC '16 (\bibinfo{publisher}{IEEE Press}, \bibinfo{year}{2016}).

\bibitem{Clark2010}
\bibinfo{author}{Clark, M.}, \bibinfo{author}{Babich, R.},
  \bibinfo{author}{Barros, K.}, \bibinfo{author}{Brower, R.} \&
  \bibinfo{author}{Rebbi, C.}
\newblock \bibinfo{title}{Solving lattice {QCD} systems of equations using
  mixed precision solvers on {GPUs}}.
\newblock \emph{\bibinfo{journal}{Computer Physics Communications}}
  \textbf{\bibinfo{volume}{181}}, \bibinfo{pages}{1517--1528}
  (\bibinfo{year}{2010}).
\newblock \urlprefix\url{https://doi.org/10.1016%2Fj.cpc.2010.05.002}.

\bibitem{JMLR:v15:hoffman14a}
\bibinfo{author}{Hoffman, M.~D.} \& \bibinfo{author}{Gelman, A.}
\newblock \bibinfo{title}{The no-u-turn sampler: Adaptively setting path
  lengths in hamiltonian monte carlo}.
\newblock \emph{\bibinfo{journal}{Journal of Machine Learning Research}}
  \textbf{\bibinfo{volume}{15}}, \bibinfo{pages}{1593--1623}
  (\bibinfo{year}{2014}).
\newblock \urlprefix\url{http://jmlr.org/papers/v15/hoffman14a.html}.

\bibitem{PhysRevD.24.2278}
\bibinfo{author}{Blankenbecler, R.}, \bibinfo{author}{Scalapino, D.~J.} \&
  \bibinfo{author}{Sugar, R.~L.}
\newblock \bibinfo{title}{Monte carlo calculations of coupled boson-fermion
  systems. i}.
\newblock \emph{\bibinfo{journal}{Phys. Rev. D}} \textbf{\bibinfo{volume}{24}},
  \bibinfo{pages}{2278--2286} (\bibinfo{year}{1981}).
\newblock \urlprefix\url{https://link.aps.org/doi/10.1103/PhysRevD.24.2278}.

\bibitem{GattLang}
\bibinfo{author}{Gattringer, C.} \& \bibinfo{author}{Lang, C.~B.}
\newblock \bibinfo{title}{{Quantum chromodynamics on the lattice}}.
\newblock \emph{\bibinfo{journal}{Lect. Notes Phys.}}
  \textbf{\bibinfo{volume}{788}}, \bibinfo{pages}{1--343}
  (\bibinfo{year}{2010}).

\bibitem{Fucito:1980fh}
\bibinfo{author}{Fucito, F.}, \bibinfo{author}{Marinari, E.},
  \bibinfo{author}{Parisi, G.} \& \bibinfo{author}{Rebbi, C.}
\newblock \bibinfo{title}{{A Proposal for Monte Carlo Simulations of Fermionic
  Systems}}.
\newblock \emph{\bibinfo{journal}{Nucl. Phys. B}}
  \textbf{\bibinfo{volume}{180}}, \bibinfo{pages}{369} (\bibinfo{year}{1981}).

\bibitem{fletcher1964}
\bibinfo{author}{Fletcher, R.} \& \bibinfo{author}{Reeves, C.~M.}
\newblock \bibinfo{title}{{Function minimization by conjugate gradients}}.
\newblock \emph{\bibinfo{journal}{The Computer Journal}}
  \textbf{\bibinfo{volume}{7}}, \bibinfo{pages}{149--154}
  (\bibinfo{year}{1964}).
\newblock \urlprefix\url{https://doi.org/10.1093/comjnl/7.2.149}.
\newblock
  \eprint{https://academic.oup.com/comjnl/article-pdf/7/2/149/959725/070149.pdf}.

\bibitem{doi:10.1137/18M1212458}
\bibinfo{author}{Gergelits, T.}, \bibinfo{author}{Mardal, K.-A.},
  \bibinfo{author}{Nielsen, B.~F.} \& \bibinfo{author}{Strako{\v s}, Z.}
\newblock \bibinfo{title}{Laplacian preconditioning of elliptic pdes:
  Localization of the eigenvalues of the discretized operator}.
\newblock \emph{\bibinfo{journal}{SIAM Journal on Numerical Analysis}}
  \textbf{\bibinfo{volume}{57}}, \bibinfo{pages}{1369--1394}
  (\bibinfo{year}{2019}).

\bibitem{DUANE1987216}
\bibinfo{author}{Duane, S.}, \bibinfo{author}{Kennedy, A.},
  \bibinfo{author}{Pendleton, B.~J.} \& \bibinfo{author}{Roweth, D.}
\newblock \bibinfo{title}{Hybrid monte carlo}.
\newblock \emph{\bibinfo{journal}{Physics Letters B}}
  \textbf{\bibinfo{volume}{195}}, \bibinfo{pages}{216--222}
  (\bibinfo{year}{1987}).
\newblock
  \urlprefix\url{https://www.sciencedirect.com/science/article/pii/037026938791197X}.

\bibitem{luscher-master}
\bibinfo{author}{Lüscher, M.}
\newblock \bibinfo{title}{Stochastic locality and master-field simulations of
  very large lattices}.
\newblock \emph{\bibinfo{journal}{EPJ Web of Conferences}}
  \textbf{\bibinfo{volume}{175}}, \bibinfo{pages}{01002}
  (\bibinfo{year}{2018}).
\newblock \urlprefix\url{http://dx.doi.org/10.1051/epjconf/201817501002}.

\bibitem{hockney2021}
\bibinfo{author}{Hockney, R.} \& \bibinfo{author}{Eastwood, J.}
\newblock \emph{\bibinfo{title}{Computer Simulation Using Particles}}
  (\bibinfo{publisher}{{CRC} Press}, \bibinfo{year}{2021}).
\newblock \urlprefix\url{https://doi.org/10.1201/9780367806934}.

\bibitem{GoodmanWeare:2010:EnsembleMCMC}
\bibinfo{author}{Goodman, J.} \& \bibinfo{author}{Weare, J.}
\newblock \bibinfo{title}{Ensemble samplers with affine invariance}.
\newblock \emph{\bibinfo{journal}{Communications in Applied Mathematic and
  Computational Science}} \textbf{\bibinfo{volume}{5}}, \bibinfo{pages}{65--80}
  (\bibinfo{year}{2010}).

\bibitem{doi:10.1137/19M1304891}
\bibinfo{author}{Garbuno-Inigo, A.}, \bibinfo{author}{N{\"u}sken, N.} \&
  \bibinfo{author}{Reich, S.}
\newblock \bibinfo{title}{Affine invariant interacting langevin dynamics for
  bayesian inference}.
\newblock \emph{\bibinfo{journal}{SIAM Journal on Applied Dynamical Systems}}
  \textbf{\bibinfo{volume}{19}}, \bibinfo{pages}{1633--1658}
  (\bibinfo{year}{2020}).

\bibitem{KENNEDY2001456}
\bibinfo{author}{Kennedy, A.} \& \bibinfo{author}{Pendleton, B.}
\newblock \bibinfo{title}{Cost of the generalised hybrid monte carlo algorithm
  for free field theory}.
\newblock \emph{\bibinfo{journal}{Nuclear Physics B}}
  \textbf{\bibinfo{volume}{607}}, \bibinfo{pages}{456--510}
  (\bibinfo{year}{2001}).
\newblock
  \urlprefix\url{https://www.sciencedirect.com/science/article/pii/S0550321301001298}.

\bibitem{CALVO2021110333}
\bibinfo{author}{Calvo, M.}, \bibinfo{author}{Sanz-Alonso, D.} \&
  \bibinfo{author}{Sanz-Serna, J.}
\newblock \bibinfo{title}{Hmc: Reducing the number of rejections by not using
  leapfrog and some results on the acceptance rate}.
\newblock \emph{\bibinfo{journal}{Journal of Computational Physics}}
  \textbf{\bibinfo{volume}{437}}, \bibinfo{pages}{110333}
  (\bibinfo{year}{2021}).
\newblock
  \urlprefix\url{https://www.sciencedirect.com/science/article/pii/S002199912100228X}.

\bibitem{EstimatorDiagonalSaad}
\bibinfo{author}{Bekas, C.}, \bibinfo{author}{Kokiopoulou, E.} \&
  \bibinfo{author}{Saad, Y.}
\newblock \bibinfo{title}{An estimator for the diagonal of a matrix}.
\newblock \emph{\bibinfo{journal}{Applied Numerical Mathematics}}
  \textbf{\bibinfo{volume}{57}}, \bibinfo{pages}{1214–1229}
  (\bibinfo{year}{2007}).

\bibitem{https://doi.org/10.1002/nla.779}
\bibinfo{author}{Tang, J.~M.} \& \bibinfo{author}{Saad, Y.}
\newblock \bibinfo{title}{A probing method for computing the diagonal of a
  matrix inverse}.
\newblock \emph{\bibinfo{journal}{Numerical Linear Algebra with Applications}}
  \textbf{\bibinfo{volume}{19}}, \bibinfo{pages}{485--501}
  (\bibinfo{year}{2012}).

\bibitem{doi:10.1080/03610919008812866}
\bibinfo{author}{Hutchinson, M.}
\newblock \bibinfo{title}{A stochastic estimator of the trace of the influence
  matrix for laplacian smoothing splines}.
\newblock \emph{\bibinfo{journal}{Communications in Statistics - Simulation and
  Computation}} \textbf{\bibinfo{volume}{19}}, \bibinfo{pages}{433--450}
  (\bibinfo{year}{1990}).

\end{thebibliography}

\begin{thebibliography}{1}
\expandafter\ifx\csname url\endcsname\relax
  \def\url#1{\texttt{#1}}\fi
\expandafter\ifx\csname urlprefix\endcsname\relax\def\urlprefix{URL }\fi
\providecommand{\bibinfo}[2]{#2}
\providecommand{\eprint}[2][]{\url{#2}}

\bibitem{PhysRevB.47.7995}
\bibinfo{author}{Scalapino, D.~J.}, \bibinfo{author}{White, S.~R.} \&
  \bibinfo{author}{Zhang, S.}
\newblock \bibinfo{title}{Insulator, metal, or superconductor: The criteria}.
\newblock \emph{\bibinfo{journal}{Phys. Rev. B}} \textbf{\bibinfo{volume}{47}},
  \bibinfo{pages}{7995--8007} (\bibinfo{year}{1993}).
\newblock \urlprefix\url{https://link.aps.org/doi/10.1103/PhysRevB.47.7995}.

\bibitem{PhysRevLett.117.097002}
\bibinfo{author}{Schattner, Y.}, \bibinfo{author}{Gerlach, M.~H.},
  \bibinfo{author}{Trebst, S.} \& \bibinfo{author}{Berg, E.}
\newblock \bibinfo{title}{Competing orders in a nearly antiferromagnetic
  metal}.
\newblock \emph{\bibinfo{journal}{Phys. Rev. Lett.}}
  \textbf{\bibinfo{volume}{117}}, \bibinfo{pages}{097002}
  (\bibinfo{year}{2016}).
\newblock
  \urlprefix\url{https://link.aps.org/doi/10.1103/PhysRevLett.117.097002}.

\bibitem{PhysRevB.50.14048}
\bibinfo{author}{Altshuler, B.~L.}, \bibinfo{author}{Ioffe, L.~B.} \&
  \bibinfo{author}{Millis, A.~J.}
\newblock \bibinfo{title}{Low-energy properties of fermions with singular
  interactions}.
\newblock \emph{\bibinfo{journal}{Phys. Rev. B}} \textbf{\bibinfo{volume}{50}},
  \bibinfo{pages}{14048--14064} (\bibinfo{year}{1994}).
\newblock \urlprefix\url{http://link.aps.org/doi/10.1103/PhysRevB.50.14048}.

\bibitem{PhysRevB.98.045102}
\bibinfo{author}{Klug, M.~J.}, \bibinfo{author}{Scheurer, M.~S.} \&
  \bibinfo{author}{Schmalian, J.}
\newblock \bibinfo{title}{Hierarchy of information scrambling, thermalization,
  and hydrodynamic flow in graphene}.
\newblock \emph{\bibinfo{journal}{Phys. Rev. B}} \textbf{\bibinfo{volume}{98}},
  \bibinfo{pages}{045102} (\bibinfo{year}{2018}).
\newblock \urlprefix\url{https://link.aps.org/doi/10.1103/PhysRevB.98.045102}.

\end{thebibliography}
\end{document}